\documentclass[aps,twocolumn,superscriptaddress,nobibnotes,showkeys, english]{revtex4-2}

\usepackage{times}
\usepackage{booktabs}
\usepackage[utf8]{inputenc}
\usepackage{amsmath}
\usepackage{amssymb}
\usepackage{graphicx}
\usepackage{bm}
\usepackage{subfigure}
\usepackage{color}
\usepackage[normalem]{ulem} 
\usepackage{dcolumn}
\usepackage{hyperref}
\usepackage{cleveref}
\usepackage{orcidlink}

\newcommand{\degree}{^{\circ}}
\newcolumntype{d}[1]{D{.}{.}{#1}}

\begin{document}

\title{Optical extinction and near-field properties	of plasmonic dimers: \\
	   Role of particle shape and separation}

\author{Gino Wegner   \orcidlink{0000-0001-6225-5269} }
\email[Corresponding author: ]{wegner@physik.hu-berlin.de}
\affiliation{Max-Born-Institut, 12489 Berlin, Germany}
\affiliation{Institut für Physik, Humboldt-Universität zu Berlin, 
	         AG Theoretische Optik \& Photonik, 
	         12489 Berlin, Germany}
\affiliation{Institute of Condensed Matter Theory and Optics,             
	         Friedrich-Schiller-University Jena, Max-Wien-Platz 1,
	         07743 Jena, Germany}

\author{Bill Antonio Bernhardt   \orcidlink{0009-0008-1973-9266}}
\affiliation{Institut für Physik, Humboldt-Universität zu Berlin, 
	         AG Theoretische Optik \& Photonik, 
	         12489 Berlin, Germany}

\author{Ulf Peschel   \orcidlink{0000-0002-3980-9224} }
\affiliation{Institute of Condensed Matter Theory and Optics,             
	         Friedrich-Schiller-University Jena, Max-Wien-Platz 1,
	         07743 Jena, Germany}

\author{Kurt Busch   \orcidlink{0000-0003-0076-8522}}
\affiliation{Institut für Physik, Humboldt-Universität zu Berlin, 
	         AG Theoretische Optik \& Photonik, 
	         12489 Berlin, Germany}
\affiliation{Max-Born-Institut, 12489 Berlin, Germany}


\date{\today}

\begin{abstract}
Localized surface plasmons (LSPs) supported by metallic nanostructures exhibit a great tunability of their resonance frequencies and associated spatial field distributions. In particular, nanoscale gaps and sharp corners exhibit promising near-field enhancements for applications such as biochemical sensing and nanoantennas. In this work, we investigate the optical properties of bowtie nanowire dimers excited by linearly polarized light, including polarization orthogonal to the dimer axis. To distinguish dimer-separation-induced and monomer-shape-induced effects, we compare bowtie, circular-cylindrical dimer and triangular wires. We emphasize the advantages of a B\'ezier-type corner parametrization with tunable curvature and employ the discontinuous-Galerkin time-domain finite-element method for numerical calculations. This enables the identification of hybrid resonances and the systematic analysis of the role of geometrical parameters. For all investigated geometries, we compare the spatially local Drude model with the spatially nonlocal Halevi model. For the specific case of silver, we demonstrate quantitative curvature-dependent LSP line shifts arising from the longitudinal nonlocality captured by the Halevi model. For higher-order LSPs, this nonlocality gives rise to a sequence of resonances that may be exploited for light-harvesting applications. Furthermore, we investigate the field enhancement associated with LSP resonances and, motivated by the occurrence of both hot and cold spots, introduce geometry-dependent measures for assessing the ability of nanostructures to enhance nonlinear optical effects, such as through the optimal positioning of SERS-active molecules. To facilitate the simultaneous study these measures, we propose the use of suitable radar charts.
\end{abstract}


\maketitle

\section{\label{sec:introduction} Introduction}

Continuous advances have been made in both the numerical and analytical investigation of the optical response of metallic nanostructures and in 
the fabrication techniques used to realize them. 
The development of these two areas has proven highly synergistic: theoretical 
and numerical studies provide predictive and interpretative insight, while experimental realizations reveal and validate the underlying physical 
phenomena. Together they lead to numerous applications.
Particular attention has been devoted to plasmons supported by metallic nanostructures, i.e., collective oscillations of the conduction-electron 
density relative to the positively charged ionic background. This is 
especially true for localized surface plasmons (LSPs) in noble metal
nanostructures, where geometric resonances lead to an enhanced coupling 
to external electromagnetic fields notably in the optical frequency range~\cite{maier2007plasmonics}.
Beyond their application in optical antenna devices -— where the focus lies 
on funneling incident radiation into sub-wavelength-localized electromagnetic 
near fields and/or the reverse process~\cite{Zhang2010} -- the LSP-generated 
near fields themselves have attracted considerable interest. In fact, these 
fields can be strongly enhanced, commonly referred to as hot spots (or 
bright spots~\cite{Zito16,Palombo2020}), and this field enhancement 
may be exploited for the amplification of various nonlinear optical 
processes.

One prominent example is surface-enhanced Raman spectroscopy (SERS)~\cite{Yu2020singlemolSERSroadmap}, 
in which plasmonic scatterers enhance the Raman response of nearby molecules. Depending  on the application, this enables either molecular identification 
or ultrasensitive sensing. 
In addition, plasmonically catalyzed inter- and intramolecular reactions~\cite{C4CC09008J} 
mediated by noble-metal nanostructures, such as spherical dimers, can be monitored on the nanoscale~\cite{C8SC04496A}, even down to the few-molecule regime~\cite{C4CC09008J}. 
Although SERS enhancement can, in principle, also arise from charge transfer between the metallic nanostructures and the molecule, thereby modifying the molecular polarizability (chemical mechanism), in this work, we focus entirely 
on the electromagnetic mechanism where the SERS enhancement originates from the increased local field intensity and the associated increase in the number of photons that interact with the molecules. 
Specifically, LSPs sustained by noble-metal nanostructures constitute promising
candidates for achieving such strong electromagnetic enhancements because their resonance frequencies and field intensities can be tuned through variations in geometry and size. For instance, plasmonic near fields have been probed experimentally using nanoscale emitters such as molecules~\cite{Palombo2020,Cang2011} and quantum dots~\cite{Xia2020}. 
Large field enhancements arising from the collective response of multiple plasmon-supporting surfaces have also been demonstrated numerically for both
single and coupled nanostructures~\cite{Toscano_optexp_2012,Teperik2013,Lei_2010,PhysRevLett.105.233901,PhysRevB.86.241110}. 
As prototypical multi-surface systems, we investigate dimer geometries~\cite{Nordlander2004,Zito16}. 
Motivated by the efficient excitation of LSPs on curved and sharply featured surfaces, we compare cylindrical and bowtie homodimers. To approximate high-aspect-ratio rods, we consider infinitely extended homogeneous nanowires. Furthermore, in order to disentangle effects related to dimer separation from those originating purely from the constituent particle shape, we additionally investigate a triangular nanowire monomer. Cylindrical monomers are not discussed separately, as they have already been studied extensively~\cite{wegner2023,RUPPIN2001205,Toscano_optexp_2012}.

Previous studies of LSPs in the quasistatic regime, particularly for cylindrical~\cite{wegner2023} and spherical~\cite{Mortensen2021} monomers as well as cylindrical dimers~\cite{Toscano_optexp_2012}, have demonstrated that realistic modeling of metallic nanostructures must account for nonlocal effects associated with the quantum-statistical nature of the conduction-electron continuum~\cite{wegner2024diss}. Moreover, shape-dependent corrections to the LSP dispersion relation arising from longitudinal nonlocality have been shown to scale with the surface-to-volume ratio. The inter-particle gap size likewise plays a central role, both in determining the validity of the quasistatic approximation and in governing the strength of plasmonic coupling~\cite{Nordlander2004}. We therefore identify the independent geometrical parameters for each class of scatterers and derive their relation to the surface-to-volume ratio. Corner rounding is introduced using B\'{e}zier curves~\cite{Farin_Curves_Surf_for_CAGD}
in a manner that permits independent variation of shape and separation while preserving the overall extent of the corresponding cylindrical reference 
geometry. This enables us to investigate curvature-dependent local and nonlocal effects and motivates a generalized notion of geometry-dependent nonlocality.

In order to describe these effects, we employ the longitudinal Boltzmann--Mermin
model at leading order in nonlocality, resulting in the bulk material description
known as the Halevi model~\cite{PhysRevB.51.7497}. We compare this approach with the classical Drude model~\cite{maier2007plasmonics,bohren2008absorption}. 
Although more general, the Halevi model differs only marginally from the linearized Euler--Drude model~\cite{bloch_bremsvermoegen} for the resonance frequencies and material parameters considered here. More pronounced differences
would arise in scenarios that exploit the additional damping channel contained 
in the Halevi model, for example at frequencies comparable to the Drude damping
rate or in systems coupled to nearby emitters where Ohmic losses become
significant~\cite{PhysRevLett.96.113002,RuppinFlour1982}. Indeed, previous
numerical studies have shown that the Halevi model introduces an additional
contribution to the induced current~\cite{wegner2023}, thereby modifying
dissipative losses.
For a detailed discussion on the current status of nonlocal material models
for plasmonic systems, we refer to a recent roadmap article
\cite{Monticone2025}.

To enforce the boundary conditions on the complex geometries while solving the macroscopic Maxwell and material equations, we employ the discontinuous Galerkin time-domain (DGTD) finite-element method~\cite{lpor_DGTD_review}.
This numerical Maxwell solver features an adaptive mesh that provides an 
accurate representation of curved surfaces and is particularly well suited for continuum material models for the plasmonic system~\cite{wegner2024diss}. In principle, it also permits the investigation of nonlinear optical effects in plasmonic scatterers~\cite{Moeferdt2018,dhuynh2016}. Using this approach, we compute extinction spectra and charge distributions, allowing us to identify the relevant spectral regions and characterize the associated plasmonic resonances. In particular, we corroborate and exploit the hybridization of LSP modes in dimer structures. Furthermore, the method yields direct access to near-field distributions. Building upon previous studies employing averaged field-enhancement measures in selected spatial subdomains (see also \cref{tab:established_field_measures_with_references}), we associate each geometry with a sufficiently large set of such measures and analyze their dependence on the geometrical parameters in order to assess the suitability of 
the structures for SERS applications.

Concerning the constituent separation, we assume sufficiently strong restoring forces that (i) suppress electron spill-out and (ii) prevent tunneling between neighboring particles. Tunneling is known to quench the fields in the gap region by effectively short-circuiting the dimer~\cite{Teperik2013}. Since tunneling effects are typically expected to become relevant only on the scale of a few 
{\AA}ngstroms, we restrict our study to gap sizes no smaller than two nanometers.

Finally, our investigation also contributes to the study of so-called cold spots (or dark spots~\cite{Zito16,Palombo2020}) ~\cite{Vernon2022,Tang2018,Haggui2012,Xia2020}. These are localized regions near a scatterer in which destructive interference causes one or several field components — and in extreme cases the total field magnitude — to nearly vanish. Such regions can exhibit stronger spatial confinement than hot spots~\cite{Haggui2012} and may be spatially manipulated using multiple incident light sources and tailored interference patterns~\cite{Vernon2022}. Interestingly, numerical and experimental studies have demonstrated transitions between hot and cold spots in asymmetric nanorod dimers through changes in the polarization of the incident field~\cite{Xia2020}, as well as around individual nanorods through variations in particle length~\cite{Palombo2020}. In the latter case, suppression of spontaneous emission within a subwavelength region was observed.

The manuscript is structured as follows. After introducing the numerical framework and the material models supplementing the macroscopic Maxwell equations in \cref{sec:numerical_scheme}, we investigate each nanostructure individually with respect to its geometry, extinction spectrum, and near-field properties. Specifically, we discuss the cylindrical dimer in \cref{sec:cyl_dimer}, the triangular monomer in \cref{sec:round_triangle}, and the bowtie dimer in \cref{sec:bowtie}. Finally, conclusions and perspectives for future work are presented in \cref{sec:conclusion_and_outlook}.

\section{\label{sec:numerical_scheme} Numerical method and material models}

In \cref{sec:cyl_dimer,sec:round_triangle,sec:bowtie}, we simulate the optical
response of metallic scatterers using a nodal Discontinuous Galerkin Time-Domain
(DGTD) finite-element method based on Ref.~\cite{hesthaven2007nodal}. In the
spirit of a finite-element method, this numerical scheme enables an accurate
representation of curved geometries through adaptive meshing. We consider an
effectively two-dimensional problem corresponding to the horizontal cross section
of a translationally invariant high-aspect-ratio wire illuminated by an in-plane
polarized plane wave. The computational domain is discretized using a triangular
mesh generated with the open-source finite-element mesh generator
GMSH~\cite{gmsh_article}. An on-the-fly Fourier transform~\cite{lpor_DGTD_review}
provides simultaneous access to temporal and spectral characteristics. Physical
observables are evaluated from field values approximated by a third-order Lagrange
nodal basis. Further details of the numerical implementation are provided in
\cref{suppl_I_sec:numerical_setup}.

The optical response is modeled by the macroscopic Maxwell equations, where the
bulk material is described either by the Drude or the Halevi model, as detailed
in Ref.~\cite{wegner2023}. Accordingly, we focus exclusively on the response of the conduction electrons, neglecting effective-mass corrections, bound-electron contributions, and the commonly employed background dielectric constant $\epsilon_{\rm BG}$. The resulting system of partial differential equations can 
be recast into conservation form (see
Refs.~\cite{hesthaven2007nodal,wegner2023,lpor_DGTD_review,wegner2024diss}), making it particularly well suited for the DGTD framework.
The material parameters are extracted from the experimental refractive-index 
data of Ref.~\cite{PhysRevB.6.4370}, yielding a plasma frequency of 
$\omega_{\rm p,Ag}\approx9.149$ eV, a Drude damping rate of 
$\gamma_{\rm Ag}\approx0.021$ eV, and a Fermi velocity of 
$v_{\rm F,Ag}\approx1.408\times10^{6}\,\mathrm{m/s}$.

The system is completed by the corresponding surface material model. We impose 
a vanishing (infinitesimally thin) surface current sheet, while the Drude model generally gives rise to an (infinitesimally thin) surface charge sheet, both of which enter the standard Maxwell boundary conditions~\cite{Jackson}. For the longitudinally nonlocal Halevi model, we additionally impose the hard-wall boundary 
condition~\cite{resonance_shifts_spill_out_hydro,PhysRevB.91.115416,Raza_2015}, which enforces a vanishing surface-normal bulk current at the interface 
and thereby eliminates the surface charge sheet. This condition still permits 
a finite surface-parallel bulk current extrapolated to the interface, commonly referred to as the slip condition~\cite{PhysRevB.84.121412}. The structures are embedded in vacuum, while the computational domain approximates an open system through a perfectly matched layer together with a Silver--Müller boundary condition~\cite{lpor_DGTD_review}.
\begin{figure}
	\centering 
	\includegraphics[scale=.55]{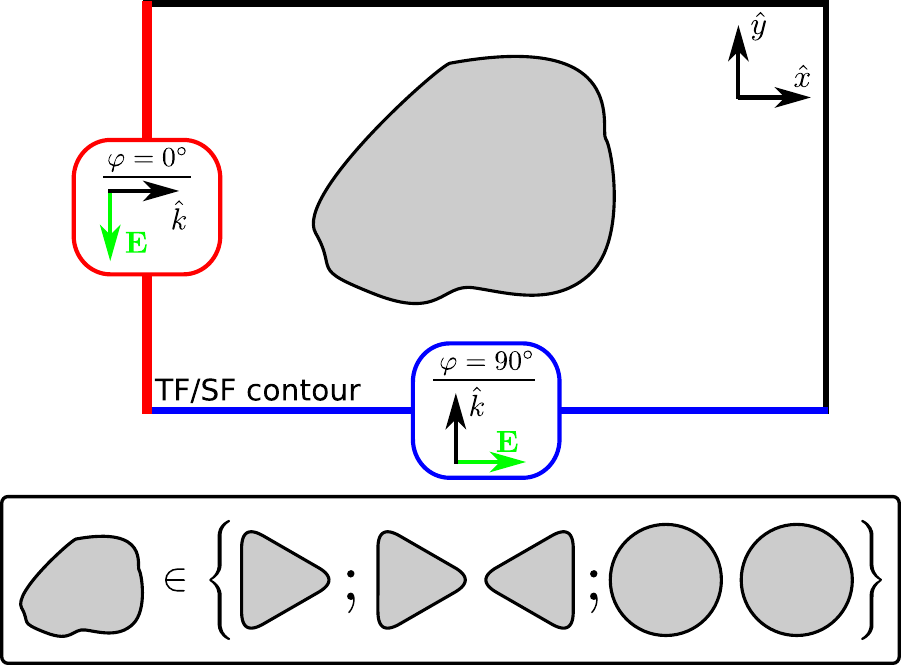}
	\caption{
		\label{fig:incidence directions}
		Schematic of the investigated setup: A plasmonic nanostructure is illuminated with linearly polarized light from different incidence directions characterized by the angle $\varphi\in\{0\degree,90\degree\}$, where $\varphi$ is measured with respect to the $x$-axis.
		As indicated in the bottom row, the nanostructure can be a nanowire 
		with triangular cross section as well as bowtie and cylindrical dimer nanowires. 
	}
\end{figure} 

The structures are excited by linearly polarized plane waves propagating in 
the horizontal plane. Its Gaussian temporal profile allows spectral information
over a finite frequency range to be extracted from a single simulation for a 
given geometry and incidence direction. The incident field is injected using 
the total-field/scattered-field (TF/SF) technique~\cite{lpor_DGTD_review}. We
choose a carrier frequency of $\omega_0=7.439$ eV ($\omega_0/\omega_{\rm p,Ag}\approx0.813$) and a temporal full width at half maximum (FWHM) of 
$1.57$ fs, covering the spectral range of the supported plasmonic resonances. Owing to the symmetry of the investigated structures, we consider two incidence directions. As illustrated in \cref{fig:incidence directions}, these are characterized by the angle $\varphi$, measured counterclockwise from the 
$x$-axis, which coincides with the dimer axis. The total simulation time is 
fixed to $333.56$ fs in all cases to suppress oscillatory artifacts in 
quantities obtained from the on-the-fly Fourier transform. Whenever the 
geometry is unchanged, the same computational mesh is employed for all 
incidence directions and material models. Representative meshes are depicted 
in \cref{suppl_fig:finest_mesh_examples_tri_bowtie_cylinder}.

To characterize the plasmonic resonances, we analyze several spectral 
observables. The simulated extinction efficiency and charge-density 
distributions at selected frequencies are used to identify the excited 
resonance modes and determine their spectral positions. While resonance frequencies are commonly associated with maxima of the extinction spectrum~\cite{bohren2008absorption}, overlapping resonances can shift 
peak positions relative to the true eigenfrequencies, produce 
shoulder-like features, or even conceal nearly degenerate modes. Such
modal superpositions are likewise reflected in the charge-density 
distributions and may require careful analysis of the relative phase 
between the contributing resonances. In the present systems, however, 
the dominant resonance can typically be identified from the real part 
of the charge density, whereas the imaginary part primarily indicates 
the strength of coupling to other resonances.

After identifying the localized surface plasmons (LSPs) and their 
spectral positions, we analyze their near-field distributions. The 
resulting frequency-dependent spatial field patterns enable us to 
classify the excited modes and to determine suitable locations for 
one or more surface-enhanced Raman scattering (SERS)-active molecules
along with the corresponding (electromagnetic) enhancements.


\section{\label{sec:cyl_dimer} Cylindrical dimer}

We consider a cylindrical dimer consisting of two identical, parallel 
cylindrical nanowires characterized by the radius $a$ and the gap size 
$g$.
Since the radius is kept fixed throughout this section, the surface-to-volume ratio remains constant.
The surface-to-volume ratio of cylindrical dimers ($\eta_{\rm cyl, dim}$) 
and monomers  ($\eta_{\rm cyl}$) are identical and given
\begin{align}
	\eta_{\rm cyl\,dim} = \eta_{\rm cyl} = \frac{2}{a}\,,
\end{align}  
while the geometrical cross sections projected onto the plane orthogonal
to the incidence direction (per unit length) read
\begin{align}
	\label{cyl_dim_eq:geom_cross_par_and_perp}
	\sigma_{\rm cyl\,dim}^{0\degree} = 2a 
	\quad \text{and} \quad
	\sigma_{\rm cyl\,dim}^{90\degree} = 4a
\end{align}
for $\varphi=0\degree$ and $\varphi=90\degree$, respectively.


\begin{figure*}
	\centering
	\includegraphics[scale=.5]{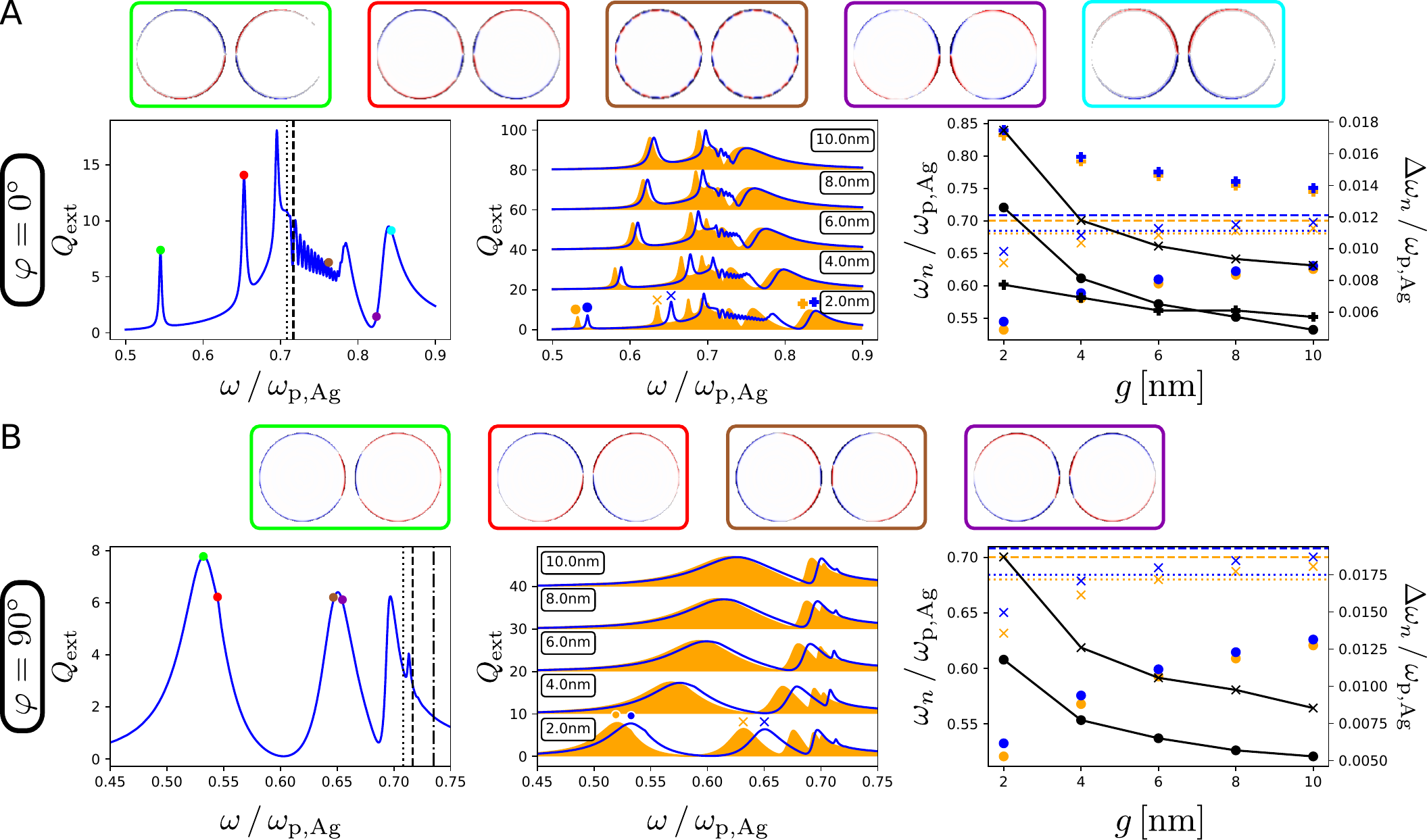}
	\caption{
		\label{cyl_dim_fig:extinction}
		Frequency-dependent extinction efficiency and charge distributions of the hybrid LSPs supported by a cylindrical dimer with radius $a=10$ nm for $\varphi=0\degree$ (Panel A) and $\varphi=90\degree$ (Panel B). The left subpanels show the extinction efficiency for the smallest gap size, $g=2$ nm, calculated with the Halevi model. The vertical lines indicate the resonance frequencies of the corresponding cylindrical monomer: dipolar ($n=1$, dotted), quadrupolar ($n=2$, dashed), and $n=6$ (dash-dotted). Colored markers denote the frequencies at which the charge distributions (shown in the correspondingly colored boxes) are evaluated from the on-the-fly Fourier transform. The displayed distributions correspond to the real part of the charge density, are normalized to their respective maximum absolute value, and are shown using a restricted color scale ranging from $-0.1$ (blue) to $+0.1$ (red) to enhance visibility. The central subpanels present waterfall plots of the extinction efficiency for different gap sizes (indicated by the labeled boxes), comparing the Drude model (filled orange curves) with the Halevi model (blue curves). Successive curves are vertically offset by 10 units for clarity. The right subpanels show the resonance frequencies of selected dipolar and quadrupolar hybrid LSPs as a function of gap size (colored symbols corresponding to the labels in the central subpanels; left ordinate) together with their partial nonlocal blueshifts (black symbols connected by lines; right ordinate).
	}
\end{figure*}

\subsection{\label{subsec:cyl_dim_extinction} Extinction}

In \cref{cyl_dim_fig:extinction}, we present the extinction efficiency for a cylindrical dimer with radius $a=10$ nm for $\varphi=0\degree$ (Panel A) and $\varphi=90\degree$ (Panel B).
The comparatively large radius allows us to distinguish gap-dependent from
radius-dependent nonlocal effects while remaining experimentally feasible.
At the same time, the overall dimensions of the structure remain sufficiently small for the quasistatic approximation to provide a useful qualitative reference, particularly for $\varphi=90\degree$. The left subpanels show the extinction efficiency for the smallest gap considered, $g=2$ nm, within the Halevi model. Compared with the isolated cylinder of the same radius, which exhibits only dipolar and quadrupolar resonances, the dimer supports a considerably richer resonance spectrum. In contrast to the monomer, which interacts only with the incident plane wave, each cylinder additionally experiences the field scattered 
by its partner, which, for an isolated cylinder under normal incidence, is 
given analytically in Ref.~\cite{bohren2008absorption} for a local bulk material model. The global maximum 
of the extinction efficiency differs by approximately a factor of two between
$\varphi=0\degree$ and $\varphi=90\degree$, reflecting the corresponding
difference in the projected geometrical cross section (see
\cref{cyl_dim_eq:geom_cross_par_and_perp}).

The excited resonances occupy a narrower spectral range for $\varphi=90\degree$. This follows from symmetry, which permits only the excitation of bonding hybrid
LSPs and therefore restricts the spectrum to lower frequencies~\cite{Moeferdt2018}.
Consistently, the corresponding charge distributions exhibit only opposite 
charges facing each other across the gap. Using the dipolar and quadrupolar
resonances as representative examples, we find that the hybrid resonances are
bounded by the corresponding monomer resonances. Moreover, the $n=6$ monomer resonance provides an upper estimate for the highest-order dimer LSPs 
that are significantly excited. The monomer resonance frequencies are obtained
from the Mie solution for an isolated cylinder, following the approach outlined 
in Ref.~\cite{wegner2023}. Specifically, we identify the peak of each individual
multipolar contribution $Q_{{\rm ext},n}$ to the extinction efficiency, where
$Q_{\rm ext}=\sum_n Q_{{\rm ext},n}$, thereby avoiding errors arising from
overlapping resonance flanks.
For $\varphi=90\degree$, two distinct bonding dipolar modes and two nearly
degenerate bonding quadrupolar modes can be identified. For the dipolar case, 
the out-of-phase mode (effective dipoles of each cylinder anti-parallel) produces the dominant peak at
$\omega\approx0.5325\,\omega_{\rm p,Ag}$, while the in-phase (effective dipoles of each cylinder parallel) mode appears only 
as a weak shoulder at $\omega\approx0.5447\,\omega_{\rm p,Ag}$. To resolve the
nearly degenerate bonding quadrupoles, we evaluate the charge distributions at
$\omega\approx0.6488\,\omega_{\rm p,Ag}$ and 
$\omega\approx0.6525\,\omega_{\rm p,Ag}$. 
As discussed in Ref.~\cite{Moeferdt2018}, the lifting of this degeneracy
originates from retardation effects.

Consistent with the symmetry arguments of Ref.~\cite{Moeferdt2018}, only the
out-of-phase bonding dipole and a single bonding quadrupole are excited for
$\varphi=0\degree$, at $\omega\approx0.5447\,\omega_{\rm p,Ag}$ and
$\omega\approx0.6528\,\omega_{\rm p,Ag}$, respectively. Apart from an 
arbitrary overall phase, their charge distributions and resonance 
frequencies coincide with those of the higher-frequency dipolar and 
quadrupolar resonances observed for $\varphi=90\degree$, demonstrating 
that they correspond to the same physical modes.

The high-frequency region for $\varphi=0\degree$ is characterized by an 
asymmetric shoulder, indicating the overlap of several antibonding hybrid
resonances. As an example of a well-defined antibonding mode, we include 
the charge distribution of the in-phase antibonding dipole at
$\omega\approx0.8476\,\omega_{\rm p,Ag}$. Its shift relative to the 
corresponding monomer dipole at $\omega\approx0.6843\,\omega_{\rm p,Ag}$ 
exceeds that of its out-of-phase counterpart. Although no frequency could 
be identified at which a purely quadrupolar antibonding mode is isolated, 
the inset at $\omega\approx0.8232\,\omega_{\rm p,Ag}$ predominantly exhibits 
an antibonding quadrupolar distribution. The right cylinder supports an 
almost pure quadrupole, whereas the quadrupole on the left cylinder is superimposed with a dipolar component, most prominently on its left-hand 
side. At still lower frequencies, we also identified mixed 
hexapolar-quadrupolar modes (not shown).

From the identification of the charge distributions with multipolar LSPs, 
we infer opposite trends in the resonance frequency with increasing mode 
order for bonding and antibonding modes. This behavior has also been 
derived analytically in the local quasistatic limit~\cite{Moeferdt2018}. Qualitatively, the convergence of both resonance branches toward the 
central spectral region with increasing mode order can be understood from 
the progressively stronger confinement of the near field to the surface of
each cylinder. Consequently, the attractive and repulsive Coulomb interactions
across the gap, which govern the bonding and antibonding resonances, 
respectively, become weaker. This tendency is also reflected in a 
characteristic feature unique to the $\varphi=0\degree$ excitation, namely 
the ripple structure in the central part of the spectrum, similar to that 
reported in Ref.~\cite{Moeferdt2018}. Each peak of this ripple sequence corresponds to a higher-order hybrid LSP; an exemplary charge distribution 
is shown for $\omega\approx0.7614\,\omega_{\rm p,Ag}$. Our analysis (not 
shown) indicates that these resonances are predominantly antibonding. 
Moreover, the order-dependent nonlocal blueshift of these higher-order 
modes appears to exceed the order-dependent redshift already present in the 
Drude model. We emphasize that these resonances remain surface plasmons. 
For the confined volume plasmons supported by this and the remaining 
structures, together with their charge distributions, we refer to \cref{suppl_I_sec:volume_plasmons} and \cref{suppl_I_fig:LVPs_all_structures_both_phi}.

For completeness, we note that the electrostatic local treatment based on 
the Laplace equation predicts a second class of antibonding modes, including 
the out-of-phase dipolar mode oriented parallel to the dimer axis. These 
modes are, however, optically dark under plane-wave 
excitation~\cite{Moeferdt2018}.

The central subpanels of \cref{cyl_dim_fig:extinction}
show the extinction efficiency for gap sizes $g\in \{2,4,6,8,10\}$ nm. The mesh corresponding to the smallest gap is 
shown in \cref{suppl_fig:finest_mesh_examples_tri_bowtie_cylinder} 
(Panel C). For every gap size, several resonances exhibit a partial 
blueshift in the Halevi model relative to the Drude model, reflecting the influence of longitudinal nonlocality. As expected, the overall direction 
of the resonance shift is opposite for bonding and antibonding modes. 
Among the two excitation geometries, $\varphi=0\degree$ appears more 
favorable for light harvesting because of its broader spectral response 
and larger extinction efficiency~\cite{Aubry_2010_NanoLett}.

The role of the gap size can also be illustrated analytically within 
the quasistatic Drude model. Introducing the geometrical parameter 
$\chi=1+g/2a$ (see also Ref.~\cite{PhysRevLett.105.233901}), the dipolar 
bonding ($\omega_1^{\rm b}$) and antibonding ($\omega_1^{\rm a}$) 
resonance frequencies are given by
\begin{align}
	\omega^{\rm b}_1 
	& = 
	\frac{\omega_{\rm p}}{\sqrt{1 + \chi / \sqrt{\chi^2-1}}}
	\stackrel{g\ll a}{\longrightarrow}
	\frac{\omega_{\rm p}}{\sqrt{1 + \sqrt{a/g}}} \\
	\omega^{\rm a}_1 
	& = 
	\frac{\omega_{\rm p}}{\sqrt{1 + \sqrt{\chi^2-1} / \chi}} \stackrel{g\ll a}{\longrightarrow} \frac{\omega_{\rm p}}{\sqrt{1 + \sqrt{g/a}}}, 
\end{align} 
as follows from Ref.~\cite[Eq.~(16)]{Moeferdt2018} after neglecting Drude 
damping ($\gamma_{\rm Ag}\ll\omega_{\rm p,Ag}$) and employing the identity~\cite[Eq.~(4.6.21)]{abramowitz_stegun} 
\begin{align}
	\operatorname{arccosh}\chi=\ln\!\left(\chi+\sqrt{\chi^2-1}\right),
	\qquad 
	\chi\ge1. 
\end{align} 
In the limit $g\ll a$, the bonding 
and antibonding resonance frequencies are identical apart from a dependence on the ratio $a/g$ and $g/a$, respectively,
consistent with the intuitive picture based on attractive and repulsive 
Coulomb interactions, respectively. Accordingly, the bonding resonance 
approaches $\omega^{\rm b}_1\rightarrow0$, whereas the antibonding resonance 
approaches $\omega^{\rm a}_1\rightarrow\omega_{\rm p}$. Notably, within this 
approximation the antibonding dipolar LSP never enters the volume-plasmon 
regime. 
The increase of the antibonding resonance frequency with decreasing gap size 
also explains the widening of the ripple structure, extending the fixed-geometry
observations reported in Ref.~\cite{Moeferdt2018}. Since this ripple sequence 
is observed only in the nonlocal model, we refer to it as the {\it nonlocal 
peak sequence}. Further evidence for this interpretation is provided 
in \cref{suppl_I_tri_fig:nonlocal_peak_sequence_ext_eff_alpha_scan} 
(Panel B), where the Fermi velocity is gradually reduced for fixed $a=10$ nm 
and $g=2$ nm, thereby continuously weakening the nonlocal response. To 
quantify the influence of longitudinal nonlocality, the right subpanels 
track the partial blueshifts of selected dipolar and quadrupolar hybrid
resonances. In every case, the monomer resonance frequency provides an 
upper bound for bonding modes and a lower bound for antibonding modes. 
For $\varphi=0\degree$, the partial blueshift increases with decreasing gap 
size, with the quadrupolar resonance exhibiting the largest shift. At 
$g=2$ nm, the antibonding dipolar resonance undergoes a smaller blueshift
than either the bonding dipolar or quadrupolar resonance. Furthermore, for 
every gap size considered, the gap-induced blueshift exceeds the corresponding
radius-dependent blueshift of the isolated cylinder. The resonance positions 
of the antibonding modes are, however, subject to larger uncertainties because 
of the pronounced shoulder structure. The lower-frequency bonding dipolar 
and quadrupolar modes excited for $\varphi=90\degree$ exhibit the same 
qualitative trends. Finally, we note that the bonding resonances excited 
for $\varphi=90\degree$ have previously been studied in the quasistatic 
Drude model~\cite{PhysRevLett.105.233901} and, approximately, within the 
linear Euler--Drude model~\cite{PhysRevB.86.241110}. The nonlocal blueshift 
of the bonding resonances can be understood by introducing an effective gap 
equal to the geometrical gap plus the distances of the induced-charge 
centroids from the opposing cylinder surfaces~\cite{Raza15}. In the local 
model, the induced-charge centroids coincide with the physical surfaces. Longitudinal nonlocality spreads the induced charge into the metal, increasing 
the effective gap and thereby reducing the Coulomb attraction between the
cylinders. The resulting increase in resonance energy gives rise to a
gap-dependent blueshift, which adds to the radius-dependent blueshift of 
the individual cylinders, yielding the overall nonlocal resonance shift.

\subsection{\label{subsec:cyl_dim_field} Near-field distributions}

Having identified the significantly excited hybrid LSPs and their spectral 
positions from the extinction spectra, we now investigate the corresponding 
near fields. Our primary objective is to classify the excited modes 
according to their spatial field distributions. Since the field depends 
on two spatial coordinates and frequency, direct visualization rapidly 
becomes difficult. To facilitate the analysis, we introduce the mean-domain 
enhancement. For each frequency, the physical domain is restricted to a 
selected subdomain, over which we evaluate the spatial average of the 
electric-field enhancement,
$\mathcal{E}=|\mathbf{E}| / |\mathbf{E}_0|$,
where $\mathbf{E}$ and $\mathbf{E}_0$ denote the total and incident electric 
fields, respectively. Details of the calculation based on the DGTD simulations 
are provided in \cref{suppl_I_sec:calc:mean_domain_enhancement}, in 
particular in \cref{suppl_I_fig:average_domain_enhancement_calculation} 
(Panel B). To place this measure into context, 
\cref{tab:established_field_measures_with_references} summarizes 
representative studies employing either spatially resolved field 
enhancements or spatial averages over selected domains.

\begin{table}[h]
	\begin{ruledtabular}	
		\begin{tabular}{lcc} 
			domain                      &   spatially resolved                    &  spatial mean   \\
			\hline
			central point of gap        &   \multicolumn{2}{c}{\{\cite{cui2010,Teperik2013,PhysRevB.86.241110,moeferdtdiss,Raza_2015,DeAbajo2008nonloc_metal_plasmons,PhysRevB.71.235420}\}}                     \\
			gap-facing point(s) on surface & \cite{PhysRevLett.105.233901} \{\cite{moeferdtdiss}\}      &                                 \\  
			horizontal gap line   & \cite{Zito16}     &  \cite{Toscano_optexp_2012}         \\
			vertical   gap line         &   \cite{Kottmann_ret_induced_plasmons}  &                    \\       
			whole surface               &                                         &  \cite{Meier83}    \\
			apices                      &   \multicolumn{2}{c}{\{\cite{PhysRevLett.103.097403}\}}    \\
			point(s) displaced from surface & \cite{PhysRevB.71.235420,PhysRevB.64.235402} & \cite{PhysRevB.64.235402}
		\end{tabular}
		\caption{\label{tab:established_field_measures_with_references} 
			Representative spatial domains used to characterize the electric-field 
			amplitude of metallic nanostructures, either through spatially resolved 
			distributions or spatially averaged measures. The gap line refers to 
			either nanoparticle dimers~\cite{Toscano_optexp_2012} or linear 
			nanoparticle chains~\cite{Zito16}. In the cited works, spatial averages 
			are evaluated for powers of the field amplitude that are not 
			necessarily equal to unity. Note that Ref.~\cite{Meier83} considers 
			only a single component of the electric field. For references enclosed 
			in curly braces, the spatial average is identical to the corresponding 
			spatially resolved quantity.
		}
	\end{ruledtabular}	  	
\end{table}

\begin{figure}
	\centering
	\includegraphics[scale=.325]{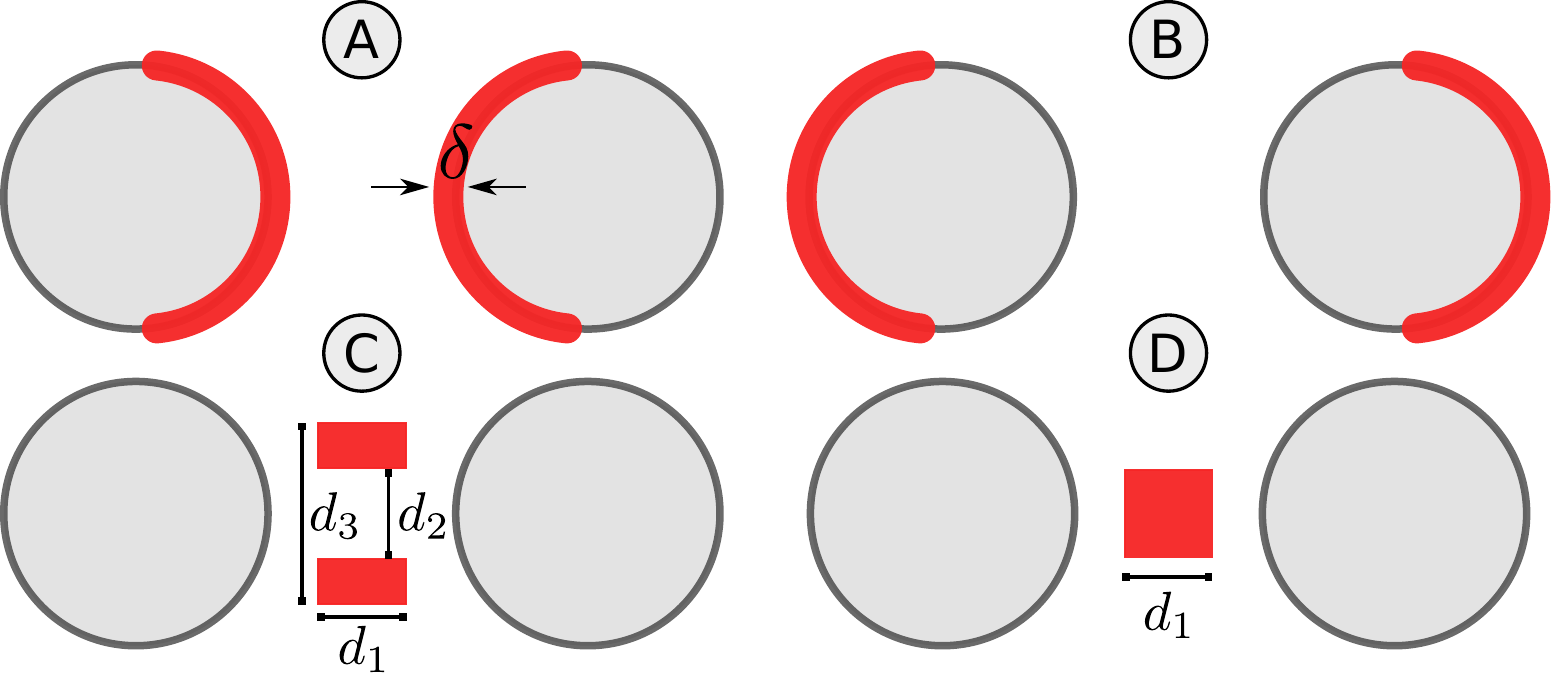}
	\caption{
		\label{cyl_dimer_fig:enh_measures_cyl_dimer_both_inc_dir}
		Depiction of the different domains $\mathcal{D}$ considered for 
		calculating the mean-domain enhancement of the cylindrical dimer. 
		We distinguish between the inner- (A) and outer-surface domains 
		(B), both defined by a sleeve width $\delta$, as well as the 
		gap-peripheral (C) and gap-centered domains (D), characterized 
		by the side lengths $d_1$, $d_2$, and $d_3$. Note that all domains 
		are independent of the gap size.
	}
\end{figure}

Spatial averaging is advantageous because the position of a target 
molecule is generally not known with nanometer
 precision~\cite{Toscano_optexp_2012,Xia2020}. Moreover, molecules 
 may diffuse over the 
 surface~\cite{Yu2020singlemolSERSroadmap,Kanehira2023,Cang2011} or 
 be randomly dispersed within a coating, as considered in 
 Ref.~\cite{Kanehira2023}. Consequently, the choice of averaging domain 
 depends strongly on the geometry under investigation. For the cylindrical 
 dimer, illustrated in \cref{cyl_dimer_fig:enh_measures_cyl_dimer_both_inc_dir}, 
 we distinguish between inner- and outer-surface domains (Panels A and B) 
 as well as gap-centered and gap-peripheral domains (Panels C and D). 
 Throughout this work, the spatial resolution is fixed to a pixel size 
 of $0.2$ nm. For the surface-related measures, this corresponds to a 
 sleeve of width $\delta\approx1.4$ nm, exceeding the characteristic 
 nonlocal length scale $v_{\rm F,Ag}/\omega_{\rm p,Ag}\sim0.1$ nm 
 associated with Thomas--Fermi screening.
 
\begin{figure*}
	\centering
	\includegraphics[scale=.65]{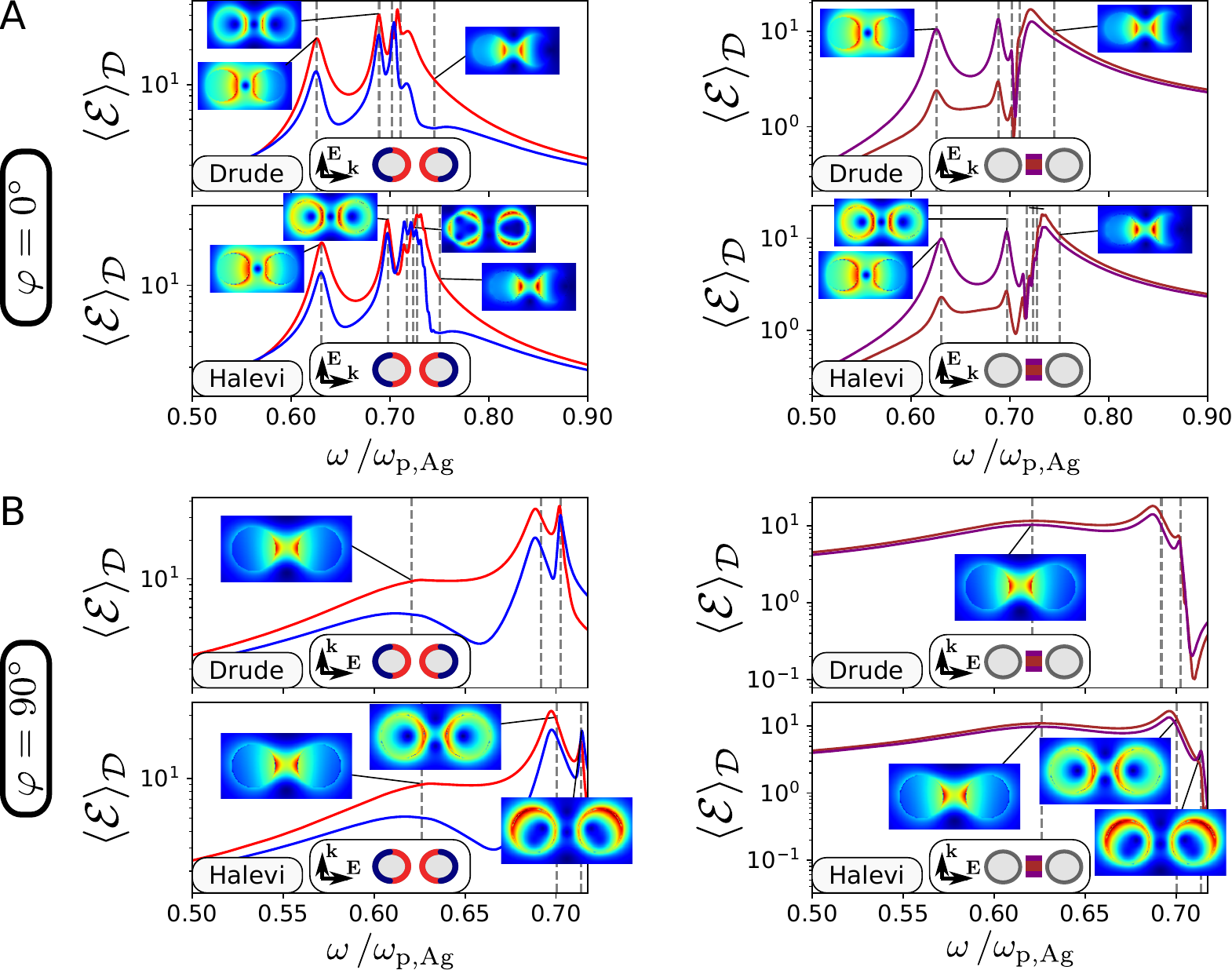}
	\caption{
		\label{cyl_dim_fig:domain_enhancement_measures}
		Mean-domain enhancement of the electric field associated with 
		LSPs sustained by the cylindrical dimer nanowire with $a=10$ nm 
		and $g=10$ nm for $\varphi=0\degree$ (Panel A) and $\varphi=90\degree$ 
		(Panel B). The inner- and outer-surface enhancements (left subpanels), 
		as well as the gap-centered and gap-peripheral enhancements (right 
		subpanels), are compared for the Drude (upper subpanels) and Halevi 
		(lower subpanels) models. The legend and incident field polarization 
		are provided in the lower central box of each subpanel. Vertical 
		dashed lines indicate the peak frequencies of the corresponding 
		extinction efficiency (see also 
		\cref{cyl_dim_fig:extinction}(central subpanels)). 
		Insets show the spatial distribution of the electric field modulus 
		at the respective frequencies, normalized to the maximum value 
		within the displayed spatial domain. Red (blue) indicates the 
		maximum (minimum) value.
	}
\end{figure*}

\Cref{cyl_dim_fig:domain_enhancement_measures} compares the 
surface-related (left subpanels) and gap-related (right subpanels)
mean-domain enhancements for a representative geometry with $a=g=10$ nm. 
Results are shown for $\varphi=0\degree$ (Panel A) and 
$\varphi=90\degree$ (Panel B), using the Drude model (upper subpanels) 
and the Halevi model (lower subpanels). Overall, the maxima of the 
enhancement measures closely follow those of the extinction spectra, 
particularly for the bonding resonances excited at $\varphi=0\degree$. 
Larger deviations occur for the antibonding resonances, whose spectral 
overlap gives rise to an asymmetric shoulder in the extinction 
spectrum. The partial nonlocal blueshift is consistently reproduced 
by all enhancement measures for both incidence directions, while 
enhancement factors exceeding 10 are obtained for both material models.

The surface-related measures reveal a clear dominance of the 
inner-surface enhancement over nearly the entire spectral range for
both incidence directions and both material models. For $\varphi=0\degree$, 
the Drude model exhibits only a single resonance for which the 
outer-surface enhancement becomes dominant. In the Halevi model, this
behavior extends to two resonances associated with representative 
peaks of the ripple structure. For $\varphi=90\degree$, outer-surface 
enhancement dominates only within the central spectral region of the 
Halevi spectrum.

These observations are corroborated by the field distributions shown 
in the insets. The low-order bonding and antibonding hybrid modes are 
strongly localized toward the gap. As the bonding modes evolve from 
dipolar to quadrupolar character, however, the field becomes 
progressively localized away from the gap center. This trend is even
more pronounced in the central spectral region, where the higher-order 
hybrid LSPs are excited. In this regime, stronger confinement of the 
field normal to the metal surface weakens the interaction across the 
gap, resulting in a more uniform charge distribution along the 
cylinder surfaces.

The predominance of the inner-surface enhancement motivates a more 
detailed analysis of the field distribution within the gap. 
For the gap-centered measure we fix $d_1=1$ nm and for the 
gap-peripheral measure we set further $d_2=4$ nm and $d_3=6$ nm.
In 
particular, we distinguish between resonances exhibiting a central
hot spot and those exhibiting a central cold spot. The case of
$\varphi=0\degree$ is especially illustrative. Based on symmetry 
considerations and the charge distributions shown in the insets 
of \cref{cyl_dim_fig:extinction} (Panel A), point-symmetric charge 
distributions are expected below the central spectral region, whereas
mirror-symmetric distributions occur above it. Accordingly, the 
enhancement changes from gap-peripheral dominance at low frequencies 
to gap-centered dominance at higher frequencies.
This behavior is confirmed by the field distributions. The bonding 
out-of-phase dipolar resonance exhibits a pronounced central cold 
spot, whereas the higher-frequency resonance displays a topology 
characterized by a central hot spot. Owing to the finite extent of 
the averaging domains and the fact that the electric field vanishes 
only at a single point in the idealized case, the gap-centered 
enhancement remains finite even for the cold-spot mode. A central 
cold spot has previously been reported for cylindrical dimers in 
Ref.~\cite{Kottmann_ret_induced_plasmons}, where the field 
enhancement was evaluated along the vertical center line of the 
gap.

For $\varphi=0\degree$, the difference between the gap-centered 
and gap-peripheral enhancements is comparable for the bonding 
dipolar and quadrupolar resonances in both the Drude and Halevi 
models. The Halevi model, however, exhibits a stronger vertical 
confinement of the enhanced field inside the gap. Consequently, 
adjusting the geometric parameters $d_2$ and $d_3$ 
(see \cref{cyl_dimer_fig:enh_measures_cyl_dimer_both_inc_dir})
provides an 
additional means of increasing the difference between the two 
gap measures at the quadrupolar resonance.

For $\varphi=90\degree$, the gap-centered enhancement dominates 
over most of the spectrum. In principle, symmetry also permits 
excitation of the out-of-phase bonding dipole associated with 
the central cold spot observed for $\varphi=0\degree$. Here, 
however, its contribution appears too weak to significantly 
modify the overall field distribution. This interpretation is 
consistent with the extinction spectra as a function of gap 
size (see the central subpanel of \cref{cyl_dim_fig:extinction} 
(Panel B)), where the corresponding resonance becomes discernible 
only for sufficiently small gaps. Although the real part of the 
charge density for $g=2$ nm and $\varphi=90\degree$ (see the left subpanel 
of \cref{cyl_dim_fig:extinction} (Panel B)) exhibits a point-symmetric 
distribution, the imaginary part (not shown) reveals that both 
bonding dipolar resonances contribute with comparable amplitudes. 
Their superposition ultimately determines the observed field 
topology.

The Halevi model exhibits one additional high-frequency resonance 
for which the gap-peripheral enhancement exceeds the gap-centered
enhancement, consistent with the central cold spot visible in the 
corresponding field distribution. More generally, and in agreement 
with Ref.~\cite{McMahon_cyl_bowtie_dimer}, the field distributions 
of all modes exhibit a slight asymmetry along the propagation 
direction of the incident wave for both incidence directions.

\begin{figure*}
	\centering
	\includegraphics[scale=.68]{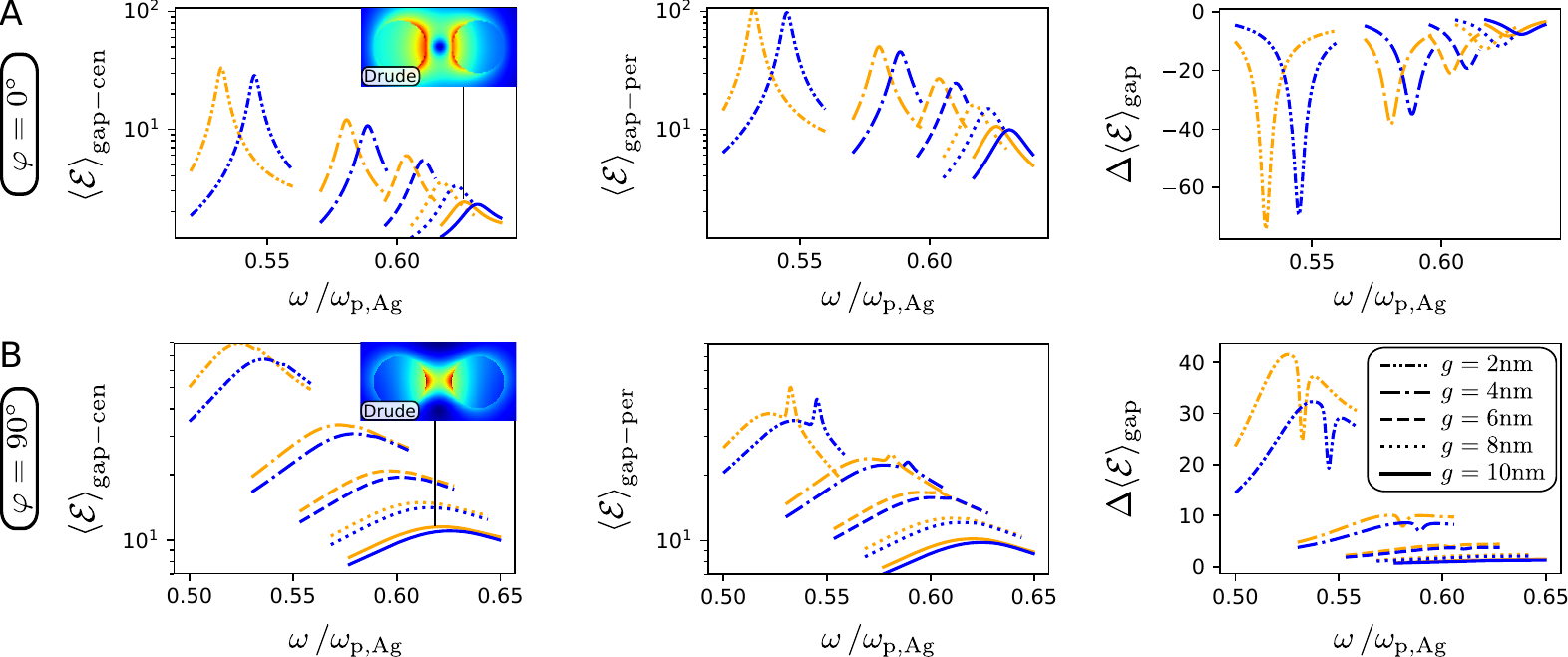}
	\caption{
		\label{cyl_dim_fig:gap_enh_measures_both_phi_gap_scan}
		Gap-size dependence of the mean gap-centered and gap-peripheral 
		enhancements of the bonding dipolar modes sustained by the 
		cylindrical dimer nanowire with $a=10$ nm and $g\in\{10;8;6;4;2\}$ nm 
		(see legend in the right subpanel of Panel B), excited for 
		$\varphi=0\degree$ (out-of-phase; Panel A) and $\varphi=90\degree$ 
		(in-phase; Panel B). Results obtained with the Drude (orange) and 
		Halevi (blue) models are compared. Shown are the gap-centered 
		(left subpanels) and gap-peripheral enhancements (central 
		subpanels), as well as their difference, 
		$\Delta \langle \mathcal{E} \rangle_{\rm gap} = \langle \mathcal{E} \rangle_{\rm gap-cen} - \langle \mathcal{E} \rangle_{\rm gap-per}$ 
		(right subpanels), over the spectral range of the dipolar LSP 
		peak(s). The left subpanels include insets of the field 
		distributions at the peak frequencies of the out-of-phase 
		dipolar LSP (Panel A), exhibiting a gap-centered cold spot, 
		and the in-phase dipolar LSP (Panel B), exhibiting a gap-
		centered hot spot both derived using the Drudel model. Both distributions are shown for $g=10$ nm 
		and the Drude model. Red (blue) denotes the maximum (minimum)
		field amplitude.
	}
\end{figure*}

To conclude the near-field analysis, we investigate the gap-size 
dependence of the central cold- and hot-spot distributions associated 
with the out-of-phase and in-phase bonding dipolar modes, respectively. 
Owing to their relatively low resonance frequencies, these modes are 
well separated from higher-order resonances, minimizing modal 
superposition. In \cref{cyl_dim_fig:gap_enh_measures_both_phi_gap_scan}, 
we compare the gap-centered (left subpanels) and gap-peripheral 
(central subpanels) enhancements, as well as their difference (right 
subpanels), for $\varphi=0\degree$ (Panel A) and $\varphi=90\degree$ 
(Panel B). Results obtained from the Drude and Halevi models are 
shown for each case. The cold-spot mode is excited for $\varphi=0\degree$, 
whereas the hot-spot mode is excited for $\varphi=90\degree$. Reference 
field distributions calculated with the Drude model for $g=10$ nm are 
provided in the insets of the left subpanels.

In addition to the partial blueshift at fixed gap size and the overall 
redshift upon decreasing the gap size, longitudinal nonlocality reduces 
the field enhancement for all investigated gap sizes and suppresses 
the difference between the two gap measures. For $\varphi=0\degree$, 
the gap-peripheral enhancement dominates for all gap sizes and 
increases as the gap narrows, leading to an increasing difference 
between the two measures. Nevertheless, for the smallest gap 
considered ($g=2$ nm), the gap-centered enhancement exceeds the 
gap-peripheral enhancement obtained for $g=8$ nm. At $g=2$ nm, both 
the Drude and Halevi models yield gap-peripheral enhancements of 
approximately two orders of magnitude, reaching enhancement levels 
relevant for single-molecule SERS.

For $\varphi=90\degree$, the gap-centered enhancement dominates, and 
the difference between the gap measures likewise increases with 
decreasing gap size. At $g=2$ nm, the maximum gap-centered enhancement 
reaches approximately 80 and 68 for the Drude and Halevi models, 
respectively. These values are still lower than the gap-peripheral 
enhancement obtained for the cold-spot mode excited at $\varphi=0\degree$. 
Furthermore, the excitation of both bonding dipolar resonances for 
$\varphi=90\degree$ results in two distinguishable peaks for 
$g\leq6$ nm, most prominently in the gap-peripheral enhancement. 
However, the larger magnitude of the gap-centered enhancement indicates 
that the overall field distribution remains dominated by the central 
hot spot. The difference between the two measures retains the 
double-peak structure, reflecting the contribution of both 
resonances.

Finally, our extinction and near-field analysis of the cylindrical 
dimer extends the results of Ref.~\cite{Kottmann01}, which employed 
tabulated frequency-dependent dielectric data and considered a larger 
radius of 50 nm. In contrast, the present study covers a broader 
spectral range and introduces spatially averaged enhancement measures 
to characterize the near-field response.

\section{\label{sec:round_triangle} Rounded triangular nanowire}

\subsection{\label{subsec:tri_construction} Geometry}

Before moving on to the bowtie dimer, we first consider its constituent 
structure, namely the triangular nanowire. We restrict ourselves to 
equilateral triangles to reduce geometrical complexity. Numerical 
simulations based on a local material model have shown that sharp 
corners can induce field singularities, a feature that can be removed 
by introducing longitudinal nonlocality~\cite{Mortensen2021}. The 
occurrence of such singularities is further supported by analytical 
treatments of the quasistatic and local response of an infinite wedge, 
where criteria for singular behavior have been derived in terms of 
problematic frequency ranges depending on the opening 
angle~\cite{surf_plas_and_singularities}. In particular, avoiding 
these frequency ranges is impractical for time-domain methods. 
Moreover, inspection of fabricated structures reveals that corners 
are typically rounded, see, e.g., Refs.~\cite{Liebig2020,Gu2011} (Ref.~\cite{Awada2012}) for TEM (SEM) images of gold triangular
nanoplates and Ref.~\cite{Nelayah2009} for a TEM image of a silver 
triangular nanoplate. In numerical studies, corners are frequently 
rounded using circular-segment parametrizations~\cite{PhysRevB.64.235402,Kottmann_2000,Wallen2008,cui2010,Toscano_optexp_2012}.

Here, we employ second-degree B\'{e}zier curves~\cite{Farin_Curves_Surf_for_CAGD} 
to smooth the sharp corners of an equilateral triangle, which is initially
circumscribed by a circle of radius $\tilde{a}$. 
\begin{figure}[h]
	\centering
    \includegraphics[scale=0.25]{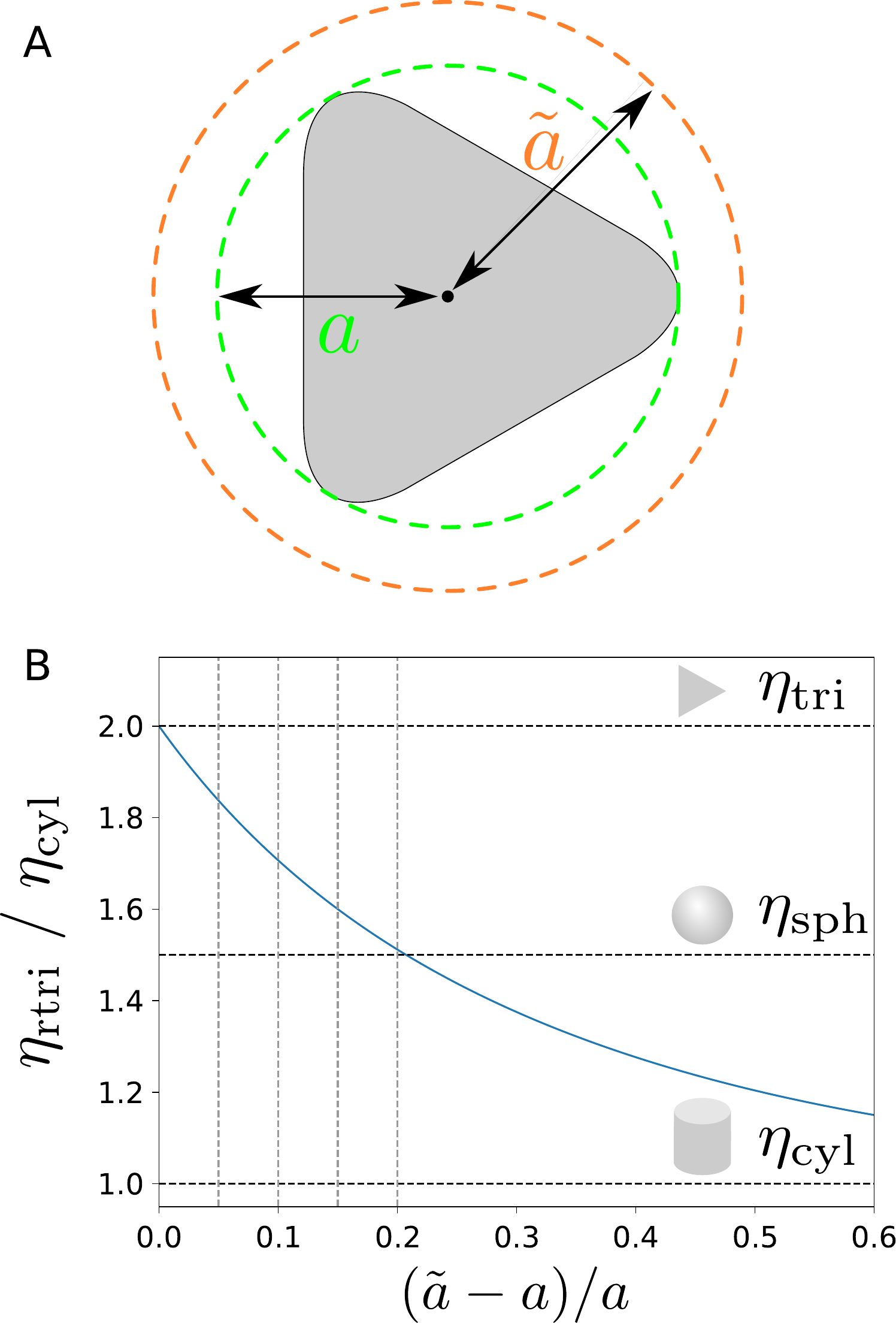}
    \caption{
    	\label{fig:rtri_geom_properties}
    	Geometrical properties of the B\'{e}zier-rounded triangular cross section.
   		Panel A: Construction of the rounded triangle via symmetry-preserving
    	curvature smoothing of an equilateral triangle (circumscribed by a 
    	circle of radius $\tilde{a}$). The resulting structure is enclosed 
    	by a circular reference boundary of radius $a$.
    	Panel B: Surface-to-volume ratio of the rounded triangle, normalized 
    	by the cylindrical value. Horizontal black dashed lines indicate 
    	the corresponding values for a cylinder ($\eta_{\rm cyl}$), sphere 
    	($\eta_{\rm sph}$), and sharp equilateral triangle ($\eta_{\rm tri}$).
    	Vertical gray dashed lines mark the relative radial differences of 
    	the structures investigated numerically in 
    	\cref{subsec:tri_extinction,subsec:tri_fields}.
    }
\end{figure}
The geometrical construction 
is shown in \cref{fig:rtri_geom_properties}(A) and described in more detail 
in \cref{suppl_I_sec:tri_geom}. The selected parametrization represents a parabola applied to each corner 
while preserving the threefold symmetry of the triangle. Furthermore, it 
enables a fully analytical derivation of the surface-to-volume ratio and 
avoids possible kinks at the transition between the curved corner segments 
and the remaining straight sides. Such kinks would otherwise introduce 
additional field enhancement through the lightning-rod 
effect~\cite{maier2007plasmonics,cui2010}, potentially shifting strong fields 
away from the corner apex (or from the gap region in the bowtie dimer). 
Similar to circular rounding, the transition point between the corner segment 
and the straight sides can be precisely controlled.

We enforce that the rounded triangle is circumscribed by a reference cylinder 
of radius $a$. The central geometrical quantities, such as the radius of 
curvature at the corner tip 
\begin{align}
	R = \frac{2}{3} (\tilde{a} - a)
	\label{tri_eq:radius_of_curv}
\end{align} 
and the surface-to-volume ratio, 
\begin{align}
	\eta_\text{rtri} 
	= 
	2 \eta_\text{cyl}  
	\frac{1 + \left[2 \sinh^{-1}\left(\sqrt{3}\right) - \sqrt{3} \right] \nu /3\sqrt{3}}{1 + 2 \nu - 7 \nu^2/9}
	\label{eq:tri_surf_to_vol}
\end{align}
with $\nu = (\tilde{a} - a) / a$
can therefore both be tuned through the radial difference $\tilde{a}-a$, even 
for a fixed reference cylinder (i.e., fixed radius $a$). Inspection of \cref{fig:rtri_geom_properties}(Panel B) reveals 
that the surface-to-volume ratio of the rounded triangle always exceeds that 
of the cylindrical reference. It eventually even surpasses the spherical 
ratio. The sharp equilateral triangle limit ($\tilde{a}\to a^+$) yields 
the maximum surface-to-volume ratio.

As a last geometrical feature, we consider the geometrical cross section defined within the plane normal to the incident wave vector. For $\varphi=0\degree$, we find (per unit length)
\begin{align}
	\sigma^{0\degree}_{\rm rtri} = \left( \sqrt{3}a + \frac{\tilde{a} -a }{3\sqrt{3}} \right)  \,,
	\label{eq:tri_geom_cross_sec_parallel}
\end{align}
while for $\varphi=90\degree$, we obtain
\begin{align}
	\sigma^{90\degree}_{\rm rtri} = \left(\frac{3a}{2} + \frac{\tilde{a} -a}{2} \right) \,.
	\label{eq:tri_geom_cross_sec_perp}   
\end{align}
In both cases, the first term corresponds to the sharp-corner limit. We note 
that both cross sections increase with the radial difference defining the
triangle, $\tilde{a}-a$ (where $\tilde{a}-a < 3a/5$), while remaining smaller than 
the geometrical 
cross section of the cylindrical monomer. 

We deliberately construct the rounded triangle from a circumscribed, rather 
than an inscribed, reference cylinder of fixed radius $a$, anticipating its 
use as the constituent of the bowtie dimer. In fact, this choice keeps the 
gap size constant when varying $\tilde{a}$ and therefore cleanly distiguishes 
shape variations from separation effects. Moreover, cylindrical and triangular nanowires occupy comparable space when substituted for one another in 
plasmonic circuits. Finally, for a given corner curvature, the inscribed triangular nanowire exhibits a smaller surface-to-volume ratio (see \cref{suppl_fig:geom_triangle} (Panel D)).
Plasmonic properties of other threefold-symmetric geometries have also been investigated, for example, the three-arm starfish nanodisk of Ref.~\cite{Babaei2018}, which is based on the Gielis superformula \cite{Gielis_parametrization_2003}. Before proceeding, we briefly note that B\'{e}zier curves can likewise be employed to round other nanostructures, 
such as nanocubes~\cite{Haggui2012} or metallic tips above planar 
substrates.

\begin{figure*}
	\centering
	\includegraphics[scale=.475]{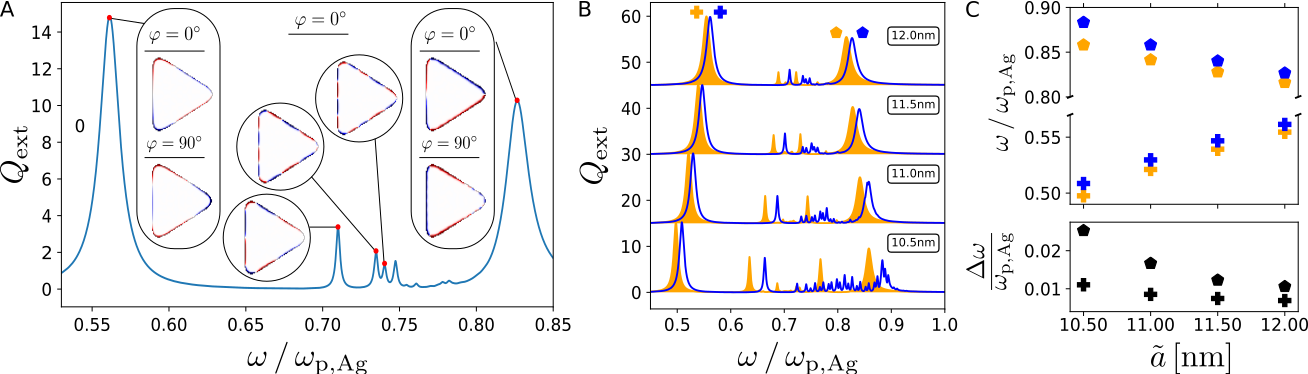}
	\caption{
		\label{tri_fig:extinction}
		Curvature impact on the extinction efficiency associated with 
		LSPs sustained by the right-oriented triangular nanowire with 
		$a=10$ nm.
		Panel A: Extinction efficiency as a function of frequency for 
		the triangular nanowire with $\tilde{a}=12$ nm and 
		$\varphi=0\degree$, calculated using the Halevi model. Peak 
		maxima are interpreted as resonance frequencies. Insets show 
		the real part of the induced charge distribution at the 
		respective resonances for $\varphi\in\{0\degree;90\degree\}$, 
		each normalized by its maximum value. The colormap is restricted 
		to the range $[-0.1,0.1]$ (blue to red) for improved visibility.	
		Panel B: Waterfall plot illustrating the curvature dependence 
		of the extinction efficiency for 
		$\tilde{a}\in\{10.5;11.0;11.5;12.0\}$ nm (see white boxes) 
		and $\varphi=0\degree$, comparing the Drude (orange) and Halevi 
		(blue) models. The curves are vertically offset by 15 units for 
		clarity.
		Panel C: Resonance frequencies (upper subpanel) and nonlocal 
		blueshifts (lower subpanel), extracted from the peak-maximum 
		frequencies in Panel B (see colored markers at the top of 
		Panel B), as functions of the respective relative radial 
		difference. The low- and high-frequency quadrupolar LSPs 
		exhibit opposite overall shifts. The nonlocal partial 
		blueshift $\Delta\omega$, defined as the difference between 
		the nonlocal and local resonance frequencies, depends on 
		the curvature.
	}    
\end{figure*}

%
%
%


\subsection{\label{subsec:tri_extinction} Extinction}

To characterize the number, spectral distribution, and geometrical dependence 
of the resonances, we study the extinction efficiency. The radius of the
circumscribing circle is fixed at $a=10$ nm, while the radial difference is 
varied over the range $\tilde{a}-a\in \{0.5, 1.0, 1.5, 2.0 \}$ nm, corresponding 
to corner radii of curvature $R\in \{1/3, 2/3, 1, 4/3\}$ nm. These values are comparable to 
those employed in Refs.~\cite{Kottmann_2000,PhysRevB.64.235402,Wallen2008,cui2010,Toscano_optexp_2012}. 
Inspection of \cref{fig:rtri_geom_properties}(Panel B) reveals that all of 
these geometries exhibit 
a surface-to-volume ratio exceeding the spherical reference value 
($\eta_{\rm sph}$). An example of the mesh used for the wire cross 
section is depicted in
\cref{suppl_fig:finest_mesh_examples_tri_bowtie_cylinder}(Panel A).

By symmetry, the extinction cross section is expected to be identical for 
both incidence directions~\cite{Awada2012}. To verify this numerically, 
we compare the simulated extinction cross sections (per unit length) for 
a triangle with $\tilde{a}=12$ nm and both material models, considering 
left- and right-oriented triangles in \cref{supp_I_tri_fig:ext_cross}. The 
degeneracy of the resonance frequencies of corresponding multipolar modes 
for the two orientations simplifies the interpretation of the hybrid 
resonances sustained by the bowtie nanowire in \cref{sec:bowtie}. 
Owing to the different geometrical cross sections, however, the
corresponding extinction efficiencies are not identical.

In the following, we therefore restrict our attention to the right-oriented
triangle and, for the extinction efficiency, to the case $\varphi=0\degree$. \Cref{tri_fig:extinction}(Panel A) shows the extinction efficiency calculated 
with the 
Halevi model for $\tilde{a}=12$ nm. Two dominant resonance peaks appear at 
the low- and high-frequency edges of the spectrum, whereas the remaining
resonances cluster toward the spectral center with progressively smaller
amplitudes, indicating weaker coupling to the incident plane wave. To classify 
the resonances, we include insets showing the induced charge distributions 
at the corresponding resonance frequencies, obtained from an on-the-fly 
Fourier transform.

As expected, all charge distributions are localized at the surface, 
confirming the absence of confined bulk plasmon modes below 
$\omega_{\rm p,Ag}$. The two outermost resonances exhibit quadrupolar 
charge distributions. Consistent with the analysis of Ref.~\cite{Awada2012}, 
the corresponding charge distributions differ for the two incidence
 directions.
\begin{figure*}
	\includegraphics[scale=0.4]{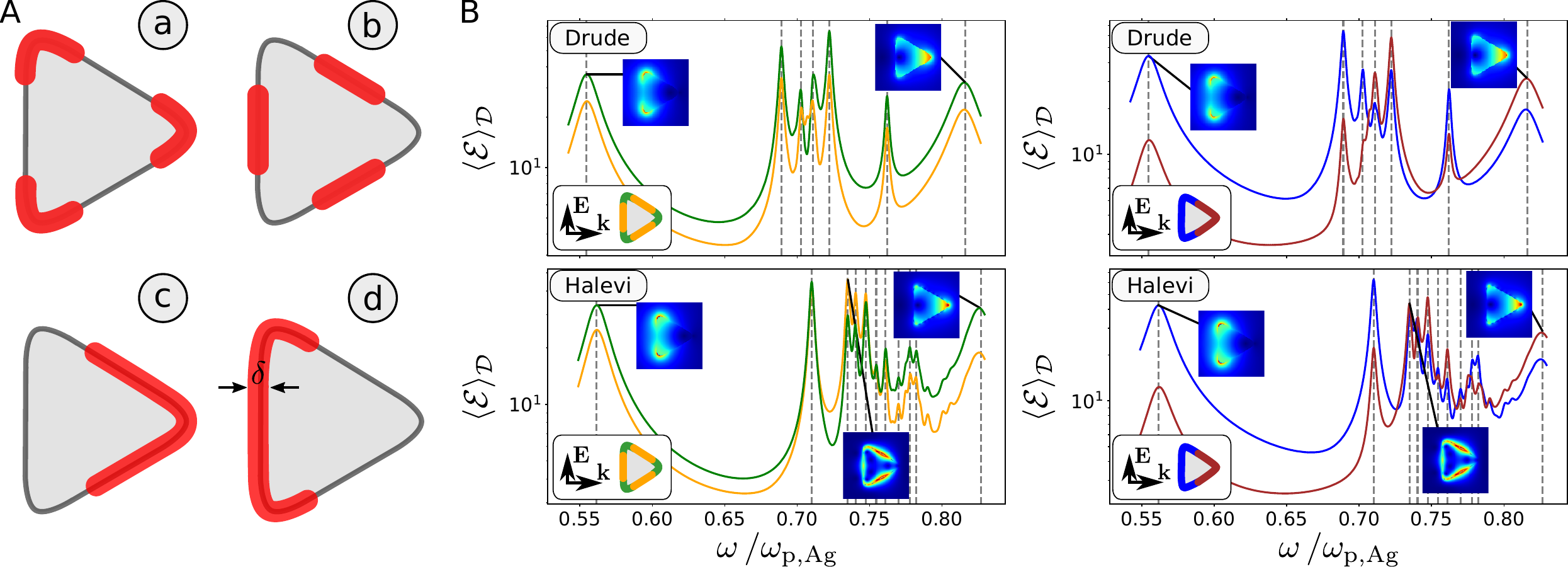}
	\caption{
		\label{tri_fig:near_field_measures_and_phi_0}
		Mean-domain enhancement of the electric field associated with LSPs
		sustained by the triangular nanowire for $a=10$ nm, $\tilde{a}=12$ nm, 
		and $\varphi=0\degree$.
		Panel A: Sketch of the selected domains $\mathcal{D}$ (red) used to
		calculate the mean-domain enhancement. We distinguish between the 
		corner domain (a; defined by the Bézier curves) and the edge 
		domain (b; the remaining straight-sided regions), both of which 
		are curvature dependent, as well as the right-half (c) and 
		left-half domains (d). The parametrization follows the description 
		in \cref{subsec:tri_construction}, where the domain width $\delta$ 
		is introduced.
		Panel B: Frequency dependence of the mean-domain enhancement. The 
		edge and corner enhancements (left subpanels), as well as the left- 
		and right-half enhancements (right subpanels), are compared, 
		each evaluated using the Drude model (upper subpanels) and the 
		Halevi model (lower subpanels). The legend and incident-field 
		polarization are provided in the lower-left box of each subpanel.
		Vertical dashed lines indicate the peak-maximum frequencies of 
		the corresponding extinction efficiency (see also \cref{tri_fig:extinction}). 
		Insets show the spatial distribution of the electric-field modulus 
		at the respective frequencies, normalized by the maximum value 
		within the displayed spatial domain. Red (blue) colors indicate 
		the maximum (minimum) value. 
	}
\end{figure*}

As regards the low-frequency quadrupolar resonances, the charge distribution 
for $\varphi=0\degree$ exhibits three nodes along the edges and one at the
right corner, whereas for $\varphi=90\degree$ all four nodes are located 
on the edges. At higher frequencies and for $\varphi=0\degree$, the nodes 
on the upper and lower edges shift to the corresponding left corners. 
In contrast, the high-frequency charge distribution for $\varphi=90\degree$
cannot be unambiguously classified because it overlaps with at least 
one higher-order LSP. Since longitudinal nonlocality primarily broadens 
the induced charge distribution, the corresponding Drude-model charge 
patterns are expected to remain qualitatively unchanged, with the main 
difference being the shifted resonance frequencies. Toward the spectral 
center, the charge distributions exhibit an increasing number of nodes,
indicating progressively higher mode order. In this context, we adopt 
the concept of an effective surface wavelength introduced in
Ref.~\cite{Christensen2014} for spherical LSPs. For the triangular cross 
section, it is defined as the ratio of the triangle circumference 
$u_{\rm rtri}$ (see \cref{suppl_eq:circumference_rtri}) to the number 
of surface charge oscillations. As the mode order increases, this 
effective wavelength becomes too short to resolve the corner curvature, 
causing the corner and straight-edge segments to be perceived 
increasingly as equivalent.

The extinction spectra and charge distributions suggest that the 
triangular nanowire also supports two branches of resonances arranged 
about a central frequency. Support for this interpretation is provided 
by transformation optics. As shown in Ref.~\cite{surf_plas_and_singularities}, 
the infinite wedge can be mapped conformally onto a reference geometry 
in which it is represented by a periodic array of slabs. In this 
transformed picture, surface plasmons supported by different slab 
interfaces couple naturally, giving rise to a hybridization 
mechanism. Consequently, the familiar Coulomb-interaction arguments 
can be invoked to explain the spectral ordering of the higher-order 
modes. We note, however, that the longitudinally nonlocal blueshift
may alter the ordering of the upper branch if the local model indeed 
predicts convergence toward the central frequency with increasing mode 
order, thereby complicating the interpretation based solely on the 
charge distributions.

Since the surface-to-volume ratio of the rounded triangle depends on 
the corner curvature, the strength of the nonlocal response is likewise 
expected to be curvature dependent. To investigate this effect,
\cref{tri_fig:extinction}(Panel B) compares the Drude and Halevi models 
in a 
waterfall plot, where the radial difference $\tilde{a}-a$ is reduced 
from $2.0$ to $0.5$ nm in steps of $0.5$ nm, thereby increasing the 
corner curvature. The most prominent feature is that, within the Halevi 
model, the cluster of resonances in the spectral center broadens toward 
higher frequencies as the curvature increases. The resulting nonlocal 
peak sequence remains bounded from below, while the high-frequency
quadrupolar resonance gradually loses spectral weight. To demonstrate  
that this behavior is indeed of nonlocal origin, we additionally reduce 
the Fermi velocity $v_{\rm F,Ag}$ while keeping the geometry fixed at 
the highest curvature, as shown in
\cref{suppl_I_tri_fig:nonlocal_peak_sequence_ext_eff_alpha_scan}
(Panel A). 
As expected, the nonlocal peak sequence progressively collapses and 
ultimately converges to the Drude spectrum. As further demonstrated in \cref{suppl_I_fig:tri_phi_0_and_90_Q_ext_nloc_peak_seq}, the same nonlocal peak sequence also appears 
for $\varphi=90\degree$. This behavior is reminiscent of the ripple 
structure reported for the cylindrical dimer in Ref.~\cite{Moeferdt2018} 
and discussed in \cref{subsec:cyl_dim_extinction}, although for this geometry it occurs only 
for $\varphi=0\degree$.

Turning to the quadrupolar resonances, we find that their frequencies
already shift in opposite directions within the Drude model. We attribute 
this behavior to the different spatial arrangements of the charge nodes 
(or poles) within the corresponding charge distributions. As shown in
\cref{suppl_I_fig:tri_charge_distr_quadrupolar_LSP_sharpness_scan}
(Panel A) for the low-frequency quadrupolar LSPs 
excited under both incidence directions, increasing the corner curvature 
leads to stronger localization and a displacement of the charge poles, 
which we interpret as a reduction of the interaction energy. Moreover, 
with increasing curvature another well-defined pair of resonances 
exhibiting opposite overall frequency shifts becomes increasingly 
prominent, although only the lower-frequency member of this pair is 
observed within the Halevi model. Finally, \cref{tri_fig:extinction}(Panel C) 
shows that each quadrupolar resonance undergoes a longitudinally 
nonlocal blueshift for every curvature considered. Importantly, this 
blueshift does not remain constant during the curvature variation, 
as would be expected if it were determined solely by the fixed reference 
radius $a$. Instead, the results clearly demonstrate a curvature-dependent
contribution to the longitudinally nonlocal blueshift. Taken together, 
the opposite curvature-induced shifts of the quadrupolar resonances and 
the widening of the nonlocal peak sequence further highlight the potential 
of these structures for broadband light-harvesting applications.

\subsection{\label{subsec:tri_fields} Near-field distributions}

Next, we study the near-field distributions of the triangular nanowire 
monomer in order to identify suitable 
locations for a SERS target molecule. For this geometry, we naturally 
focus on the mean surface enhancement. Among the selected domains, we 
distinguish between the corner and edge enhancements, where the 
former is defined by the B\'{e}zier-rounded corners, 
as well as the left- and right-half surface enhancements, as illustrated 
in \cref{tri_fig:near_field_measures_and_phi_0} (Panel A).

In \cref{tri_fig:near_field_measures_and_phi_0} (Panel B), we compare the 
edge and corner enhancements 
(left subpanels) with the left- and right-half surface enhancements 
(right subpanels) for $a=10$ nm, $\tilde{a}=12$ nm, and 
$\varphi=0\degree$. Results are shown for both the Drude model (upper 
subpanels) and the Halevi model (lower subpanels). The insets display 
the normalized modulus of the electric field at selected resonance 
frequencies. As for the cylindrical dimer, the local maxima of the 
enhancement measures coincide closely with those of the extinction 
efficiency (cf.~\cref{tri_fig:extinction} (Panel B)). Consequently, 
the resonance 
peaks predicted by the Halevi model are blueshifted relative to their 
Drude counterparts.

In addition, the Halevi model exhibits a modulation of the enhancement 
spectra at higher frequencies, coinciding with the spectral region of 
the nonlocal peak sequence discussed in \cref{subsec:cyl_dim_extinction}. 
Within this 
frequency range, the Halevi model reveals several edge-dominated 
resonances, as illustrated by the central inset of the lower left 
subpanel. This contrasts with the Drude model, which generally favors 
corner enhancement, except for a single edge-dominated resonance in the 
spectral center, as evident from the insets of the upper left subpanel. 
Owing to the connection with the nonlocal peak sequence, and hence with higher-order modes, we attribute this behavior to the decreasing effective 
surface wavelength. Comparing the field distributions predicted by the 
Drude and Halevi models, we further observe that the low-frequency 
quadrupolar mode becomes smeared toward the interior of the triangle, 
whereas the high-frequency quadrupolar mode exhibits pronounced ripples 
along the boundary. We attribute the latter to spectral overlap between 
the high-frequency quadrupolar resonance and neighboring higher-order 
nonlocal resonances.

The right subpanels of \cref{tri_fig:near_field_measures_and_phi_0}
(Panel B) show that, at low 
frequencies, the strongest field enhancement is localized at the pair 
of corners aligned with the incident field polarization, in agreement 
with the numerical studies of Refs.~\cite{PhysRevB.64.235402,Babaei2018}. 
This behavior is expected because, at sufficiently low frequencies, the quasistatic approximation remains valid, implying that the incident 
electric field is nearly uniform across the particle and aligned with 
the line connecting these two corners. At higher frequencies, however, 
the incident field becomes increasingly inhomogeneous over the particle 
during a given optical cycle, allowing charge accumulation at other 
corners. Consequently, for frequencies above approximately
$\omega\approx0.8\,\omega_{\rm p,Ag}$, the strongest field enhancement 
shifts to the corner oriented perpendicular to the incident polarization. 
An intermediate situation is illustrated by the inset in the lower right 
subpanel of \cref{tri_fig:near_field_measures_and_phi_0}(Panel B).

The same overall behavior is observed for $\varphi=90\degree$ (see \cref{suppl_I_fig:tri_a_10_a_tilde_12_phi_90_aver_dom_enh}), except 
that the transition between the left- and 
right-half enhancements is reversed. In this case, the right-half 
enhancement dominates at lower frequencies, whereas the left-half 
enhancement becomes stronger above $\omega\approx0.75,\omega_{\rm p,Ag}$, 
again consistent with the validity range of the quasistatic 
approximation. Consequently, both the number and the locations of 
the corner hotspots depend on the excitation frequency and the 
incident polarization. Such knowledge is essential when surface-enhanced 
Raman spectroscopy is to be performed simultaneously at multiple 
locations on the same nanoparticle.

\begin{figure*}
	\centering
	\includegraphics[scale=.28]{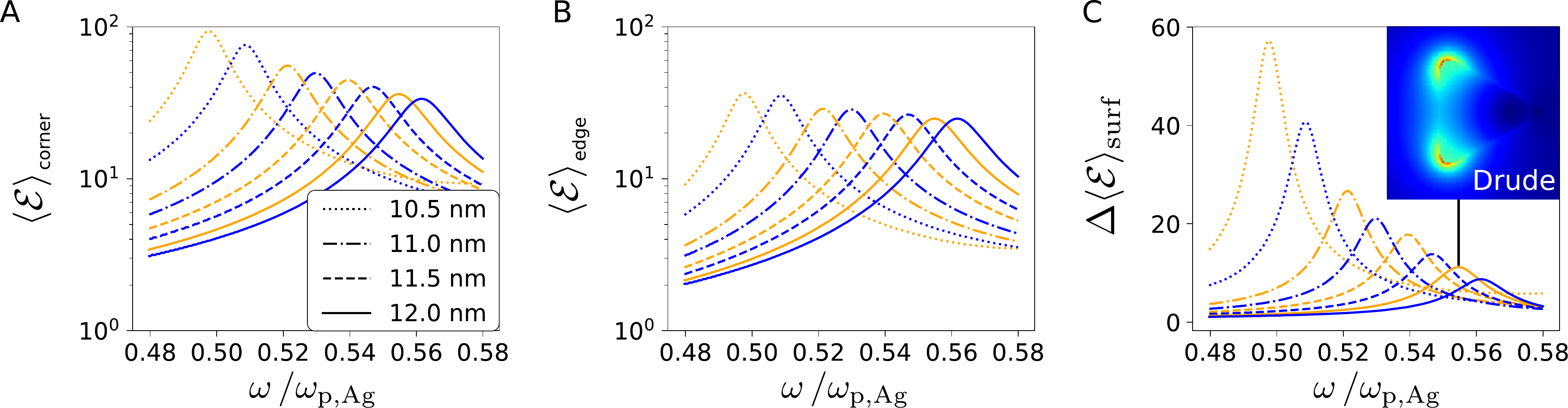}
	\caption{
		\label{tri_fig:curv_scan_near_field_low_quadrupole}
		Curvature dependence of the mean-corner and mean-edge enhancements 
		for the triangular nanowire with $a=10$ nm and $\tilde{a}\in\{12.0;11.5;11.0;10.5\}$ nm (see legend in Panel A) 
		for $\varphi=0\degree$. The Drude (orange) and Halevi (blue) models 
		are considered. Shown are the mean-corner enhancement (Panel A) 
		and mean-edge enhancement (Panel B), as well as their difference 
		$\Delta \langle \mathcal{E} \rangle_{\rm surf} = \langle \mathcal{E} \rangle_{\rm corner} - \langle \mathcal{E} \rangle_{\rm edge}$ 
		(Panel C), all as functions of frequency over the spectral range 
		of the low-frequency quadrupolar LSP. In Panel C, an inset displays 
		the field distribution at the peak maximum frequency of the 
		quadrupole for $\tilde{a}=12$ nm, obtained using the Drude 
		model. Red (blue) colors indicate the maximum (minimum) 
		value.
	}
\end{figure*}

To conclude the investigation of the triangular nanowire monomer, we 
examine the influence of curvature on the edge and corner enhancement
 associated with the low-frequency quadrupolar LSP, which is spectrally 
 well isolated from neighboring resonances. As a candidate mode for 
 multi-site SERS, we focus on the case of $\varphi=0^{\circ}$, shown 
 in \cref{tri_fig:curv_scan_near_field_low_quadrupole}. 
 Based on \cref{tri_fig:near_field_measures_and_phi_0}(Panel B) and 
 the field distribution shown in the inset of \cref{tri_fig:curv_scan_near_field_low_quadrupole}(Panel C) for the 
 Drude model, we expect the field to become increasingly concentrated 
 at the corners as the curvature increases. Indeed, in both the Drude 
 and Halevi models, the corner enhancement (Panel A) and the edge 
 enhancement (Panel B) increase monotonically with increasing 
 curvature. For a given curvature, however, the Drude model consistently 
 predicts larger peak enhancement factors for both measures.
 Panel C shows the difference between the two enhancement measures, 
 revealing an increasing dominance of corner enhancement with increasing 
 curvature for both material models, with the effect being more 
 pronounced in the Drude model.

Interestingly, for the sharpest structure ($a=10$ nm, $\tilde{a}=10.5$ nm), 
the mean corner enhancement reaches approximately $94$ ($76$) in the 
Drude (Halevi) model. These values approach the enhancement levels 
commonly associated with single-molecule 
SERS ~\cite{Yu2020singlemolSERSroadmap,Kanehira2023}.

\section{\label{sec:bowtie} Bowtie-shaped nanowires}

\subsection{\label{subsec:bowtie_geometry} Geometry}

Based on the construction introduced in \cref{subsec:tri_construction}, the bowtie consists of two tip-to-tip, B\'{e}zier-rounded equilateral triangular nanowires arranged in parallel. We consider an unconnected dimer with a finite gap size. Consequently, while the total surface area and volume are doubled relative to the monomer, the surface-to-volume ratio \cref{eq:tri_surf_to_vol} remains unchanged, as does the corner radius of curvature \cref{tri_eq:radius_of_curv}. The corresponding geometrical cross sections (per unit length) are 
\begin{align}
	\sigma^{0\degree}_{\rm bow} = \sqrt{3}a + \frac{\tilde{a} -a }{3\sqrt{3}} \quad \text{and} \quad \sigma^{90\degree}_{\rm bow} = 3a + \tilde{a} -a \,.
	\label{eq:bowtie_geom_cross_sec_parallel_and_perp}   
\end{align}
Since $\tilde{a}\leq 8a/5$ [see the discussion surrounding \cref{eq:tri_geom_upper_bound_radial_diff}], both cross sections are always smaller than those of the corresponding cylindrical dimer.

We emphasize that, although the radii of the reference cylindrical dimer are kept fixed at $a=10$ nm, the parametrization allows us to vary independently either the constituent shape (through $\tilde{a}$) or the interparticle separation (through $g$). This separation of shape and gap effects provides a clear framework for analyzing their respective influences on the optical response.


%
\subsection{\label{subsec:bowtie_extinction} Extinction}

\begin{figure*}
	\centering
	\includegraphics[scale=.43]{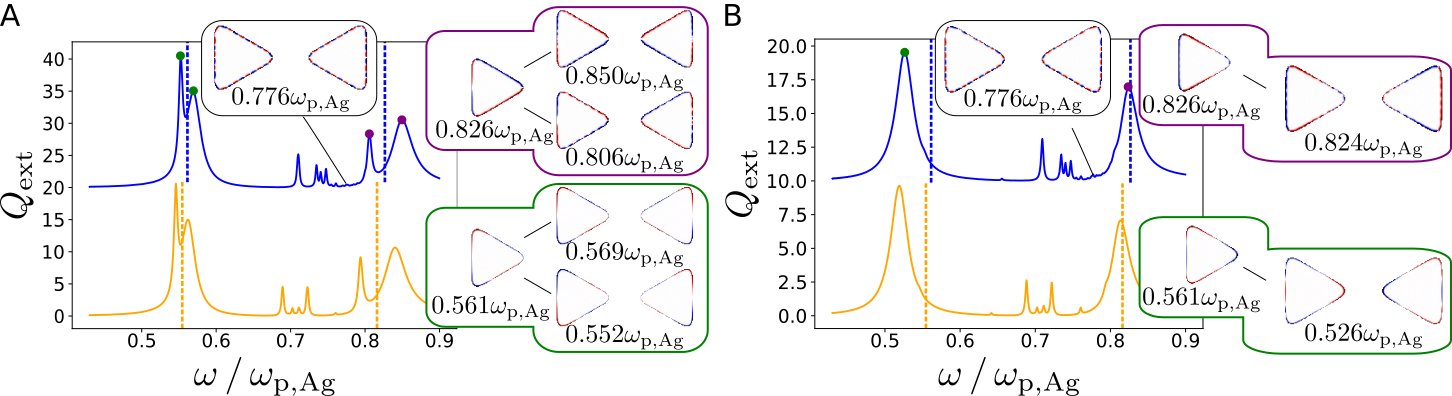}
	\caption{
		\label{bow_fig:Q_ext_one_setup_phi_0_90_Halevi_Drude}
		Extinction efficiency and charge distributions at selected resonances 
		of the bowtie nanowire with radii $a=10$ nm and $\tilde{a}=12$ nm for 
		a gap size of $g=10$ nm.
		Panel A: Extinction efficiency for $\varphi=0\degree$ calculated using 
		the Drude model (orange) and the Halevi model (blue). The curves are
		vertically shifted by 20 units for clarity. Vertical dashed lines 
		indicate the quadrupolar resonance frequencies of the corresponding
		triangular nanowire monomer. Insets show the real part of the charge
		distributions at the resonance frequencies of the bonding and 
		antibonding quadrupolar LSPs, obtained via an on-the-fly Fourier
		transform. The charge distributions are normalized to their 
		respective maximum values, and the colormap is restricted to the range $[-0.1,0.1]$ (blue to red) for improved visibility. For comparison, 
		the corresponding charge distribution of the monomer resonance is 
		also shown.
		Panel B: Same as Panel A, but for $\varphi=90\degree$ and with an
		artificial shift of 10 units. The insets display only bonding dimer 
		LSPs. 
	}
\end{figure*}

We begin the spectral analysis of the bowtie with a representative 
geometry defined by $a=10$ nm, $\tilde{a} =12$ nm, and $g=10$ nm. 
The extinction efficiencies obtained with the Drude and Halevi models 
are displayed in \cref{bow_fig:Q_ext_one_setup_phi_0_90_Halevi_Drude} 
for $\varphi = 0 \degree$ (Panel A) and $\varphi = 90 \degree$ (Panel B). 
The geometrical parameters are identical to those of the triangular 
nanowire studied in \cref{tri_fig:extinction}(Panel A). As for the 
monomer, we observe nonlocal blueshifts together with a nonlocal 
sequence of resonances that emerges near the center of the spectrum 
and extends toward higher frequencies. A comparable sequence has 
previously been reported from numerical simulations based on the 
linear Euler--Drude model for an isosceles bowtie of similar dimensions 
under $\varphi = 90 \degree$ excitation~\cite{dhuynh2016}. 
We particularly emphasize the central insets in both panels, which 
display the charge distributions of representative high-order LSPs. 
In agreement with the triangular monomer, and unlike the cylindrical 
dimer, this nonlocal sequence is present for both excitation
 directions.

For $\varphi = 0 \degree$ (Panel A), two resonance doublets appear, 
each straddling the corresponding quadrupolar LSP resonance of the
monomer. This behavior is characteristic of plasmon
hybridization~\cite{Nordlander2004,Prodan_2003_hybrid_first}. 
Indeed, the induced charge distributions identify the members of each 
doublet as bonding and antibonding hybrid quadrupolar LSPs located on 
either side of the corresponding monomer resonance. For the high-frequency doublet, both hybrid resonances overlap with at least one weaker 
high-order mode.
The hybridization picture also applies for $\varphi = 90 \degree$ 
(Panel B), although only the bonding hybrid resonances are excited 
efficiently, similar to the cylindrical dimer. Owing to the absence of a charge node at the gap-facing 
corners, the low-frequency bonding resonance 
$\omega \sim  0.526 \omega_{\rm p,Ag}$ 
is symmetric under point inversion followed by charge-sign reversal, 
whereas the bonding resonances excited for $\varphi = 0 \degree$ are 
strictly point-symmetric because such a node is present. Unlike the 
cylindrical dimer, the overall spectral range containing the dominant
resonances changes only weakly with the excitation polarization. We 
attribute this behavior not to the common dimer symmetry, but rather 
to the fact that the triangular monomer itself supports distinct 
low- and high-frequency quadrupolar resonances for each polarization.

\begin{figure*}
	\centering
	\includegraphics[scale=.5]{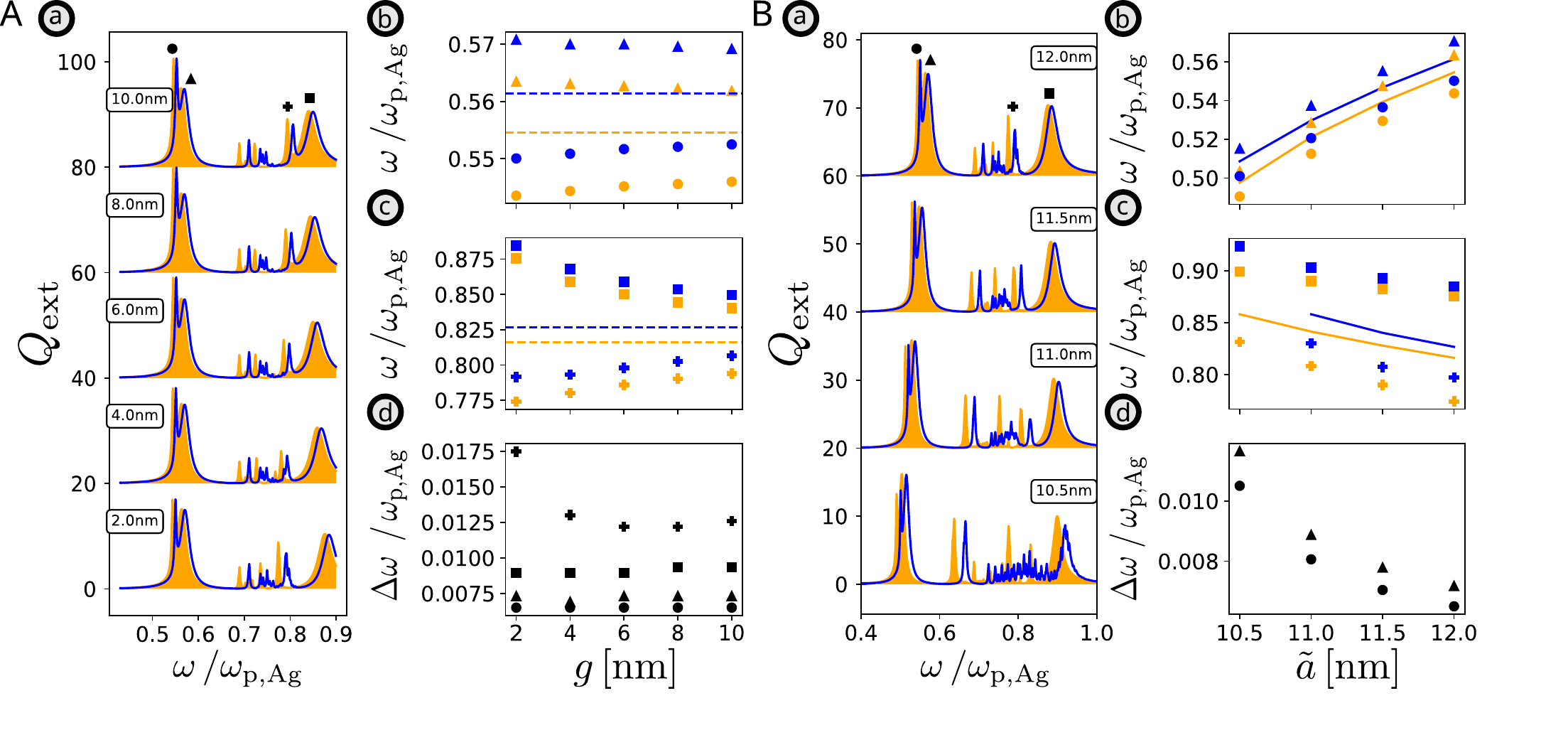}
	\caption{
		\label{fig_bow:curv_and_gap_scan_of_Q_ext_bowtie_Drude_Haleiv_phi_0}
		Study of the geometric-parameter dependence of the extinction 
		efficiency of a bowtie nanowire for $a=10$ nm and $\varphi=0\degree$.
		Panel A: Gap-size dependence for $\tilde{a}=12$ nm and 
		$g\in\{10;8;6;4;2\}$ nm. The extinction efficiency is calculated 
		using the Drude (orange) and Halevi (blue) models (subpanel a). 
		The boxes in subpanel (a) indicate the respective gap sizes, and 
		the curves are artificially offset by 20 units for clarity. 
		Subpanel (b) and (c) show, respectively, the resonance frequencies 
		of the low-frequency and high-frequency quadrupolar dimer LSPs 
		(see markers at the top of subpanel (a)). For a subset of the 
		resonances, the corresponding nonlocal blueshifts are displayed 
		in subpanel (d).
		Panel B: Same as Panel A, but illustrating the curvature dependence for $g=2$ nm and $\tilde{a}\in\{10.5;11.0;11.5;12.0\}$ nm.		
	}
\end{figure*}

To further exploit the Coulomb picture underlying the hybridization
model~\cite{Nordlander2004}, we next investigate the dependence of 
the hybrid quadrupolar LSPs on the gap size. The separation is varied 
over $g \in \{10,8,6,4,2 \}$ nm. The mesh corresponding to the smallest 
gap is shown in 
\cref{suppl_fig:finest_mesh_examples_tri_bowtie_cylinder}
(Panel B). Since the radii are kept fixed, the constituent shape 
remains unchanged. This therefore represents a purely separation-driven, 
or dimer-related, study of plasmon hybridization. Owing to the richer 
spectrum, we focus in \cref{fig_bow:curv_and_gap_scan_of_Q_ext_bowtie_Drude_Haleiv_phi_0}
(Panel A) on the case $\varphi = 0 \degree$; the corresponding results 
for $\varphi = 90 \degree$ are presented in
\cref{suppl_I:bow_Q_ext_curve_and_gap_scan_phi_90}.

As expected, the bonding and antibonding quadrupolar resonances undergo 
overall red- and blueshifts, respectively, as the gap decreases 
(subpanel a). This behavior is corroborated by the resonance frequencies 
shown in subpanels b and c, where the monomer quadrupolar frequencies 
remain fixed because the monomer geometry is unchanged. The low-frequency 
hybrid quadrupoles (subpanel b) are notably less sensitive to variations 
in the gap size, similar to observations reported for selected resonances 
of square dimers.\cite{cui2010} Such gap-insensitive resonances may 
prove advantageous in bowtie arrays, where fabrication inevitably 
introduces variations in the interparticle separation~\cite{Yu2020singlemolSERSroadmap}. 
Finally, subpanel d shows that the nonlocal blueshift remains nearly 
constant for the low-frequency hybrid resonances and the high-frequency
antibonding quadrupole, indicating that it is governed primarily by the 
monomer geometry. We omit the interpretation of the high-frequency bonding quadrupole because 
it increasingly overlaps with the nonlocal resonance sequence. 
Interestingly, the spectral width of this sequence changes only marginally 
with gap size, consistent with the strong confinement of the associated 
near fields to the particle surfaces.

To examine the constituent-based perspective of the hybridization picture, 
we next vary the corner curvature, as shown in Panel B. The gap size is 
fixed at $g=2$ nm, while the curvature is varied through 
$\tilde{a} \in \{ 12.0,11.5,11.0,10.5 \}$ nm.

First, we observe a pronounced widening of the nonlocal resonance 
sequence, mirroring the behavior of the triangular monomer, see \cref{tri_fig:extinction}(Panel B). This demonstrates that the
high-order resonance sequence is considerably more sensitive to 
variations in the constituent shape than to changes in the interparticle
separation. As anticipated from the behavior of the monomer quadrupolar
resonances, the two hybrid resonance doublets exhibit opposite overall 
frequency shifts (subpanel a). Inspection of subpanels b and c further 
reveals that the local and nonlocal monomer quadrupolar resonance 
frequencies, which serve as reference frequencies for the hybridized 
modes, shift with curvature. The corresponding dimer resonances closely 
follow these shifts. Consequently, by varying either the constituent 
shape or the interparticle separation, one can tune pairs of bonding 
and antibonding hybrid resonances either in the same spectral direction 
or in opposite directions. This behavior is qualitatively distinct from 
that of the cylindrical dimer, where the fixed cylinder radius permits 
only separation-induced tuning.

The curvature scan also reveals an increasing nonlocal blueshift, as 
shown in subpanel d. Owing to the progressive broadening of the nonlocal 
resonance sequence, we only report the blueshifts of the low-frequency 
hybrid resonances. In particular, for 
$\tilde{a} =10.5$ nm, the high-frequency bonding quadrupolar LSP is 
completely masked by the overlapping high-order resonances.
A further noteworthy feature of the curvature scan is the evolution of 
the relative amplitudes of the low-frequency bonding and antibonding 
quadrupolar resonances (subpanel a). From the geometrical construction, 
increasing curvature corresponds to a reduction in the particle volume. 
The accompanying decrease in the amplitude of the bonding resonance 
therefore suggests that this mode is more strongly influenced by 
retardation effects than its antibonding counterpart.
For completeness, we note that plasmons in equilateral triangular and 
bowtie graphene nanostructures, as well as the concept of plasmon 
hybridization in these systems, have been investigated in 
Ref.~\cite{Wang2015}.

We conclude the extinction study by emphasizing that, compared with 
the cylindrical dimer, the additional geometrical parameter 
$\tilde{a}$ provides an extra degree of freedom for tailoring the 
spectral response. In particular, curvature tuning produces substantially 
larger spectral changes than variations in the gap size. Consequently 
and under the premise of the geometrical construction given in \cref{subsec:tri_construction}, curvature dispersion in fabricated bowtie arrays is expected to have 
a more pronounced impact on their optical properties than gap-size 
dispersion alone.

\subsection{\label{subsec:bowtie_fields} Near-field distributions}

We begin the near-field analysis by examining the localization of the field 
at the particle surface, employing the surface domains introduced for the
triangular nanowire, see \cref{tri_fig:near_field_measures_and_phi_0} (Panel A). Building on the insights gained from this analysis, we subsequently 
investigate the field localization within the gap, adopting the domain 
definitions introduced for the cylindrical dimer, see \cref{cyl_dimer_fig:enh_measures_cyl_dimer_both_inc_dir}.

\begin{figure*}
	\centering
	\includegraphics[scale=.27]{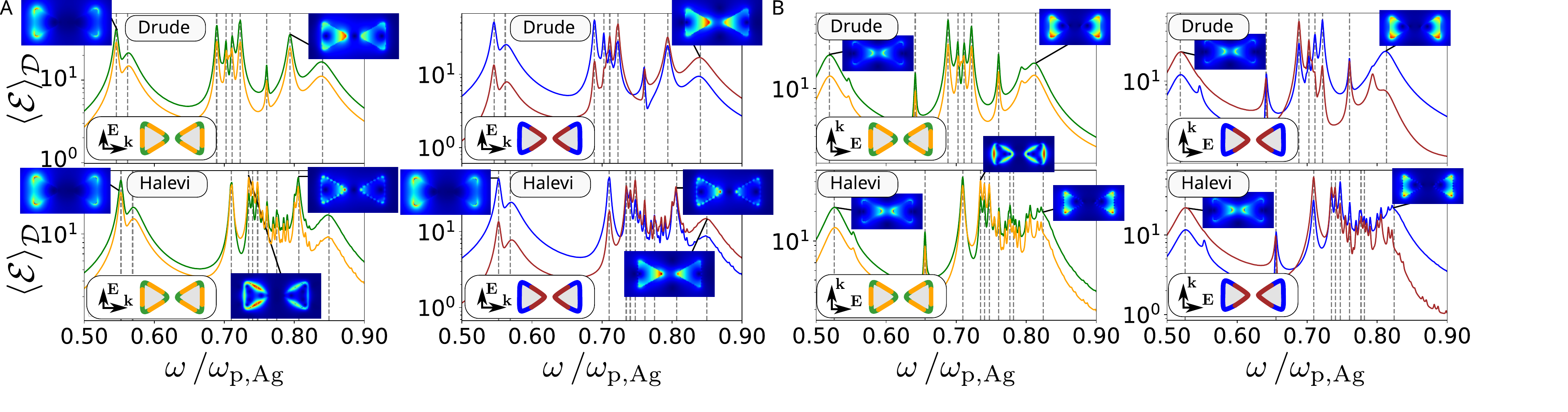}
	\caption{
		\label{bow_fig:aver_surf_doman_enhan_bowtie_both_phi_fixed_geom}
		Mean-surface-domain enhancement of the electric field associated 
		with LSPs sustained by the bowtie nanowire for $a=10$ nm, 
		$\tilde{a}=12$ nm, and $g=10$ nm.
		Panel A: Frequency dependence of the mean-domain enhancement for
		$\varphi=0\degree$. The edge and corner enhancements (left subpanels), 
		as well as the inner- and outer-surface enhancements (right subpanels),
		are compared, each evaluated using the Drude model (upper subpanels) 
		and the Halevi model (lower subpanels). The legend and incident-field
		polarization are provided in the lower-left box of each subpanel. 
		Vertical dashed lines indicate the peak-maximum frequencies of 
		the corresponding extinction efficiency (see also \cref{fig_bow:curv_and_gap_scan_of_Q_ext_bowtie_Drude_Haleiv_phi_0}). 
		Insets show the spatial distribution of the electric-field modulus 
		at the respective frequencies, normalized by the maximum value 
		within the displayed spatial domain. Red (blue) colors indicate 
		the maximum (minimum) value.
		Panel B: Same as Panel A, but for $\varphi=90\degree$. 
	}
\end{figure*}

\subsubsection{Surface measures}

As shown in \cref{bow_fig:aver_surf_doman_enhan_bowtie_both_phi_fixed_geom}, 
we compare the edge and corner enhancement (left subpanels) as well as the 
inner- and outer-surface enhancement (right subpanels) for the Drude 
(upper subpanels) and Halevi (lower subpanels) models, considering
$\varphi=0\degree$ (Panel A) and $\varphi=90\degree$ (Panel B). In all 
cases, the maxima of the extinction efficiency and the enhancement 
measures are closely aligned. This includes the  $\varphi$-dependent number 
of quadrupolar dimer resonances, the nonlocal blueshift, and the 
high-frequency modulation induced by the nonlocal resonance sequence. 
Similar to the triangular nanowire, the Drude model exhibits only a 
single peak with dominant edge enhancement for both excitation 
directions. In contrast, the Halevi model displays edge-dominated 
enhancement over a broader spectral range around the center of the 
spectrum. Representative edge-localized field distributions are shown 
in the central insets of the lower left subpanels of Panels A and B.

Turning to the comparison of inner- and outer-surface enhancement, we 
observe for $\varphi=0\degree$ a dominance of outer-surface enhancement 
at low frequencies. This corresponds to hotspots located at the corners 
along a line parallel to the incident electric-field polarization, as 
illustrated by the inset showing the low-frequency bonding quadrupole 
for the Halevi model. Gap-averted hotspots have likewise been reported 
for numerically investigated two-cut semi-circular and square nanodimers~\cite{cui2010}. At higher frequencies, the hotspots shift 
toward the gap, as demonstrated by the field distributions of the 
high-frequency quadrupoles. Similar to the triangular nanowire, this 
trend is reversed for  $\varphi=90\degree$ (Panel B), as exemplified by 
the field distributions of the bonding quadrupolar LSPs. As observed for 
the cylindrical dimer, the field distributions of LSPs for both excitation 
directions exhibit a slight asymmetry along the propagation direction 
of the incident field.

\begin{figure*}
	\centering
	\includegraphics[scale=.6]{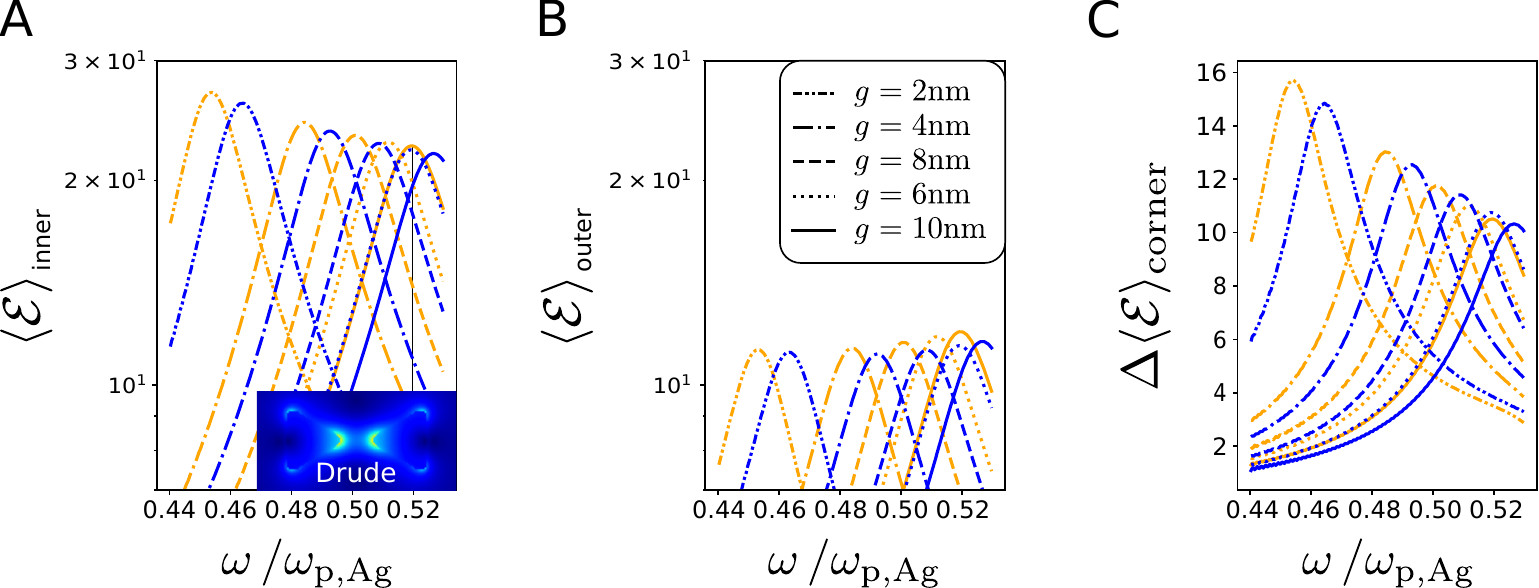}
	\caption{
		\label{bow_fig:gap_scan_low_quads_gap_fac_avert_corner_measures}
		Gap-size dependence of the mean inner- and outer-corner enhancements 
		of the low-frequency bonding quadrupole sustained by the bowtie 
		nanowire with $a=10$ nm and $\tilde{a}=12$ nm for $g\in\{10;8;6;4;2\}$ nm (see legend in Panel B), excited with $\varphi=90\degree$. The 
		Drude (orange) and Halevi (blue) models are considered. Shown are 
		the inner-corner (Panel A) and outer-corner enhancements (Panel B), 
		as well as their difference $\Delta \langle \mathcal{E} \rangle_{\rm corner} = \langle \mathcal{E} \rangle_{\rm inner} - \langle \mathcal{E} \rangle_{\rm outer}$ (Panel C), all as functions of frequency over 
		the spectral range of the quadrupolar LSP peak. In Panel A, an 
		inset displays the field distribution at the peak maximum frequency 
		of the quadrupole for (g=10) nm, obtained using the Drude model. 
		Red (blue) colors indicate the maximum (minimum) value.
	}
\end{figure*}

Since none of the considered amplitude enhancement measures exceed the 
threshold associated with single-molecule SERS, we next investigate the 
gap-size dependence of selected resonances. To more clearly identify 
suitable specimen positions, we restrict the inner- and outer-surface 
enhancement to the corner domains. Furthermore, to avoid interference 
from the nonlocal resonance sequence, we focus on the low-frequency 
quadrupolar modes. 
In \cref{bow_fig:gap_scan_low_quads_gap_fac_avert_corner_measures}, we 
consider the low-frequency bonding quadrupole for $a=10$ nm and 
$\tilde{a} = 12$ nm. We restrict the analysis to $\varphi=90\degree$,
since the corresponding resonance under $\varphi=0\degree$ excitation is 
approximately insensitive to the gap size. The gap size is varied 
over $g \in \{ 10,8,6,4,2 \}$ nm. 
We observe a stronger enhancement localization toward the gap (Panel A) 
than toward the outer corners (Panel B), consistent with the field 
distribution shown in the inset of Panel A. The extent of this 
localization increases monotonically with decreasing gap size, as 
does the difference between the two measures (Panel C). In all cases, 
relative to the Drude model,
the Halevi model leads to a reduction of the the near-field enhancement 
and introduces a 
partial blueshift in addition to the overall redshift associated with 
decreasing gap size.

\begin{figure*}
	\centering
	\includegraphics[scale=.5]{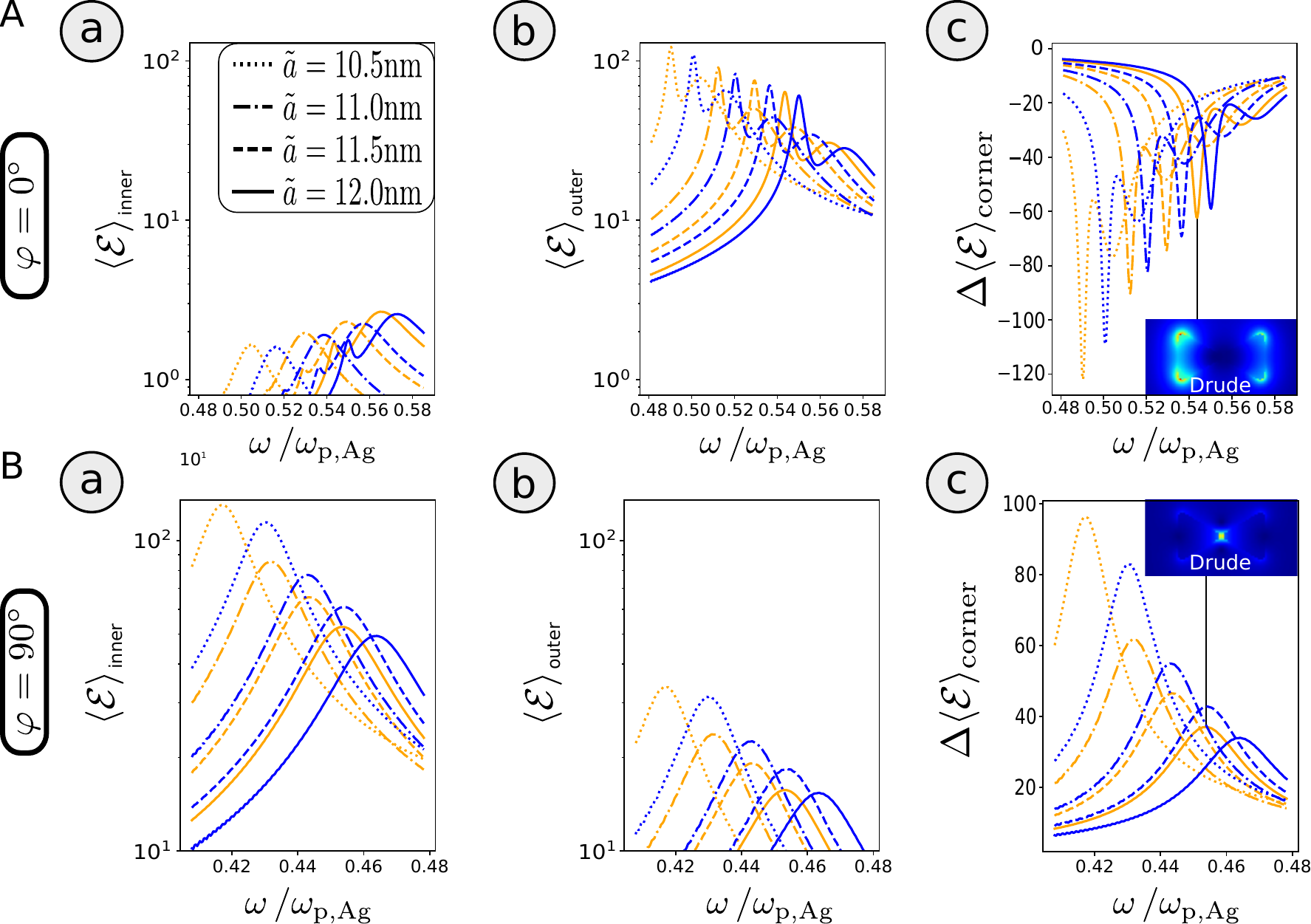}
	\caption{
		\label{bow_fig:curv_scan_low_quads_gap_fac_avert_corner_measures}
		Curvature dependence of the mean inner- and outer-corner enhancements 
		of the low-frequency quadrupoles sustained by the bowtie nanowire 
		with $a=10$ nm and $g=2$ nm for $\tilde{a}\in\{12.0;11.5;11.0;10.5\}$ nm (see legend in Panel A, subpanel (a)).	
		Panel A: Curvature dependence of the low-frequency quadrupoles for
		$\varphi=0\degree$. The Drude (orange) and Halevi (blue) models 
		are considered. Shown are the inner-corner (subpanel a) and 
		outer-corner enhancements (subpanel b), as well as their difference
		$\Delta \langle \mathcal{E} \rangle_{\rm corner} = \langle \mathcal{E} \rangle_{\rm inner} - \langle \mathcal{E} \rangle_{\rm outer}$ 
		(subpanel c), all as functions of frequency over the spectral range 
		of the respective LSP peaks. In subpanel (c), an inset displays the 
		field distribution at the peak maximum frequency of the bonding 
		quadrupole for $\tilde{a}=12$ nm, obtained using the Drude model. 
		Red (blue) colors indicate the maximum (minimum) value.
		Panel B: Same as Panel A, but for $\varphi=90\degree$.
	}
\end{figure*}

Finally, we investigate the curvature dependence for $a=10$ nm, $g=2$ nm, 
and $\tilde{a} \in \{ 2.0,11.5,11.0,10.5 \}$ nm. Owing to the influence 
of curvature on all corners, we consider both excitation directions
in \cref{bow_fig:curv_scan_low_quads_gap_fac_avert_corner_measures}:
$\varphi=0\degree$ (Panel A) and $\varphi=90\degree$ (Panel B). The 
corresponding Drude-model field distributions of the low-frequency 
bonding quadrupole for the least curved structure are shown in 
subpanel c of Panel A and Panel B, respectively. Similar to the 
extinction characteristics, curvature tuning produces a stronger effect 
than gap-size variation.

For parallel excitation ($\varphi=0\degree$, Panel A), we first observe
a decrease of the inner-corner enhancement (subpanel a), accompanied 
by a degradation of the low-frequency bonding quadrupole relative to 
the antibonding quadrupole. For the strongest curvature, only the 
latter remains clearly discernible. At the same time, the outer-corner 
enhancement (subpanel b) increases, with the bonding mode exceeding an 
enhancement factor of 100 for both material models. Accordingly, 
subpanel c reveals an increasing difference between the two measures 
for the bonding resonance for all curvatures and both material models.

In contrast, for the bonding resonance excited with $\varphi=90\degree$ 
(Panel B), the inner-corner enhancement (subpanel a) increases more 
strongly and eventually exceeds an enhancement factor of 100 for both 
material models. The outer-corner enhancement (subpanel b) also 
increases, but less pronouncedly, resulting in monotonous increase of the difference 
between the two measures. For both excitation directions, the resonance 
shifts follow the expected trends, while the Halevi model consistently
 quenches the near-field enhancement.

\subsubsection{Gap measures}

\begin{figure*}
	\centering
	\includegraphics[scale=.5]{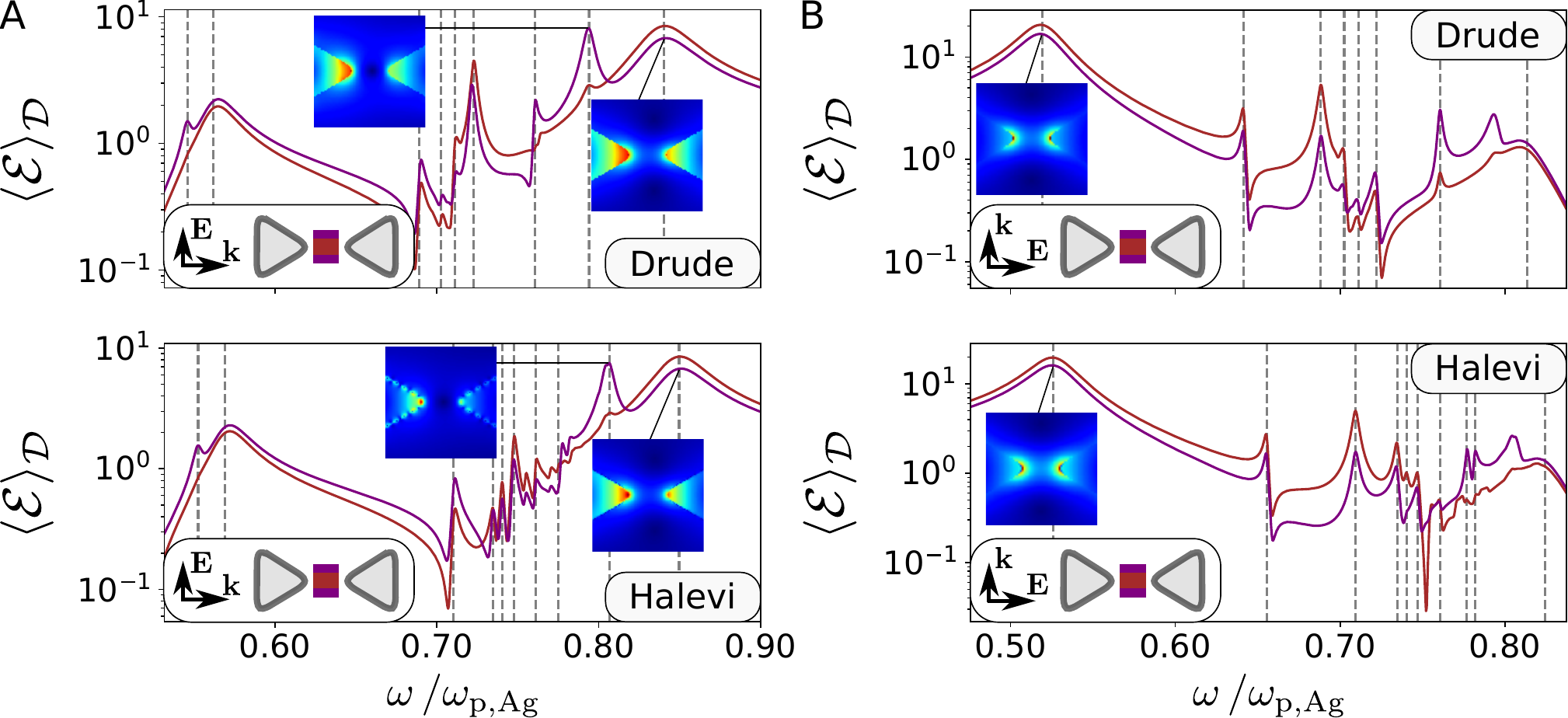}
	\caption{
		\label{bowtie_fig:gap_measures_both_phi_fix_geom}
		Mean-gap-domain enhancement of the electric field associated with 
		LSPs sustained by the bowtie nanowire for $a=10$ nm, 
		$\tilde{a}=12$ nm, and $g=10$ nm.	
		Panel A: Frequency dependence of the mean-gap-domain enhancement for $\varphi=0\degree$. The gap-centered and gap-peripheral enhancements 
		are compared, each evaluated using the Drude model (upper subpanel) 
		and the Halevi model (lower subpanel). The legend and incident-field
		polarization are provided in the lower-left box of each subpanel. 
		Vertical dashed lines indicate the peak-maximum frequencies of the
		corresponding extinction efficiency (see also \cref{fig_bow:curv_and_gap_scan_of_Q_ext_bowtie_Drude_Haleiv_phi_0}). Insets 
		show the spatial distribution of the electric-field modulus at the respective frequencies, normalized by the maximum value within the displayed spatial domain $(x,y\in[-15,15]\,\mathrm{nm})$. Red (blue)
		denotes the maximum (minimum) value.
		Panel B: Same as Panel A, but for $\varphi=90\degree$. 
		}
\end{figure*}

Having demonstrated gap localization for both incidence directions, we now 
aim to distinguish resonances featuring hotspots exactly at the gap center 
from those exhibiting a cold spot there. 
For this purpose, we consider the gap-centered ($d_1=1$ nm) and gap-peripheral measures ($d_1=1$ nm; $d_2=4$ nm; $d_3=6$ nm), respectively (analogous to 
the cylindrical dimer case illustrated in \cref{cyl_dimer_fig:enh_measures_cyl_dimer_both_inc_dir} Panels C and D). 
The results are compiled in  \cref{bowtie_fig:gap_measures_both_phi_fix_geom}.

As a general trend for both incidence directions, and in accordance with 
the association between extinction and enhancement maxima, we observe the 
nonlocal blueshift and the emergence of the nonlocal peak sequence. 
Furthermore, since the outer flanks of the central spectral range are 
expected to be occupied by high-order resonances with strong surface 
localization, it is reasonable that the gap measures locally decrease 
in these regions and subsequently increase again towards the respective
quadrupolar resonance peaks. Note that, for each incidence direction, the 
spectral regions with large gap enhancement coincide with those exhibiting 
strong inner-surface enhancement in
\cref{bow_fig:aver_surf_doman_enhan_bowtie_both_phi_fixed_geom}.

For $\varphi=0\degree$ (see 
\cref{bowtie_fig:gap_measures_both_phi_fix_geom}) (Panel A)), this correspondence 
concerns the bonding and antibonding high-frequency quadrupolar LSPs. Due 
to their different charge distributions (see insets of
\cref{bow_fig:Q_ext_one_setup_phi_0_90_Halevi_Drude}), 
the antisymmetrically charged bonding resonance exhibits a coldspot at 
the gap center, whereas the mirror-symmetrically charged antibonding
mode features a hotspot there. This is reflected in the different 
relative contributions of the respective gap measures for the two 
resonances (see also the displayed field distributions), highlighting 
the utility of distinguishing between gap-peripheral and gap-centered 
measures. A direct comparison of the field distributions of the 
high-frequency bonding LSP in the Drude and Halevi models further 
reveals the influence of at least one higher-order resonance. 
However, this overlap does not eliminate the field minimum at the 
gap center.

Towards lower frequencies, the gap-peripheral measure dominates. Moreover, 
for the low-frequency bonding quadrupolar LSP, the gap-centered enhancement 
is smaller than that of the corresponding antibonding mode, consistent
with the point-symmetric charge distribution of the former. For $\varphi=90\degree$ (see \cref{bowtie_fig:gap_measures_both_phi_fix_geom} 
(Panel B)), both measures are 
shifted towards lower frequencies. In the case of the low-frequency 
bonding quadrupolar LSP, the absence of a charge-density node at the 
gap-facing corners prevents point symmetry and results in a stronger 
gap-centered enhancement (see the field distributions for both the 
Drude and Halevi models).

\begin{figure*}
	\centering
	\includegraphics[scale=.7]{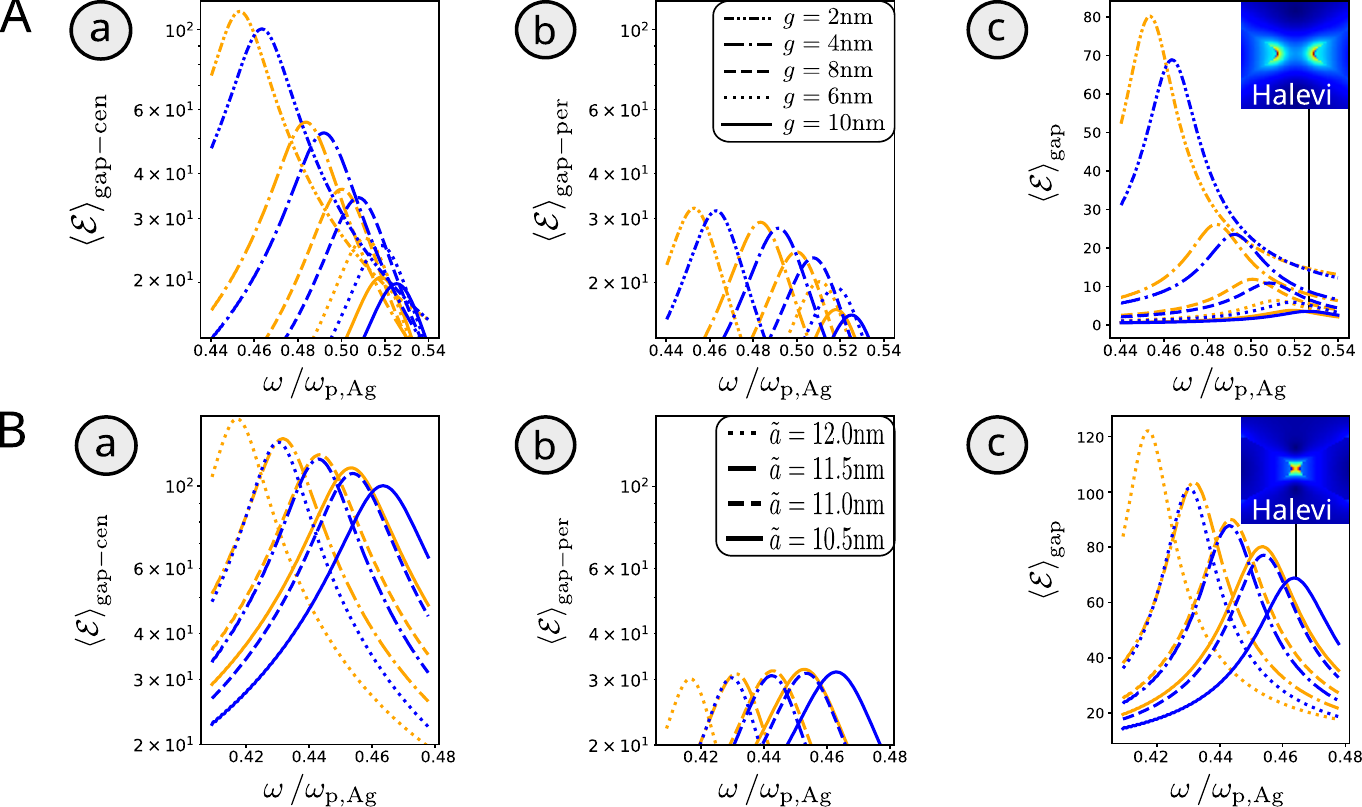}
	\caption{
		\label{bowtie_fig:gap_measures_both_scans_phi_90_low_freq_range}
          Geometrical-parameter dependence of the gap-centered and 
          gap-peripheral mean enhancements of the low-frequency bonding 
          quadrupole sustained by the bowtie nanowire with $a=10$ nm for
          $\varphi=90\degree$.         
          Panel A: Gap-size dependence of the low-frequency quadrupole 
          for $g\in\{10;8;6;4;2\}$ nm (see legend in subpanel (b)) and
          $\tilde{a}=12$ nm. The Drude (orange) and Halevi (blue) models 
          are considered. Shown are the gap-centered (subpanel a) and gap-peripheral enhancements (subpanel b), as well as their 
          difference $\Delta \langle \mathcal{E} \rangle_{\rm gap} = \langle \mathcal{E} \rangle_{\rm gap-cen} - \langle \mathcal{E} \rangle_{\rm gap-per}$ (subpanel c), all as functions of frequency over the 
          spectral range of the respective LSP peaks. In the inset of 
          subpanel (c), the field distribution at the peak maximum frequency 
          of the bonding quadrupole is shown for $\tilde{a}=12$ nm and 
          $g=10$ nm, obtained using the Halevi model. Red (blue) colors 
          indicate the maximum (minimum) values. 
          Panel B: Same as Panel A, but illustrating the curvature dependence 
          for $\tilde{a}\in\{12.0;11.5;11.0;10.5\}$ nm and $g=2$ nm. The field distribution in the inset of subpanel (c) is evaluated using the
           Halevi model for $g=2$ nm.
	}
\end{figure*}

With reference to \cref{bowtie_fig:gap_measures_both_scans_phi_90_low_freq_range}, we conclude this section by examining 
the influence of gap (Panel A) and curvature variations (Panel B) on 
a gap-centered resonance. To avoid overlap with high-order resonances 
and to retain a pronounced sensitivity to gap-size variations, we focus 
on the bonding low-frequency quadrupolar LSP for $\varphi=90\degree$. 
Overall, we observe the characteristic field quenching and blueshifting 
effects of the Halevi model. In all cases, the gap-centered measure 
dominates both in magnitude and in its increase, consistent with the 
field distributions shown for the Halevi model with $a=10$ nm, 
$\tilde{a}=12$ nm, and either $g=10$ nm (Panel A, subpanel c) or 
$g=2$ nm (Panel B, subpanel c).

Reducing the gap size results in a redshift of the resonance, consistent 
with its bonding character. Due to the particular charge distribution, 
increasing the curvature also leads to a redshift, see the insets of \cref{bowtie_fig:gap_measures_both_scans_phi_90_low_freq_range} 
(Panel B). A reduction of the gap size alone already
increases the gap-centered enhancement beyond a factor of 100 for both 
material models. This value is further enhanced by increasing the 
curvature, reaching factors of $152$ ($132$) in the Drude (Halevi) 
model. However, the absolute change of each measure is larger for gap 
variations, whereas the gap-peripheral measure decreases slightly with 
increasing curvature. Consequently, the difference between the two
measures reaches larger positive values for curvature variations.

\begin{figure*}
	\centering
	\includegraphics[scale=.5]{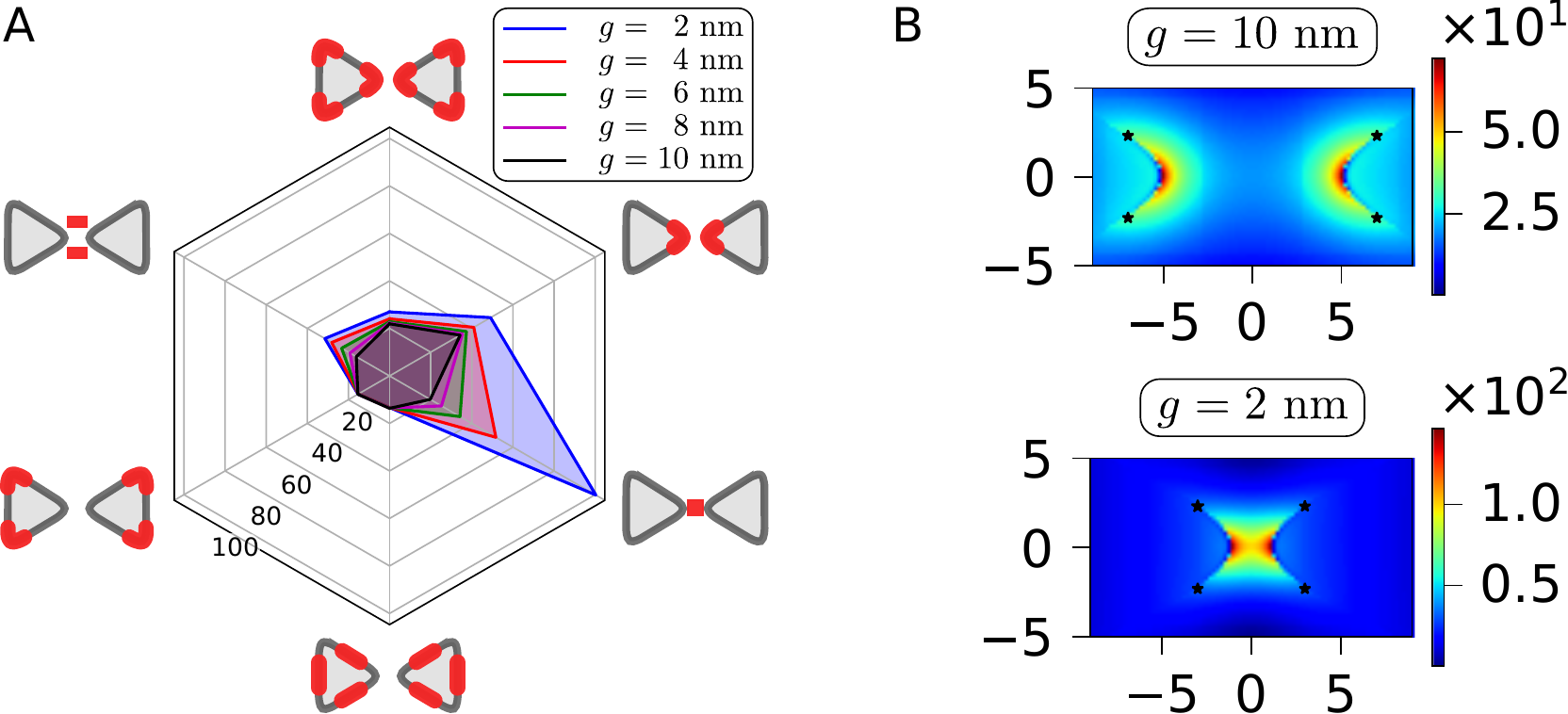}
	\caption{
		\label{bowtie_fig:star_plot}
		Gap-size dependence of all domain measures evaluated for the low-frequency bonding quadrupolar mode supported by the bowtie structure for $\varphi=90\degree$. Panel A: Radar chart summarizing all domain-enhancement measures defined for the bowtie geometry (see sketches at the top of the corresponding axes). The parameters are $a=10$ nm, $\tilde{a}=12$ nm, and $g\in\{10;8;6;4;2\}$ nm (see legend in the upper right corner). For the surface-domain measures, a sleeve width of $\delta=1.4$ nm is employed, while the gap-domain measures are calculated using $d_1=1$ nm, $d_2=4$ nm, and $d_3=6$ nm. Panel B: Field-amplitude distributions at resonance for $g=10$ nm (top) and $g=2$ nm (bottom), normalized by the incident field amplitude and obtained using the Halevi model. The black stars indicate the transition points between the Bézier-defined corners and the straight edges, providing an estimate of the extent of the inner-corner domain. 
	}
\end{figure*}

\subsubsection{Systematic analysis of mean-domain enhancement measures}

To conclude the discussion of mean-domain enhancement, we focus on the low-frequency bonding quadrupolar LSP excited in the bowtie for
 $\varphi=90\degree$. In \cref{bowtie_fig:star_plot}(Panel A), we compile all 
 surface- and gap-domain measures for different gap sizes while keeping 
 $a=10$ nm and $\tilde{a}=12$ nm fixed. For the largest gap size 
 ($g=10$ nm), we observe a dominance of the inner-corner enhancement 
 over all other measures. 
 This is consistent with the amplitude enhancement shown at the top 
 of \cref{bowtie_fig:star_plot} (Panel B).

As the gap size decreases, the inner corners, and consequently their 
associated fields, approach each other. Simultaneously, the regions 
of enhanced field amplitude along the corners become increasingly 
confined towards the $x$-axis, while the fixed curvature preserves 
the spatial extent of the corner domains (see the stars in the field 
distributions marking the endpoints of the B\'{e}zier paths). Consequently, 
Panel A reveals a transition towards a dominant gap-centered enhancement, 
which becomes apparent from $g=4$ nm onward. This dominance is further 
enhanced for $g=2$ nm, in agreement with the amplitude enhancement 
displayed at the bottom of \cref{bowtie_fig:star_plot}(Panel B).

\section{\label{sec:conclusion_and_outlook}Conclusion and outlook}

Within this study, we have investigated the optical extinction and near-field
response of cylindrical and bowtie nanowire dimers under plane-wave excitation. 
By additionally considering a triangular monomer nanowire, we were able to
distinguish shape-related (non)local effects from separation-related
contributions. We have considered two orthogonal
polarization states that are sufficient to excite all bright modes under
plane-wave illumination with reference to symmetry arguments.

To the best of our knowledge, this work provides the first demonstration 
of curvature-dependent longitudinally nonlocal effects in localized surface 
plasmons (LSPs). These findings emphasize the necessity of a complete 
analytical description of the scattering geometry, even in fully numerical
studies. In particular, all geometrical parameters and their relations to
geometric observables must be explicitly provided. Furthermore, their 
connection to relevant physical quantities, such as dispersion relations,
should be analyzed. The parametrization of corners via B\'{e}zier curves 
facilitates a $C_1$-continuous boundary while simultaneously allowing 
for an analytical expression of the surface-to-volume ratio, which 
constitutes a central quantity governing (quasistatic) longitudinally 
nonlocal effects in LSPs.

Based on a combined analysis of extinction spectra and charge distributions, 
we identified the majority of resonances as LSPs, while demonstrating that 
each structure also supports longitudinal volume plasmon oscillations. In
addition, we confirmed the applicability of the LSP hybridization concept. 
While this concept is intuitive for dimers, a complete analytical justification 
of an analogous description for the triangular monomer nanowire remains an
open question. Among the hybrid resonances, the bowtie bonding and antibonding 
LSP pairs exhibited the strongest geometrical tunability. For cylindrical 
and bowtie dimers with comparable linear dimensions, only the bowtie 
geometry exhibits resonance doublets allowing for shifts of opposite and similar direction, resulting from the combined variation of gap size 
and curvature. This highlights the distinction between a separation-based 
and a constituent-based interpretation of plasmon hybridization.

In particular, we identified low-order LSPs in both dimers that are highly
sensitive to the gap size. This raises the question of how effective 
piezoelectric elements may be for dynamically tuning nanogap dimensions 
in experimental setups. At the same time, the gap-insensitive LSPs supported 
by the bowtie structure could serve as stable reference signals for 
modulation techniques.

To systematically investigate the near-field response of a given scatterer 
and simplify the identification of characteristic hot- and cold-spot
distributions, we introduced a set of mean-domain-enhancement measures. 
For dimers, these measures comprise surface- and gap-related domains. 
Overall, the enhancement maxima were found to occur close to the 
extinction maxima, further supporting their association with LSPs, 
while longitudinal nonlocality was shown to generally reduce the resonant 
field amplitudes. Compared to the cylindrical dimer, the bowtie geometry 
exhibits a clearer distinction between inner- and outer-surface 
enhancement, originating from the symmetry breaking induced by the corners
 and the comparatively large effective surface wavelength of the low-order 
 modes. The transition between inner- and outer-surface enhancement 
 dominance with increasing frequency can be reversed by switching the 
 incident field polarization. We envision this property as a means to 
 control plasmon–emitter coupling for emitters located at different
 positions~\cite{Tang2018}.

Furthermore, both dimers support resonances with gap-centered cold spots, 
which have previously been proposed as non-diffraction-limited alternatives 
to Laguerre–Gaussian beams for fluorescence switching in microscopy. In 
the cylindrical dimer, we additionally identified a pair of gap-centered 
hot- and cold-spot modes with similar resonance frequencies and relative
 weights controlled by the incident polarization. Consequently, a single 
 laser source combined with a linear polarizer and mirror system could 
 enable fluorescence switching of one or multiple molecules located at 
 identical or distinct positions.

Given the complexity of hot- and cold-spot landscapes, these distributions 
may also be exploited for the fabrication of nanostructure templates using
photo-sensitive and photo-resistant materials. Moreover, the possibility of
locally deforming polymer coatings through controlled spatial variations 
 of the optical near field, as demonstrated in Ref.~\cite{Haggui2012}, 
 deserves further investigation. Owing to the combined variation of curvature 
 and gap size, the bowtie geometry supports a gap-centered hot spot exceeding 
 the threshold required for single-molecule SERS.
 
The gap-centered cold spots identified in this work are directly surrounded 
by a hot-spot halo. For cylindrical dimers, we demonstrated that this 
surrounding region also contains areas suitable for single-molecule 
SERS. Furthermore, an appropriate choice of corner curvature enables 
sufficiently strong hot spots at all four outer corners of the bowtie 
structure. This motivates further exploration of hot spots outside the 
nanogap and highlights the potential of dimer-axis-parallel excitation 
using plane waves. These corner sites could enable simultaneous 
investigation of multiple specimens, for example via SERS, allowing 
similarities and differences between them to be identified and 
classified.

To facilitate systematic near-field comparisons, we advocate the use of 
radar charts. For selected resonances and fixed geometrical parameters, 
all mean-domain-enhancement measures can be summarized within a single
representation. Multiple charts may correspond to variations in resonance 
order, geometry, material model, or excitation conditions such as 
polarization. As an extended application, this approach could be applied 
to nanoparticle chains, such as those studied in Ref.~\cite{Zito16},
where multiple enhancement measures associated with individual gaps 
could be compiled to rapidly identify gap-specific hot- and cold-spot
characteristics of different chain resonances.

A common feature of the extinction and near-field analyses of all three 
structures is the emergence of non-degenerate high-order LSP spectra 
induced by longitudinal nonlocality. The width of these nonlocal peak 
sequences was found to depend primarily on the gap size for the cylindrical 
dimer, whereas it was more strongly influenced by curvature for the 
bowtie structure due to similar characteristics of its monomer 
constituent. The widening of these sequences suggests potential 
applications in light harvesting. Furthermore, the sequence appears 
in the cylindrical dimer only for excitation along the dimer axis, 
a configuration that is frequently neglected in the literature. For 
the other two structures, both incidence directions provide access 
to these modes.

The corresponding resonances possess short effective surface wavelengths,
resulting in field distributions extending along larger portions of the 
surface. The increasing surface confinement with mode order is associated 
with enhanced field intensities and may therefore lead to larger effective
SERS-active areas. Following experimental confirmation of the nonlocal 
peak sequences, future studies could investigate the coupling of these 
modes to ensembles of emitters embedded in dielectric coatings, similar 
to the fabrication approaches of Refs.~\cite{Kanehira2023,SteteHeurQuantumModell,doi:10.1021/acsphotonics.8b00766,Rothe_2019}, for all geometries considered here.

We note that additional loss mechanisms not captured by the Halevi model 
may distort these sequences within the spectrum, potentially transforming 
them into shoulder-like features. Compared to low-order LSPs, high-order
resonances couple less efficiently to plane waves. Future work should 
therefore explore more efficient excitation mechanisms, such as electron 
beams, following the discussion in Ref.~\cite{Christensen2014}, where 
differences in incident wave fronts and additional characteristic length 
scales may enable improved coupling. Such studies would also provide 
a route toward identifying dark bowtie modes beyond the plane-wave
excitation investigated here.

Finally, we emphasize that the Halevi model neglects transverse 
nonlocality associated with shear effects in the electron 
fluid~\cite{PhysRevB.60.7966,wegner2024diss,universe7040108,Monticone2025}. 
Future  investigations should therefore address the influence 
of transverse nonlocality, including a study of the existence of transverse 
shear plasmon waves extending throughout the material volume, analogous 
to longitudinal volume and localized plasmon modes.

\begin{acknowledgements}
	G.W., U.P., and K.B. acknowledge funding by the German Research Foundation (DFG) in the framework of the CRC 1375 (Project ID 398816777 - Project A06).
\end{acknowledgements}

\appendix 

\section{\label{suppl_I_sec:tri_geom}Geometry of the triangular nanowire}

\begin{figure*}
	\centering
	\includegraphics[scale=0.5]{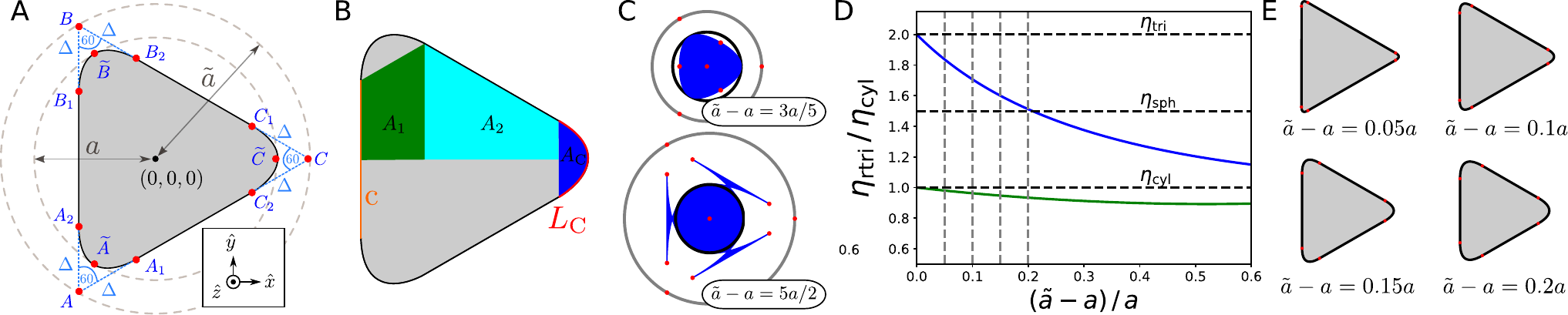}
	\caption{
		\label{suppl_fig:geom_triangle}
		Geometrical properties of the B\'{e}zier-rounded equilateral triangle.
		Panel A: Construction of the rounded triangle, circumscribed by
		a reference circle of radius $a$, from an initially sharp 
		equilateral triangle circumscribed by a circle of radius 
		$\tilde{a}$. Panel B: Partial curves and areas used to derive 
		the circumference and total area. Panel C: Illustration of the 
		bluntest triangle with $\tilde{a}-a=3a/5$ (upper subpanel) and 
		of a highly singular geometry with $\tilde{a}-a>3a/5$ (lower 
		subpanel). Red dots denote the control points, while the grey 
		(black) curves indicate the outer (inner) reference cylinders. 
		Panel D: Surface-to-volume ratio of the B\'{e}zier-rounded triangle 
		with a circumscribing (solid blue line) and an inscribing (solid 
		green line) reference cylinder of radius $a$, normalized by the cylindrical 
		value. The vertical grey dashed lines mark the radial differences 
		used in the numerical simulations. The horizontal black dashed 
		lines indicate the reference ratios of the cylinder 
		($\eta_{\rm cyl}$), sphere ($\eta_{\rm sph}$), and sharp 
		triangle ($\eta_{\rm tri}$). Panel E: B\'{e}zier-rounded 
		triangles investigated in this manuscript, corresponding to 
		the radial differences marked in Panel D.
	}
\end{figure*}

In the main text, we have studied the optical properties of triangular and 
bowtie nanowires. Here, we provide the details of the geometrical 
parametrization of their horizontal cross sections. We start from an 
equilateral triangle circumscribed by a circle of radius $\tilde{a}$. 
To round each corner, we employ a second-degree B\'{e}zier 
parametrization~\cite{Farin_Curves_Surf_for_CAGD}. One of the three 
required control points per corner coincides with the original apex, while 
the remaining two are displaced by an equal distance $\Delta$ along the 
two intersecting triangle edges.

Taking the horizontal corner ($\mathbf{C}$) in
\cref{suppl_fig:geom_triangle}(Panel A) as 
an example, we define the parametrization
\begin{align}
	\mathbf{r}_C(t) = (1-t)^2 \mathbf{C}_1 + 2t(1-t)\mathbf{C} + t^2 \mathbf{C}_2 ,
\end{align}
with the control points
\begin{align}
	\mathbf{C}_1 & = (\tilde{a}- \sqrt{3}\Delta/2, \Delta/2)^{\tau}, \\
	\mathbf{C} & = (\tilde{a},0)^{\tau}, \\
	\mathbf{C}_2 & = (\tilde{a}- \sqrt{3}\Delta/2,-\Delta/2)^{\tau}.
\end{align}

Because
\begin{align}
	\partial_t \mathbf{r}_C(t=0) = 2(\mathbf{C}-\mathbf{C}_1)
\end{align}
and
\begin{align}
	\partial_t \mathbf{r}_C(t=1) = 2(\mathbf{C}_2-\mathbf{C}),
\end{align}
the tangents at the endpoints of the B\'{e}zier curve are parallel to 
the remaining straight triangle segments (from $B_2$ to $C_1$ and from 
$C_2$ to $A_1$). Consequently, no kinks arise at the transition between 
the rounded corners and the straight edges, thereby avoiding artificial 
sources of field enhancement localized at the displaced control points.

Let $\mathbf{\tilde{C}}$ denote the point at which the parametrization 
touches the circumscribing reference cylinder. Due to symmetry, we 
obtain
\begin{align}
	\tilde{C}_x = \left(\mathbf{r}_{C}\right)_x(t=1/2)=a
	\quad\Rightarrow\quad
	\Delta = \frac{4(\tilde{a}-a)}{\sqrt{3}} .
\end{align}
This relation illustrates how the limit $\tilde{a}\rightarrow a^+$ 
recovers the sharp equilateral triangle, where all three control points 
coincide. The control points of the remaining two corners follow from 
rotations by $\pm 2\pi/3$.

The choice of a circumscribing reference cylinder results in a horizontal 
cross section smaller than that of the corresponding cylindrical wire.
Accordingly, reduced volume-related losses, such as those arising from 
Drude damping, are expected.

The general expression for the radius of curvature is
\begin{align}
	{R}(t)=\frac{2}{3}\sqrt{3(2t-1)^2+1}^{\,3}(\tilde{a}-a).
\end{align}
The minimum curvature radius is obtained at the apex ($t=1/2$), whereas 
the maximum occurs at the outer control points.

To obtain the circumference, it suffices to consider only one edge and 
one corner parametrization, as exemplified in \cref{suppl_fig:geom_triangle} 
(Panel B). We obtain
\begin{align}
	c & = B_{1,x} - A_{2,x} = \sqrt{3}a - 5 \frac{\tilde{a}-a}{\sqrt{3}} ~~~~~ \text{and} ~~\\ \mathcal{L}_C & = \int_0^1 dt ~\sqrt{\Big[\partial_t \left(\mathbf{r}_{C}\right)_{x}(t)\Big]^2 + \Big[\left(\mathbf{r}_{C}\right)_{y}(t)\Big]^2} \\
	& = 2\frac{\tilde{a} -a}{3} \left[2\sqrt{3} + \operatorname{arsin} \sqrt{3}\right]
\end{align} 
such that
\begin{align}
	u_{\rm rtri} = 3 (c + \mathcal{L}_{C}) = 3 \sqrt{3} a - [\sqrt{3} - 2 \operatorname{arsinh} \sqrt{3}](\tilde{a} -a)
	\label{suppl_eq:circumference_rtri}
\end{align}
In addition, the (horizontal) area is obtained from two trapezoidal and one 
corner area according to
\begin{align}
	\mathcal{A}_1 &= \frac{1}{2}\left(B_{2,x}-B_{1,x}\right)
	\left(B_{2,y}+B_{1,y}\right) \\
	&= \sqrt{3} \left(\tilde{a}-a\right)\left(2a-\tilde{a}\right)~, \\
	\mathcal{A}_2 &= \frac{1}{2}\left(C_{1,x}-B_{2,x}\right)
	\left(C_{1,y}+B_{2,y}\right)\\
	&= \frac{\sqrt{3}}{8} 
	\left[ a+ \left(\tilde{a}-a\right)\right]
	\left[3a-5\left(\tilde{a}- a\right)\right] ~\text{and} \\
	\mathcal{A}_C &= \int_0^1 dt ~ \partial_t r_{C,y}(t) \left[r_{C,x}(t) - C_{1,x}\right] \\ 
	&= \frac{8}{3\sqrt{3}} \left(\tilde{a}-a\right)^2                             
\end{align}
such that
\begin{align}
	A_{\rm rtri} &= 2(\mathcal{A}_1+\mathcal{A}_2) + 3\mathcal{A}_C \\
	&= \frac{3\sqrt{3}}{4}a^2 + \frac{3\sqrt{3}}{2}a\left(\tilde{a}-a\right) - \frac{7\sqrt{3}}{12}\left(\tilde{a} -a\right)^2
	\label{suppl_eq:area_rtri}             
\end{align}
\Cref{suppl_eq:circumference_rtri,suppl_eq:area_rtri} both reproduce the 
limit of a sharp equilateral triangle of side length $\sqrt{3}a$. 
The surface-to-volume ratio for a wire of height $H$ can be written as 
\begin{align}
	\eta_{\rm rtri} & = \frac{u_{\rm rtri}H}{A_{\rm rtri}H} = 2 \eta_{\rm cyl} \, \chi\left(\frac{\tilde{a} -a}{a}\right) \\
	\text{where} ~ \chi(\nu) & = \frac{1 - (3\sqrt{3})^{-1} \left[\sqrt{3} - 2 \operatorname{arsinh}\sqrt{3}\right] \nu}{1 + (3\sqrt{3})^{-1} \left[6\sqrt{3} \nu - 7\nu^2 /\sqrt{3}\right]}   
\end{align}
with the cylindrical value of $\eta_{\rm cyl} = 2 / a$. An inspection of \cref{suppl_fig:geom_triangle} (Panel D) reveals, that the spherical value of $\eta_{\rm cyl} = 3 / a$ is eventually exceeded. This is true for the 
realizations which we consider in the main text 
(see \cref{suppl_fig:geom_triangle}(Panel E)). Note, the upper limit for 
the relative radial difference. It is obtained, when two control points 
on one edge of the initial triangle of length $\sqrt{3}\tilde{a}$ coincide, 
such that 
\begin{align}
	\sqrt{3}\tilde{a} = 2 \Delta ~~\Rightarrow ~~ \tilde{a}-a = 3 a / 5
	\label{eq:tri_geom_upper_bound_radial_diff}
\end{align} 
For larger relative differences, a singular structure is obtained as in \cref{suppl_fig:geom_triangle}(Panel C).

We note, that alternatively, we could have forced the reference cylinder 
of radius $a$ to inscribe the rounded triangle. This would fix the 
circumscribing circle of the sharp equilateral triangle to a radius 
of $2a$. The displacement $\Delta^{\rm alt}$ can be fixed via the 
corresponding radius of curvature according to $\Delta^{\rm alt} = (2\sqrt{3}) {R}^{\rm alt}(t=1/2)$. In order to be able to compare with the 
initial parametrization, we enforce
\begin{align}
	{R}^{\rm alt}(t=1/2) \stackrel{!}{=} {R}(t=1/2) = \frac{2}{3} \left(\tilde{a} -a\right)
\end{align}
From \cref{suppl_fig:geom_triangle}(Panel D) we deduce, that in the range 
of interest with $\tilde{a}-a \lesssim 3a/5$, the new parametrization 
falls below even cylindrical surface-to-volume ratio.

In a final step, we derive the geometrical cross sections. These are defined 
as the projection of the scatterer onto the plane, which is orthogonal to the incident plane wave propagation. For $\varphi=0\degree$, we consider the maximum $y$-component of the B\'{e}zier-curve near point $\mathbf{B}$. Consequently, 
we have $\sigma^{0\degree}_{\rm geom, rtri} = 2 r_{B,y}(t=t_{\rm max}) H$ and
\begin{align}
	& 0 \stackrel{!}{=} \partial_t r_{B,y}(t_{\rm max}) \\
	\Rightarrow ~& \sigma_{\rm geom, rtri}^{0\degree} = \left(\sqrt{3} a + \frac{\tilde{a} -a }{3\sqrt{3}}\right)H ~.
\end{align}
For $\varphi=90\degree$, we find 
\begin{align}
	\sigma_{\rm geom, rtri}^{90\degree} = \left(\frac{3 a}{2} + \frac{\tilde{a} -a }{2}\right)H~.
\end{align}
Both results comply with the expected limit of a sharp triangle.

\section{\label{suppl_I_sec:numerical_setup} Numerical setup}

\begin{table*}
	\centering
	\begin{minipage}{0.45\textwidth}      	
		\begin{tabular}{llc}
			\hline\hline  
			gap size [nm] & smallest insphere radius [nm]       & time step [as] \\
			$10$          & ~~~~~~$0.0449024$ (in right cylinder)    & $0.248$ \\
			$8$           & ~~~~~~$0.0449024$ (in right cylinder)    & $0.248$ \\
			$6$           & ~~~~~~$0.0449024$ (in right cylinder)    & $0.248$ \\
			$4$           & ~~~~~~$0.0449024$ (in right cylinder)    & $0.248$ \\
			$2$           & ~~~~~~$0.0449024$ (in right cylinder)   & $0.248$ \\
			\hline\hline
		\end{tabular}\\~\\
		\centering{ (a) ~Data of cylindrical dimer}
	\end{minipage}      		      	      	   		
	\begin{minipage}{0.5\textwidth}      	
		\begin{tabular}{llc}
			\hline\hline  
			outer radius [nm]       & smallest insphere radius [nm] & time step [as] \\
			\hline $12.0$ (right-orient.)  & ~~~~~~$0.045$ (in triangle)         & $0.248$ \\
			$11.5$ (right-orient.)  & ~~~~~~$0.043$ (in triangle)         & $0.237$ \\
			$11.0$ (right-orient.)  & ~~~~~~$0.044$ (in TF-domain)        & $0.246$ \\
			$10.5$ (right-orient.)  & ~~~~~~$0.045$ (in triangle)         & $0.250$ \\
			$12.0$ (left-orient.)   & ~~~~~~$0.045$ (in triangle)         & $0.250$ \\									
			\hline\hline
		\end{tabular}\\~\\
		\centering{ (b) ~Geometry data of triangular monomers}
	\end{minipage}  ~\\[0.5cm]
	\begin{minipage}{0.6\textwidth}      	
		\begin{tabular}{lllc}
			\hline\hline  
			outer radius [nm] & gap size [nm] & smallest insphere radius [nm]       & time step [as] \\
			\hline $12.0$     & $10$          & ~~~~~~$0.044$ (in left triangle)    & $0.241$ \\
			$12.0$            & $8$           & ~~~~~~$0.044$ (in left triangle)    & $0.241$ \\
			$12.0$            & $6$           & ~~~~~~$0.045$ (in left triangle)    & $0.248$ \\
			$12.0$            & $4$           & ~~~~~~$0.043$ (in left triangle)    & $0.240$ \\
			$12.0$            & $2$           & ~~~~~~$0.042$ (in right triangle)   & $0.233$ \\
			$11.5$            & $2$           & ~~~~~~$0.043$ (in TF domain)        & $0.240$ \\
			$11.0$            & $2$           & ~~~~~~$0.040$ (in left triangle)    & $0.219$ \\
			$10.5$            & $2$           & ~~~~~~$0.043$ (in left triangle)         & $0.236$ \\
			\hline\hline
		\end{tabular}\\~\\
		\centering{ (c) ~Geometric data of bowtie dimers}
	\end{minipage}  
	\begin{minipage}{0.39\textwidth}
		\begin{tabular}{lcc}
			\hline\hline  notation    &  quantity                & value         \\
			\hline $E_0$              &  field amplitude            & $1$    \\
			$\varphi$          & incidence angle                  & $\in \{0^{\circ} ; 90^{\circ}\}$ \\ 
			$\sigma_t$    &  temporal width             & $200$         \\
			$\omega_0$  &  carrier frequency  & $2\pi(0.006)$ \\
			$t_0$       &  pulse time shift           & $1400$ 		\\
			\hline\hline
		\end{tabular} \\~\\
		\centering{ (d) ~Data of Gaussian time dependence }	
	\end{minipage}    		      	
	\caption{
		\label{suppl_I_tab:gauss_pulse_in_sphere_all_setups}
		Listing of the parameters defining the minimum in-sphere radius 
		and simulation time step for the cylindrical dimer (a), triangular 
		monomer (b), and bowtie dimer (c), together with the Gaussian 
		temporal profile of the excitation pulse. The data of the latter are
		based on the length scale $\ell_0=1$ nm as well as the time scale $\ell_0 / c_0$ where $c_0$ is the vacuum speed of light.
	}
	
\end{table*}

Within the main text, we have studied the optical response of infinitely 
extended silver nanowires excited by a pulse of given carrier wavelength
and Gaussian envelope, see 
\cref{suppl_I_eq:incident_field_def} and 
\cref{suppl_I_tab:gauss_pulse_in_sphere_all_setups}(d) for 
the corresponding parameters. The infinite extent eliminates wire-end effects 
and provides an approximation for high-aspect-ratio wires (see, e.g., the 
electron-optical micrographs in
Refs.~\cite{Rothe_2019,doi:10.1021/acsnano.0c05240}). 
Due to the translational invariance along the wire axis and the polarization 
of the incident electric field within the horizontal plane, the 
electromagnetic problem can be reduced to two dimensions (the 
$xy$-plane).

The incidence direction is defined relative to the $x$-axis (which coincides 
with the dimer axis) by the angle $\varphi$, measured counter-clockwise. 
Based on the symmetry properties discussed for all three structures in 
the manuscript, we distinguish between incidence parallel 
($\varphi=0\degree$) and perpendicular ($\varphi=90\degree$) to the 
$x$-axis.

The complex geometries of the considered wires require a numerical solution 
of the boundary-value problem based on the macroscopic Maxwell and material
equations. Here, we employ the nodal Discontinuous Galerkin Time-Domain 
(DGTD) finite-element method~\cite{lpor_DGTD_review}. This approach retains 
the geometrical flexibility of conventional finite-element methods while 
being tailored to conservative-form equations. Consequently, it can 
handle continuum models describing dispersive (via auxiliary 
differential equations), lossy, and nonlocal bulk material 
responses \cite{wegner2024diss}, in addition to the Maxwell curl 
equations \cite{hesthaven2007nodal}. The dynamics governed by the latter 
preserve Maxwell's divergence constraints, which remain fulfilled due 
to the initially vanishing fields and charge 
densities~\cite{lpor_DGTD_review}.

The computational mesh is constructed as a conformal, unstructured, and 
adaptive collection of straight-sided triangular elements. Each element 
is equipped with a local interpolating Lagrange basis of global polynomial 
order $p=3$. Due to the element-local nature of the basis functions, the 
matrices associated with individual elements remain small, thereby 
facilitating efficient matrix operations, particularly inversion. Moreover,
these matrices are obtained from those of a single reference element via 
affine transformations. Consequently, only the corresponding Jacobian
matrix elements need to be stored for each element, reducing memory 
requirements. All element-local operations can be performed in parallel 
across the mesh, thereby improving computational efficiency.

Field discontinuities (according to the Maxwell boundary conditions at material interfaces) and the incident field values are introduced through numerical fluxes 
at the interfaces between neighboring elements. Note, that the interfacial basis nodes provided by each of the two neighboring elements coincide and the numerical flux connects the elements at these nodes. In contrast to finite-difference time-domain 
methods (FDTD), the electric and magnetic field nodes are co-located, 
allowing material interfaces to be represented sharply. The spatial 
discretization is refined near the metal-vacuum interface to resolve 
both the geometry and the field variations associated with the (non)local 
material response. Within a $0.2$ nm-wide layer surrounding the interface, 
we impose the smallest mesh-element edge length. Over a distance of
$2$ nm extending from this layer into the scatterer, the edge lengths 
increase linearly towards the bulk value of $0.6$ nm. Since the fields 
are evaluated at the basis nodes, the smallest physically resolved 
spatial scales are even smaller than the local element size.

For mesh generation, we employ the open-source finite-element and three-dimensional mesh generator Gmsh \cite{gmsh_article}, 
which also supports the implementation of the B\'{e}zier parametrization used in 
this work.

\begin{figure*}
	\includegraphics[scale=0.4]{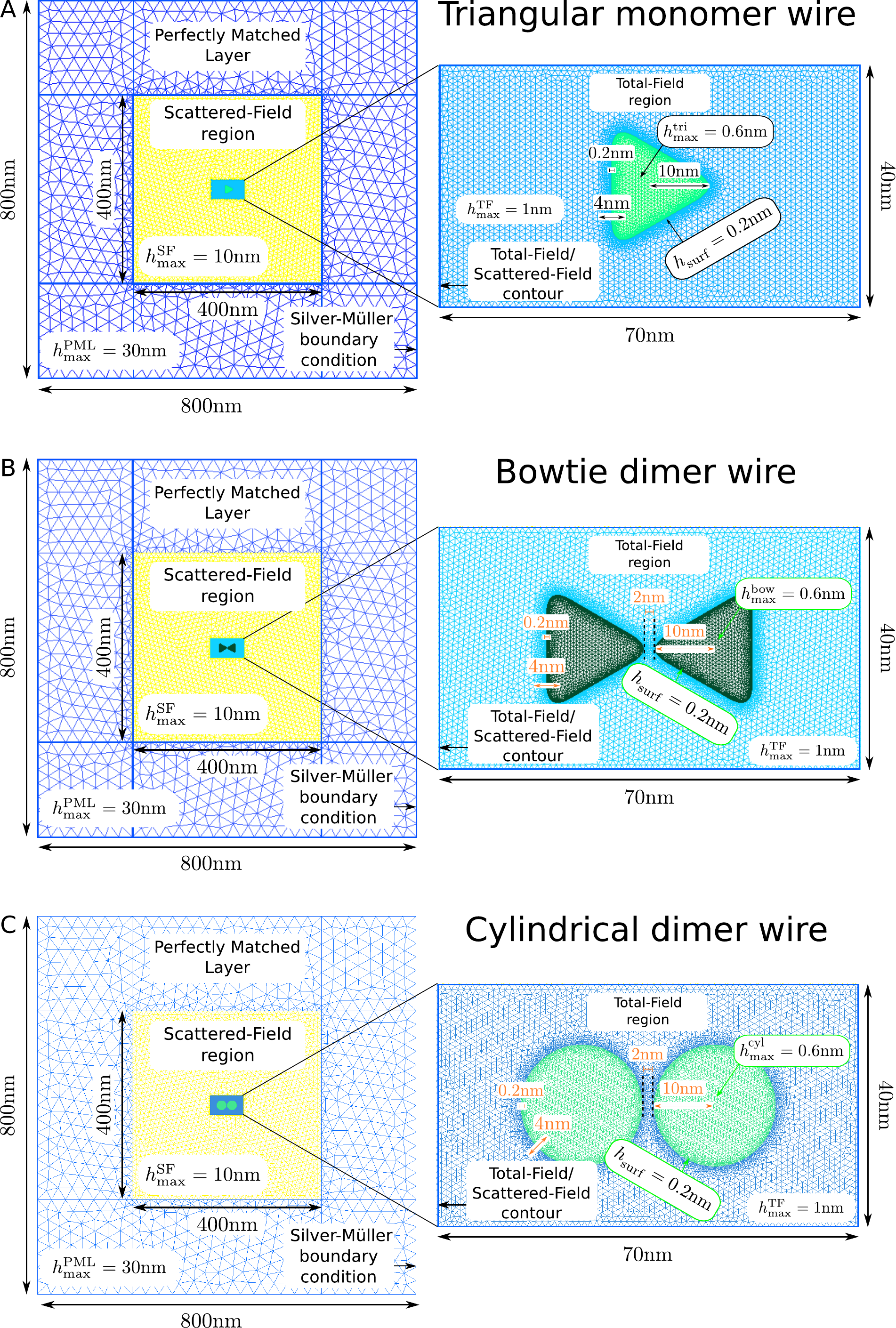}
	\caption{
		\label{suppl_fig:finest_mesh_examples_tri_bowtie_cylinder}
		Illustration of the finest mesh employed for the triangular 
		monomer (Panel A), bowtie dimer (Panel B), and cylindrical dimer 
		nanowire (Panel C) for representative sets of geometrical parameters. 
		For each geometry, the different subdomains (and contours) of the computational domain are labeled, and the optimal element edge 
		lengths ($h_{\rm max}$) are specified together with the 
		corresponding subdomain extents. The actual mesh element 
		sizes may deviate slightly from these prescribed values to 
		conform to the local geometry.
	}
\end{figure*}

\begin{figure*}
	\centering 
	\includegraphics[scale=0.44]{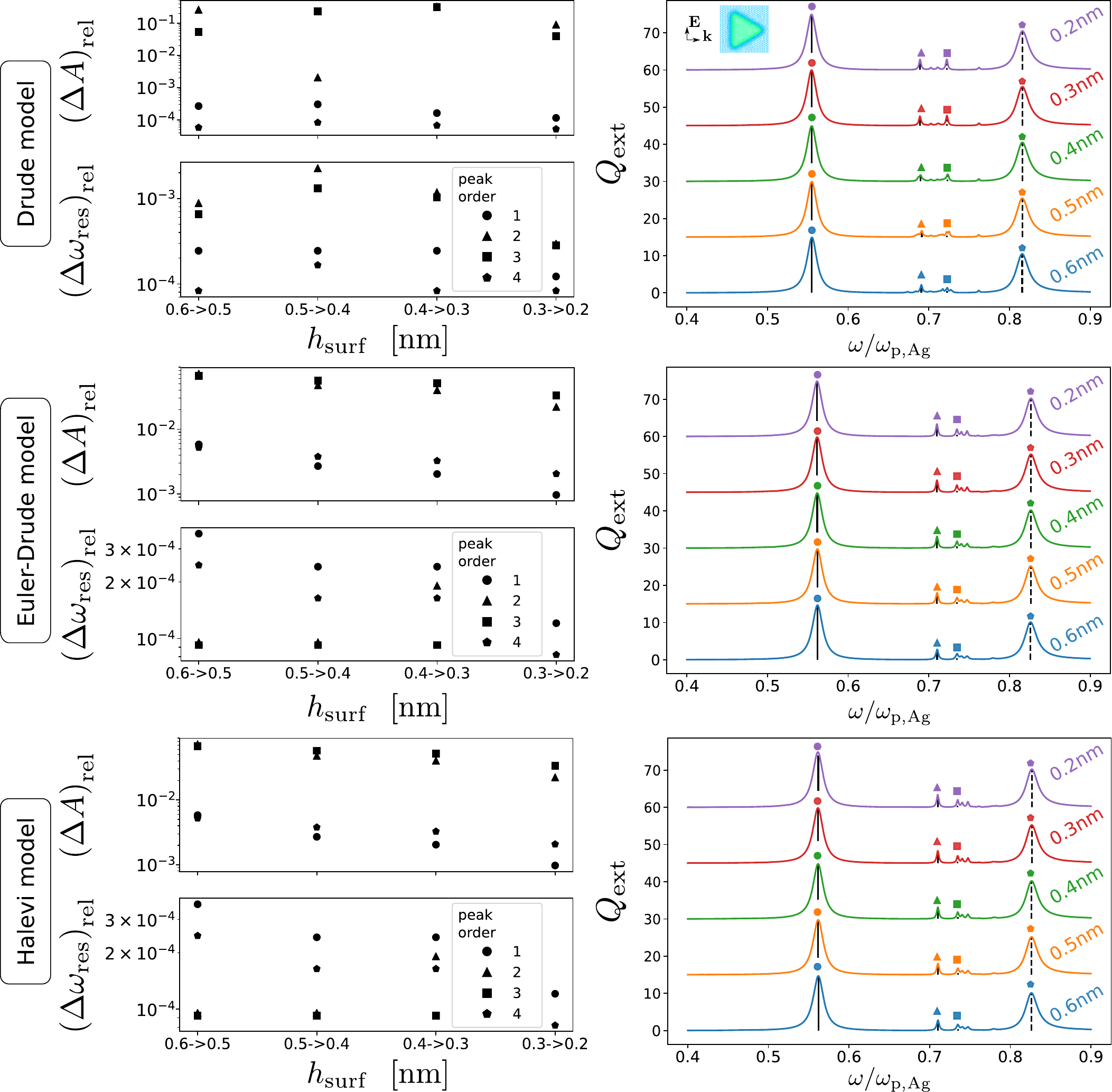}
	\caption[Convergence study of a B\'{e}zier-rounded and triangular nanowire monomer]{
		\label{suppl_fig:left_tri_convergence_peak_measures}
		Illustration of the final stage of a convergence study for a
		right-oriented triangular nanowire with geometrical parameters 
		$a=10$ nm and $\tilde{a}=12$ nm, excited by a plane wave with
		Gaussian temporal dependence and polarization angle 
		$\varphi=0\degree$. The optimal surface element edge length 
		is varied according to $h_{\rm surf}\in \{0.6;0.5;0.4;0.3;0.2 \}$ nm. 
		The simulation error is quantified using \cref{suppl_eq:rel_amplitude_error,suppl_eq:rel_frequency_error} 
		for the relative deviations in peak amplitude, 
		$\left(\Delta A\right)_{\rm rel}$, and resonance frequency, 
		$\left(\Delta \omega_{\rm res}\right)_{\rm rel}$ (left panels). 
		The right panels show waterfall plots of the extinction efficiency 
		for each value of $h_{\rm surf}$, with an artificial vertical offset 
		of 15 units. Results are presented for the Drude model (upper 
		panels), the Euler–Drude model ($\beta=\sqrt{3/5}v_{\rm F}$; 
		central panels), and the Halevi model (lower panels).}
\end{figure*}

In contrast to conventional finite-element methods, the DGTD formulation 
results in a semi-discrete system that is amenable to explicit time 
integration schemes. Among these, we employ an explicit low-storage 
Runge–Kutta scheme with 14 stages and fourth-order accuracy\cite{niegemann_efficient_2012} to complement 
the high spatial accuracy (of order $h^p$ for regular elements and $h^2$ 
for curved surfaces) with a corresponding temporal accuracy. The explicit 
nature of the time integrator imposes a conditional stability criterion, 
which we satisfy by selecting a global time step based on the smallest 
element of the mesh. The relatively large number of stages compared to 
the formal order increases the maximum stable time 
step~\cite{lpor_DGTD_review}.

\Cref{suppl_I_tab:gauss_pulse_in_sphere_all_setups}(a-c) summarizes, for each geometry and parameter 
configuration, the minimum insphere radius of the smallest element 
in the corresponding finest mesh and the resulting time step. For the 
cylindrical dimer, the size and location of the smallest element remain 
unchanged with varying gap size. In contrast, for the triangular wire 
and bowtie structures, the smallest element size depends on both 
$\tilde{a}$ and $g$. As can be seen, all element sizes are below 
the nonlocal length scale given by the Thomas–Fermi screening length 
($\ell_{\rm TF}\sim0.1$ nm for Ag).

Using the total-field/scattered-field (TF/SF) technique, we inject 
an excitation pulse of finite temporal duration at the TF/SF interface 
(see \cref{suppl_fig:finest_mesh_examples_tri_bowtie_cylinder}). This enables 
the extraction of spectral 
information over a finite frequency range from a single simulation. 
Furthermore, the finite pulse energy ensures that all excited 
resonances eventually decay. The spectral response is obtained via
an on-the-fly Fourier transform based on a time-step-wise update 
scheme, thereby reducing memory requirements. The total simulation 
time is $333.56$ fs, chosen to minimize oscillatory artifacts in 
the discrete Fourier transform.

To emulate an open system, the physical part of the computational 
domain is surrounded by a uniaxial perfectly matched layer (PML),
which is terminated by a contour on which a Silver–Müller boundary 
condition is applied. For each geometry, \cref{suppl_fig:finest_mesh_examples_tri_bowtie_cylinder} shows the 
tessellated computational domain corresponding to the finest mesh, 
including the individual subdomains, geometrical parameters, and 
mesh sizes. These meshes result from dedicated convergence studies. 
We note that, for a given geometry and set of geometrical parameters,
the same mesh is used for simulations with both incidence directions 
and all three material models.

\begin{figure*}
	\centering 
	\includegraphics[scale=0.44]{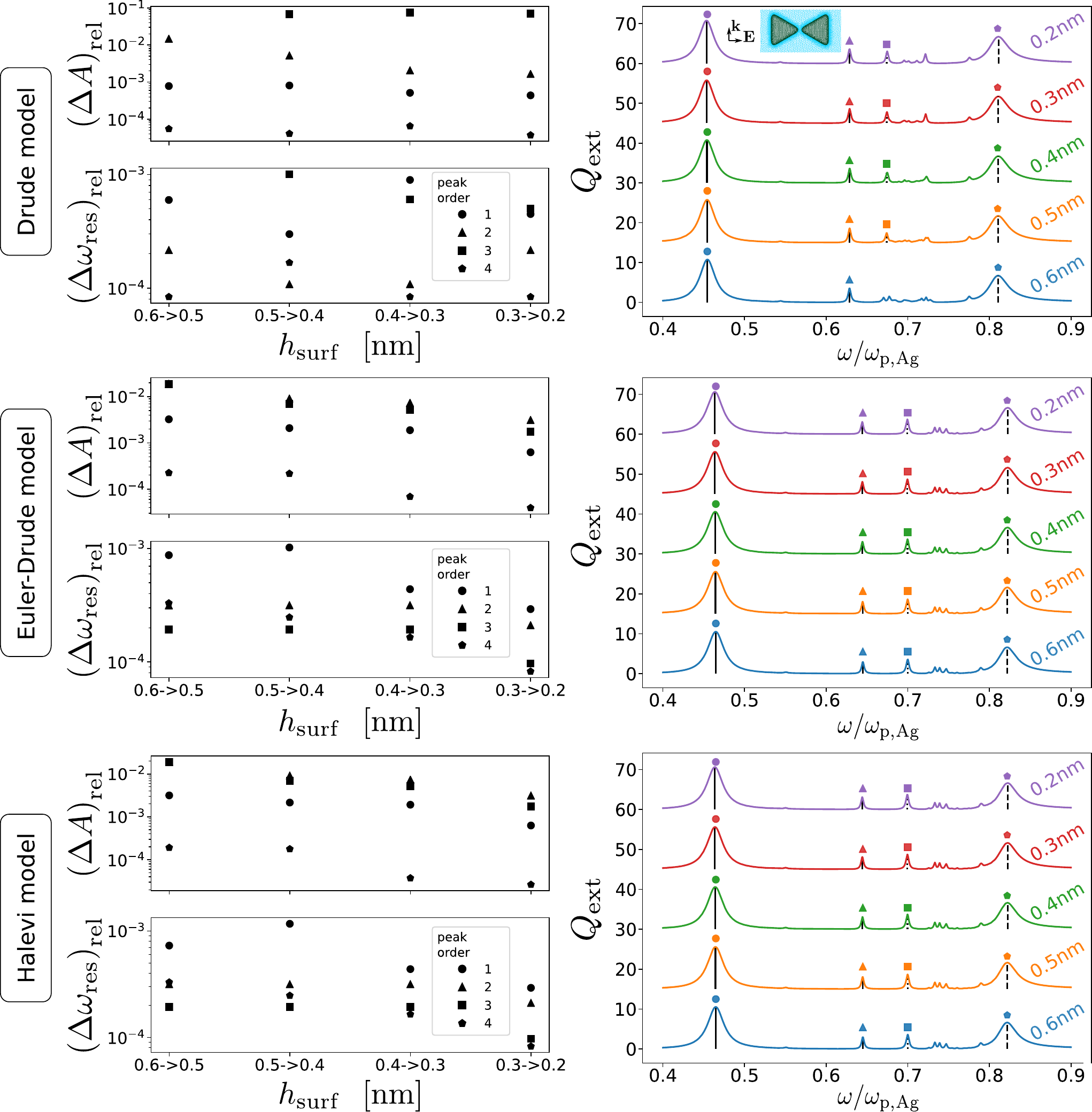}
	\caption[Convergence study of a B\'{e}zier-rounded bowtie nanowire dimer]{
		\label{suppl_fig:bowtie_convergence_peak_measures}
		Illustration of the final stage of a convergence study for the 
		simulation of a bowtie dimer with geometrical parameters 
		$a=10$ nm, $\tilde{a}=12$ nm, and $g=2$ nm, excited by a plane 
		wave with Gaussian temporal dependence and polarization angle $\varphi=90\degree$. The optimal surface element edge length 
		is varied according to $h_{\rm surf}\in \{0.6;0.5;0.4;0.3;0.2 \}$ nm. 
		The simulation error is quantified using \cref{suppl_eq:rel_amplitude_error,suppl_eq:rel_frequency_error} 
		for the relative deviations in peak amplitude, 
		$\left(\Delta A\right)_{\rm rel}$, and resonance frequency, 
		$\left(\Delta \omega_{\rm res}\right)_{\rm rel}$ (left panels). 
		The right panels show waterfall plots of the extinction efficiency 
		for each value of $h_{\rm surf}$, with an artificial vertical 
		offset of 15 units. Results are presented for the Drude model 
		(upper panels), the Euler–Drude model 
		($\beta=\sqrt{3/5}v_{\rm F}$; central panels), and the Halevi 
		model (lower panels). 
	}
\end{figure*}

\begin{figure*}
	\centering 
	\includegraphics[scale=0.44]{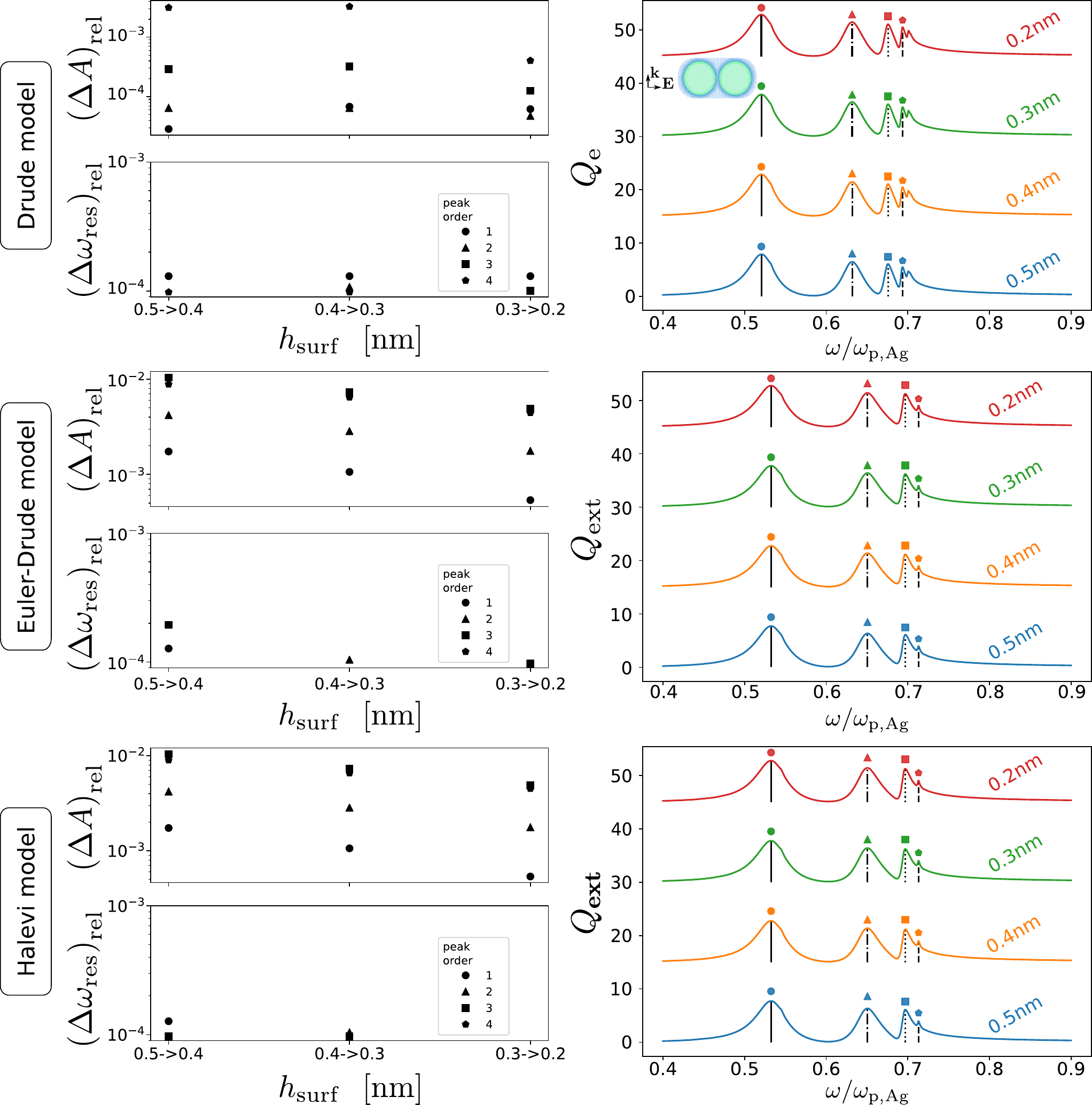}
	\caption[Convergence study of a B\'{e}zier-rounded bowtie nanowire dimer]{
		\label{suppl_fig:cyl_dim_convergence_peak_measures}
		Illustration of the final stage of a convergence study for the 
		simulation of a cylindrical dimer with geometrical parameters
		$a=10$ nm and $g=2$ nm, excited by a plane wave with Gaussian 
		temporal dependence and polarization angle $\varphi=90\degree$. 
		The optimal surface element edge length is varied according to 
		$h_{\rm surf}\in \{0.5;0.4;0.3;0.2 \}$ nm. The simulation error is 
		quantified using \cref{suppl_eq:rel_amplitude_error,suppl_eq:rel_frequency_error} 
		for the relative deviations in peak amplitude, 
		$\left(\Delta A\right)_{\rm rel}$, and resonance frequency, 
		$\left(\Delta \omega_{\rm res}\right){\rm rel}$ (left panels).
		The right panels show waterfall plots of the extinction efficiency 
		for each value of $h_{\rm surf}$, with an artificial vertical offset 
		of 15 units. Results are presented for the Drude model (upper 
		panels), the Euler–Drude model ($\beta=\sqrt{3/5}v_{\rm F}$; 
		central panels), and the Halevi model (lower panels). 
	}
\end{figure*}

Here, we assess the numerical error of the simulations for all three 
geometries and material models based on the extinction efficiency. The 
latter is obtained by integrating the Poynting flux along both sides of 
the TF/SF contour and subsequently normalizing by the analytically known 
incident irradiance and the geometrical cross section. In particular, we 
consider selected resonance peaks. For each peak, we determine the 
amplitude $A$ and resonance frequency $\omega_{\rm res}$ at the maximum 
in order to calculate the corresponding relative differences
\begin{align} 
	\left(\Delta A\right)_{\rm rel}(h^i_{\rm surf}) ~= ~\quad~\quad~\quad~\quad~\quad~\quad~\quad\quad\quad~\nonumber \\ 
	\left|
	\frac{Q_{\rm ext}(h^i_{\rm surf}; \omega_{\rm res}(h^i_{\rm surf}))   
		- Q_{\rm ext}(
		h^{i+1}_{\rm surf}; \omega_{\rm res}(h^{i+1}_{\rm surf})
		)}
	{Q_{\rm ext}(
		h^{i+1}_{\rm surf}; \omega_{\rm res}(h^{i+1}_{\rm surf})
		)}\right|
	\label{suppl_eq:rel_amplitude_error} \\
	\left(\Delta \omega_{\rm res}\right)_{\rm rel}(h^i_{\rm surf}) 
	= 
	\left|
	\frac{\omega_{\rm res}(h^i_{\rm surf}) 
		- \omega_{\rm res}(h^{i+1}_{\rm surf})}
	{\omega_{\rm res}(h^{i+1}_{\rm surf})}
	\right| 
	\label{suppl_eq:rel_frequency_error}
\end{align} 
Following the Cauchy criterion, we evaluate the convergence behavior by 
comparing successive refinements rather than calculating the relative 
difference with respect to an analytical reference value, which is 
unavailable for the considered geometries. We summarize the (relative) 
error measures for all material models, evaluated for selected peaks of 
the triangular monomer ($\varphi=0\degree$) and the cylindrical and bowtie 
dimers ($\varphi=90\degree$), in \cref{suppl_fig:left_tri_convergence_peak_measures,suppl_fig:bowtie_convergence_peak_measures,suppl_fig:cyl_dim_convergence_peak_measures}, respectively. These figures also include waterfall plots of the 
extinction efficiencies obtained for each mesh refinement. For each 
combination of material model and geometry, we consider the four 
resonances with the largest amplitudes.

\begin{figure*}
	\centering
	\includegraphics[scale=.825]{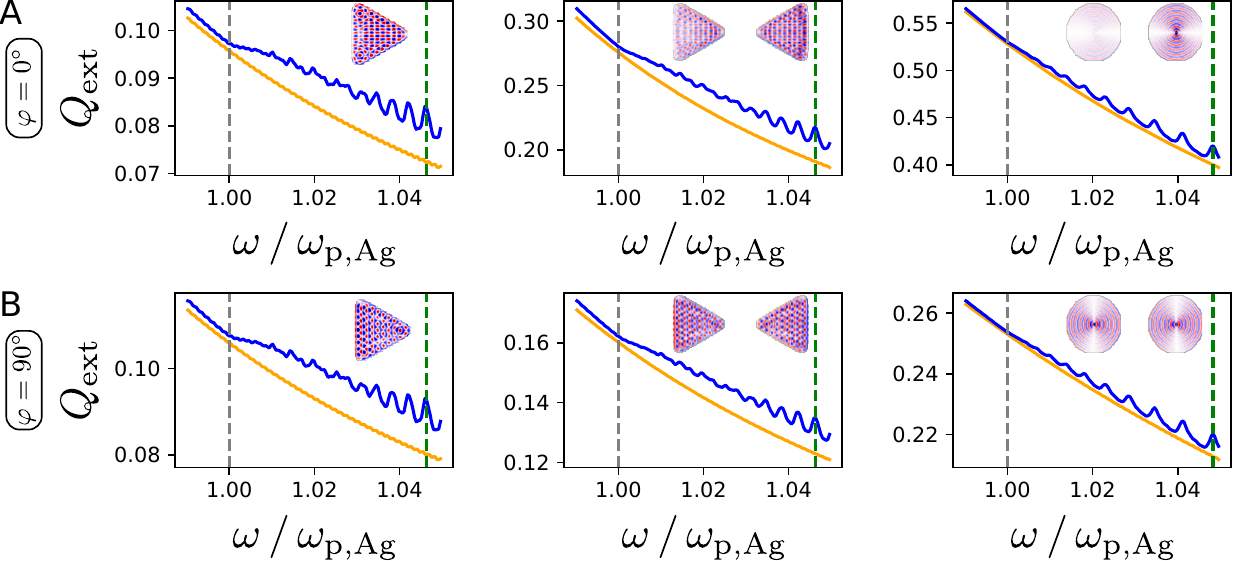}
	\caption{
		\label{suppl_I_fig:LVPs_all_structures_both_phi}
		Extinction efficiency of the considered geometries for $a=10$ nm, 
		$\tilde{a}=12$ nm, and $g=10$ nm. Results are shown for 
		$\varphi=0\degree$ (Panel A) and $\varphi=90\degree$ (Panel B) 
		for the triangular nanowire (left subpanels), bowtie dimer 
		(central subpanels), and cylindrical dimer nanowire (right 
		subpanels). The Drude (solid orange line) and Halevi (solid blue 
		line) models are considered. The vertical dashed gray line marks 
		the onset of localized volume plasmons (LVPs), while the vertical 
		dashed green line indicates the resonance frequency at which the 
		real part of the charge distribution is evaluated (see inset in 
		each panel). The charge distributions are normalized to unity, 
		with red corresponding to $1$ and blue to $-1$. The pixel side 
		length is $0.1$ nm.
	}
\end{figure*}

Generally, we observe more pronounced errors in the amplitudes than 
in the resonance frequencies. For the triangular monomer and bowtie 
dimer, the central peaks generally exhibit larger amplitude errors 
than the outer resonances, with the Drude model showing the largest 
deviations. For the cylindrical dimer, the amplitude error likewise 
increases with mode order (apart from an interchange of the errors 
associated with the third- and fourth-order modes). In contrast to 
the other geometries, however, the Drude model yields smaller 
amplitude errors for the cylindrical dimer.

Furthermore, the triangular and bowtie nanowires exhibit comparable 
values for both error measures when employing the two nonlocal material 
models. The same applies to the amplitude errors of the cylindrical
dimer, whereas the frequency errors differ slightly, particularly 
with respect to the vanishing errors at the employed resolution.

For the transition to the finest mesh ($0.3\mapsto0.2$ nm), the 
amplitude error remains well below one percent for the first and 
fourth peaks (and for all but the third peak in the Drude model) 
of the triangular monomer and bowtie dimer for all material models. 
For the cylindrical dimer, it remains below one percent (one 
permille) for the nonlocal (local) models. The corresponding 
frequency error for this mesh refinement step is well below one 
permille for all geometries and material models.

For the triangular and cylindrical dimer geometries, the nonlocal 
material models exhibit a decreasing trend in the amplitude error, 
whereas the Drude model shows either saturation or partially 
irregular behavior. For the bowtie, only the frequency error 
measures and the amplitude error of the third peak for the Drude 
model do not exhibit a clear decreasing tendency.

We note that error measures associated with different peaks may 
coincide for a given surface discretization and material model. 
For the triangular nanowire, the frequency errors of the second 
and third peaks for the nonlocal models at the finest refinement 
step ($0.3\mapsto0.2$ nm) vanish within the numerical resolution. 
Likewise, the third peak of the bowtie is unresolved in the Drude 
model for $h_{\rm surf}=0.6$ nm. Several frequency error measures 
of the cylindrical dimer also vanish at the employed resolution.

For the triangular and bowtie nanowires (for both incidence 
directions), we therefore focus primarily on the quadrupolar 
resonances located in the outer spectral range for the analyses 
presented in the main text, as these exhibit comparatively 
smaller numerical errors. For the cylindrical dimer, the explicit 
discussions mainly concern the first- and second-order resonances, 
corresponding to the dipolar and quadrupolar modes, respectively.

\section{Extinction properties}
\begin{figure*}
	\centering
	\includegraphics[scale=.275]{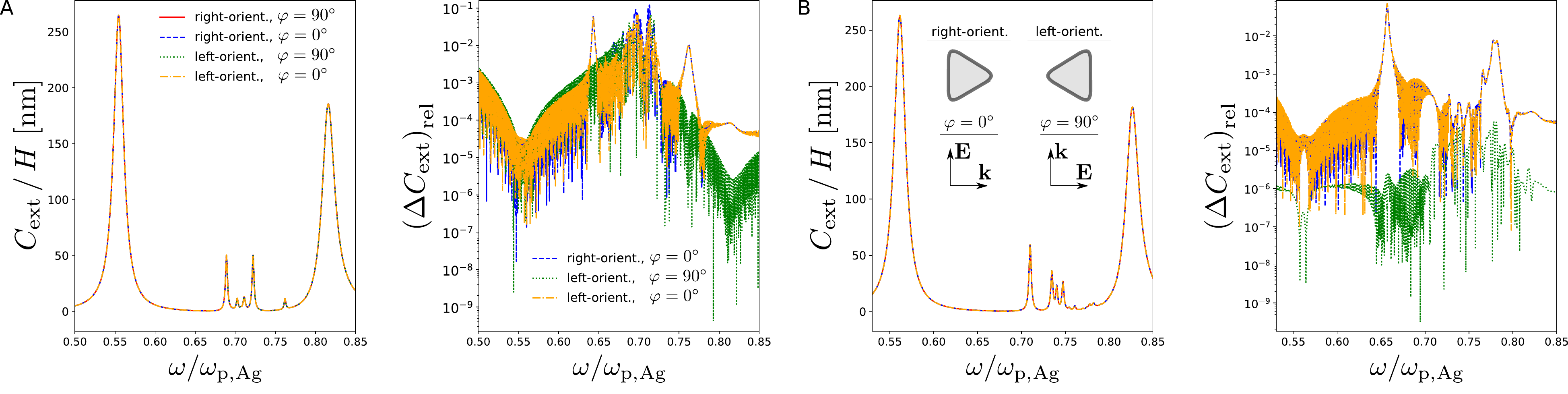}
	\caption{
		\label{supp_I_tri_fig:ext_cross}	
		Degeneracy of the extinction cross section for the Drude model 
		(Panel A) and the Halevi model (Panel B). The left subpanels 
		show the extinction cross section for each configuration (see 
		the legend in the left subpanel of Panel A for line styles and 
		colors). The orientations of the nanowire and the incident 
		field polarizations are illustrated in the inset of the left 
		subpanel of Panel B. The right subpanels display the relative 
		difference with respect to the reference case of a 
		right-oriented triangle excited with $\varphi=90^\circ$.
	}
\end{figure*}

\subsection{\label{suppl_I_sec:volume_plasmons} Localized volume plasmons}

Within the main text, we have discussed numerous features of LSPs 
sustained by triangular, bowtie, and cylindrical dimer nanowires. 
For completeness, we provide here the extinction efficiencies of all 
structures for the Drude and Halevi models and both incidence directions 
beyond $\omega_{\rm p, Ag}$. As shown in \cref{suppl_I_fig:LVPs_all_structures_both_phi}, 
we observe a 
multitude of peaks beyond this threshold only for the longitudinally 
nonlocal model, irrespective of whether $\varphi=0\degree$ (Panel A) 
or $\varphi=90\degree$ (Panel B). This behavior is expected, since 
only the longitudinal model supports localized volume plasmons (LVPs), 
which correspond to standing-wave-like resonances. For these modes, 
the associated wavenumber, here the longitudinal wavenumber $k_{\rm L}$,
must become predominantly real. This is the case because 
$\omega\ge\omega_{\rm p, Ag}\gg\gamma_{\rm Ag}$ and
$k_{\rm L, Ag}\approx\sqrt{\omega^2-\omega_{\rm p, Ag}^2}/\beta_{\rm HF}\equiv\operatorname{Re}k_{\rm L, Ag}$,
where $\beta_{\rm HF}=\sqrt{3/5}v_{\rm F}$. The peak amplitudes are 
substantially larger than the weak oscillations observed in the Drude
model. The latter originate from errors in the on-the-fly Fourier transform due to 
the finite simulation time.

From the charge-density insets at selected resonance (peak maximum) 
frequencies, we infer that these resonances extend throughout the 
entire cross section. For the cylindrical dimer (right subpanels), 
we observe different nodal patterns depending on the incidence 
direction. The triangular nanowire (left subpanels) and bowtie dimer 
(central subpanels) exhibit one or two vortices in the charge density 
located at the corners aligned with the incident field polarization.

Due to the numerical accuracy of our simulations, we cannot conclusively 
determine whether the localized volume resonances of the three structures 
can be classified according to odd and even multipole orders (or peak 
indices). For the cylindrical monomer, such a classification can be 
established, with odd-order peaks exhibiting lower amplitudes (see,
e.g., the graphical derivation in Sec.~3.2.4 of \cite{moeferdtdiss} 
and the analytical derivation in Sec.~A.4 of \cite{wegner2024diss}). 
We merely note that, for the cylindrical dimer beyond 
$\omega=1.04~\omega_{\rm p, Ag}$, a small peak appears to emerge 
between two larger peaks, which may indicate a similar multipolar 
ordering.

\subsection{Degeneracy of triangular extinction cross sections}

Within the main text, we have discussed the extinction properties 
of a triangular nanowire associated with the excitation of LSPs. 
Based on symmetry considerations, the extinction cross section of 
the triangle is expected to be degenerate for isotropic material 
models. This is demonstrated in \cref{supp_I_tri_fig:ext_cross} for a left- 
and right-oriented wire employing the Drude (Panel A) and Halevi 
(Panel B) models and considering both incidence directions.

We observe that the resonance frequencies essentially coincide for 
all cases. The main differences arise in the point-wise amplitudes, 
which we quantify by considering the point-wise relative difference 
with respect to the right-oriented triangle excited by a plane wave
with $\varphi=90^{\degree}$. The relative difference increases 
towards the spectral center, where higher-order resonances are 
excited, and is generally smaller for the Halevi model. We note, 
however, that the extinction efficiency itself also decreases towards 
the spectral center, resulting in a smaller denominator in the 
relative-difference measure.

The two dominant resonances, corresponding to the quadrupolar LSPs, 
exhibit differences well below one permille for both material models 
and incidence directions. Therefore, we focus primarily on these 
resonances in the main text.

\begin{figure}
	\centering
	\includegraphics[scale=.75]{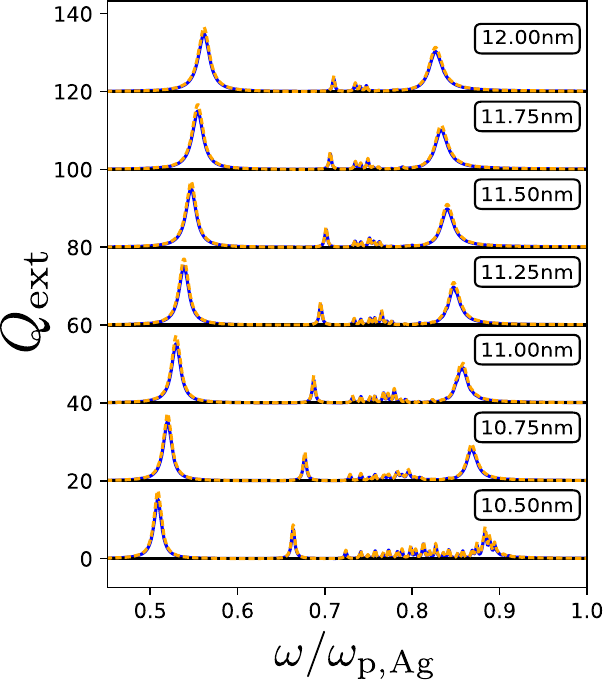}
	\caption{
		\label{suppl_I_fig:tri_phi_0_and_90_Q_ext_nloc_peak_seq}
		Curvature dependence of the extinction efficiency associated with
		LSPs sustained by a right-oriented triangular nanowire 
		($a=10$ nm), calculated using the Halevi model and excited by 
		a linearly polarized plane wave. The curvature is varied 
		according to $\tilde{a}\in \{10.5;10.75;11.0;11.25;11.5;11.75;12.0 \}$ nm
		(see white boxes). 
		Results for $\varphi=0\degree$ ($\varphi=90\degree$) are shown by 
		the solid blue (dashed orange) curves. Peak maxima are interpreted as resonance
		frequencies. The curves are vertically offset by 20 units for clarity.
	}
\end{figure}
\subsection{Nonlocal peak sequences}

\subsubsection{Degeneracy for triangular nanowire monomer}

In the manuscript, we presented the nonlocal peak sequence only 
for $\varphi=0\degree$ for the sake of conciseness. Here, we compare 
the extinction efficiencies for $a=10$ nm for both incidence directions 
while performing a curvature variation to enhance the visibility and 
widening of the sequence. Accordingly, we restrict the discussion to 
the Halevi model. As shown in \cref{suppl_I_fig:tri_phi_0_and_90_Q_ext_nloc_peak_seq}, the 
resulting spectra 
differ only marginally, with the main distinction arising from the 
different normalization by the geometrical cross section.

\subsubsection{Dependence on Fermi velocity}
\begin{figure*}
	\includegraphics[scale=.6]{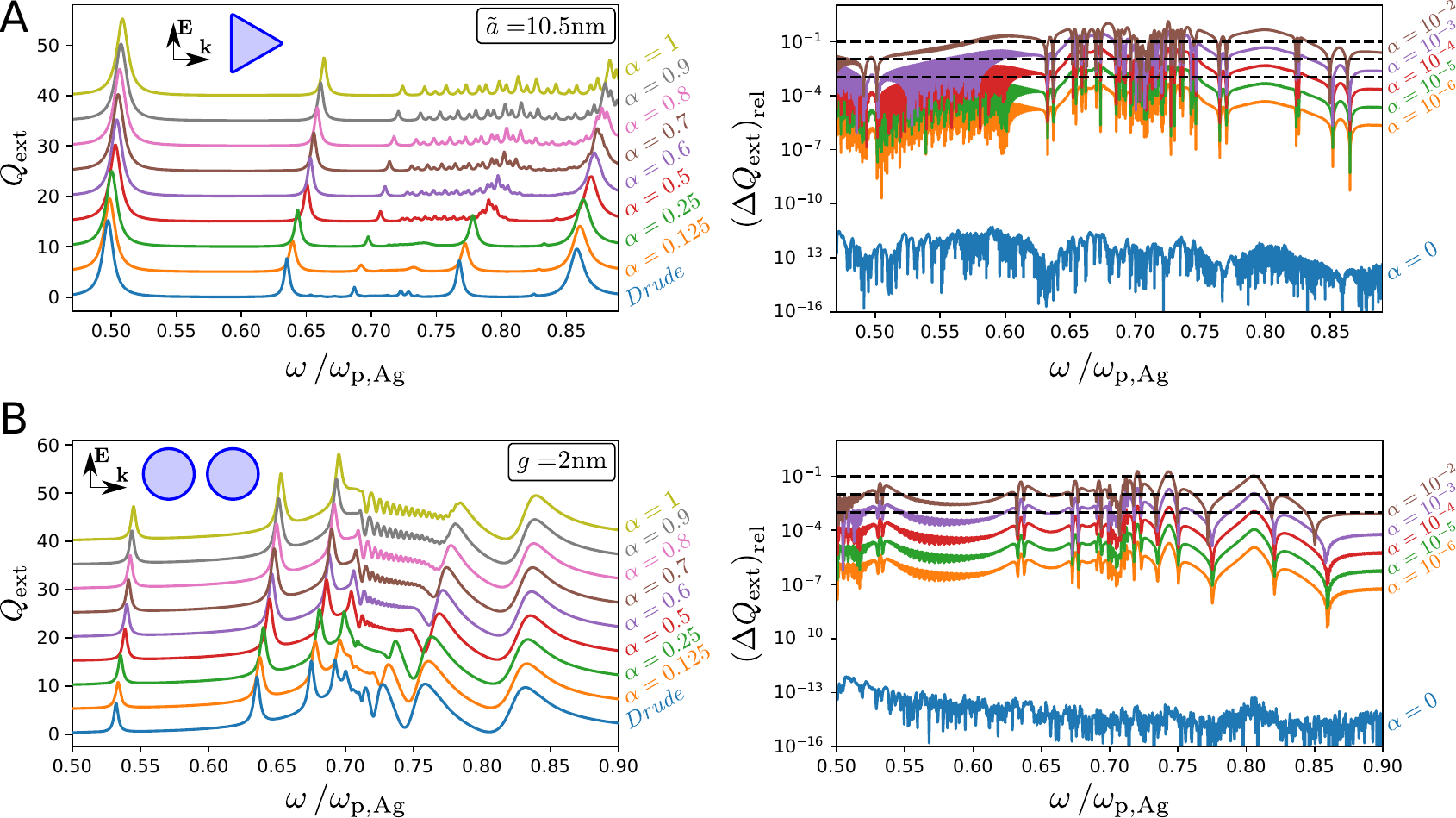}
	\caption{
		\label{suppl_I_tri_fig:nonlocal_peak_sequence_ext_eff_alpha_scan}
		Depiction of the extinction efficiency as a function of frequency 
		for the triangular nanowire ($a=10$ nm and $\tilde{a}=10.5$ nm; 
		Panel A) cylindrical dimer nanowire ($a=10$ nm and $g=2$ nm; 
		Panel B) excited with $\varphi=0\degree$. When employing the Halevi 
		model, the silver Fermi velocity is scaled according to the mapping $v_{\rm F,Ag}\mapsto \alpha v_{\rm F,Ag}$, with $0\leq \alpha\leq1$. 
		The left subpanels compare the extinction efficiency obtained from 
		the Halevi model for $0.125\leq\alpha\leq1$ with the corresponding Drude-model result (blue curve in the same subpanel). The curves 
		are vertically offset by 5 units for both the triangular nanowire 
		and cylindrical dimer to create waterfall plots. The right subpanels 
		show the absolute relative difference between the extinction 
		efficiency calculated with the Halevi model for a given $\alpha$ 
		and the Drude-model result, defined as $\left(\Delta Q_{\rm ext}\right)_{\rm rel}=|(Q^{\rm Halevi}_{\rm ext}-Q^{\rm Drude}_{\rm ext})/Q^{\rm Drude}_{\rm ext}|$. The horizontal black dashed lines 
		indicate relative differences of $10\%$, $1\%$, and $0.1\%$.
	}
\end{figure*}

Within the manuscript, we have shown that longitudinal nonlocality 
introduces nonlocal peak sequences into the spectral center of the 
extinction efficiency for all considered nanowire geometries. These 
sequences are associated with high-order resonances and are biased
towards the high-frequency side of the spectrum. For the triangular 
and bowtie nanowires, the sequences widen with increasing curvature, 
whereas for the cylindrical dimer they broaden with decreasing gap 
size. Although a gap-dependent effect is also present for the bowtie, 
it is comparatively weak.

To highlight these sequences for $a=10$ nm,
\cref{suppl_I_tri_fig:nonlocal_peak_sequence_ext_eff_alpha_scan} 
displays the extinction efficiency of the triangular nanowire for 
the largest curvature ($\tilde{a}=10.5$ nm; Panel A) and of the 
cylindrical dimer for the smallest gap size ($g=2$ nm; Panel B). 
The bowtie is omitted, since its sequence is qualitatively similar 
to that of the triangular nanowire.
For the cylindrical dimer, the sequence emerges only for 
$\varphi=0\degree$. For the triangular nanowire, we likewise 
restrict the discussion to $\varphi=0\degree$, owing to the 
degeneracy of the extinction cross sections, see also 
\cref{suppl_I_fig:tri_phi_0_and_90_Q_ext_nloc_peak_seq}.

We initiate the analysis of the role of nonlocality for the 
triangular nanowire. The strength of the nonlocal response scales 
with the Fermi velocity $v_{\rm F}$, as it originates from the 
quantum-statistical nature of the many-electron conduction system. 
Using the mapping 
$v_{\rm F}\mapsto\alpha v_{\rm F}$ with $0\leq\alpha\leq1$, we 
gradually reduce the nonlocal contribution by decreasing $\alpha$, 
while retaining the Halevi model and keeping the geometrical 
parameters and incident field unchanged.
As shown in the waterfall plot of 
\cref{suppl_I_tri_fig:nonlocal_peak_sequence_ext_eff_alpha_scan}
(left subpanel of Panel A), the number of peaks forming the nonlocal 
peak sequence decreases with decreasing $\alpha$. At the same 
time, the more pronounced individual resonances gradually lose 
their nonlocal blueshift. Eventually, the Drude spectrum is 
recovered. This trend is further supported by the relative 
difference between the Halevi spectrum for a given $\alpha$ and 
the corresponding Drude spectrum, see 
\cref{suppl_I_tri_fig:nonlocal_peak_sequence_ext_eff_alpha_scan} 
(right subpanel of Panel A).
The results for the cylindrical dimer are qualitatively similar, 
as shown in 
\cref{suppl_I_tri_fig:nonlocal_peak_sequence_ext_eff_alpha_scan} (Panel B).

\begin{figure*}
	\centering
	\includegraphics[scale=.5]{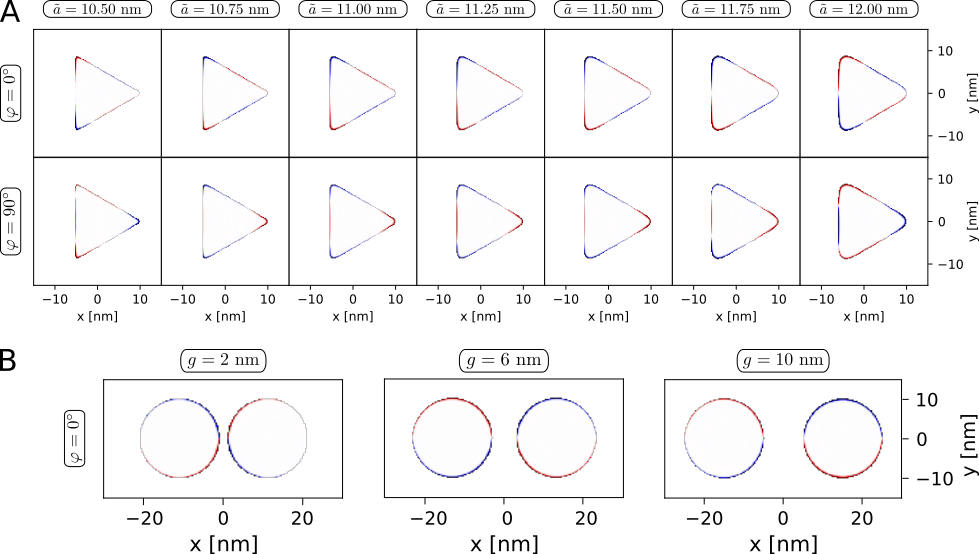}
	\caption{
		\label{suppl_I_fig:tri_charge_distr_quadrupolar_LSP_sharpness_scan}
		Geometric-parameter dependence of the real part of the charge 
		density for selected LSPs sustained by the triangular and cylindrical
		dimer nanowires ($a=10$ nm), excited by a linearly polarized plane 
		wave and modeled using the Halevi model. 
		Panel A: Charge distributions of the low-frequency quadrupolar 
		LSP sustained by the triangular nanowire for varying curvature 
		(see subpanel headers) and for $\varphi=0\degree$ (upper subpanels) 
		and $\varphi=90\degree$ (lower subpanels). Each charge distribution 
		is normalized to its maximum value within the displayed spatial 
		domain. The color scale is restricted to the range from $-0.1$ (blue) to $+0.1$ (red). 
		Panel B: Same as Panel A, but for the cylindrical dimer nanowire, considering only $\varphi=0\degree$ and varying the gap size 
		(see subpanel headers) for the low-frequency bonding dipolar LSP.   
	}
\end{figure*}
\subsection{Origin of overall lineshifts for selected resonances}

\subsubsection{Triangular nanowire}

Within the main text, we observed that the quadrupolar LSPs of the 
triangular nanowire exhibit opposing overall lineshifts as the curvature 
is varied. Here, we propose a possible explanation based on the 
corresponding charge-density distributions. Since the high-frequency 
quadrupoles are superimposed with higher-order resonances for both 
incidence directions (not shown here), we focus exclusively on the 
low-frequency quadrupoles for $\varphi\in \{0\degree;90\degree \}$ in \cref{suppl_I_fig:tri_charge_distr_quadrupolar_LSP_sharpness_scan} (Panel A).

For $\varphi=0\degree$, we observe that, with increasing curvature, the 
charge poles on the right surface half gradually vanish, while the 
remaining two poles become increasingly localized at the corners. These 
two poles then approximately form a dipolar configuration. The 
corresponding effective wavelength increases with increasing sharpness, 
which translates into a redshift in frequency space.

For $\varphi=90\degree$, the charge density progressively localizes at 
all three corners. Although the effective dipole argument cannot be 
transferred completely to this case, it appears reasonable to apply 
it locally across the two horizontal edges. In this context, we note 
that the fixed reference cylinder ($a=10$ nm) keeps the positions of 
the corner apices unchanged.

\subsubsection{Cylindrical dimer}

Within the manuscript, we explained using a simple Coulomb picture that 
the charge distributions associated with bonding and antibonding LSPs 
lead to red- and blueshifts, respectively. Inspection of the extinction 
efficiency further revealed that the bonding dipolar LSP excited for
$\varphi=0\degree$ undergoes a stronger redshift as the gap size 
decreases. To corroborate this observation, we consider the 
corresponding charge-density distributions for $g=10$ nm and $g=2$ nm, 
shown in 
\cref{suppl_I_fig:tri_charge_distr_quadrupolar_LSP_sharpness_scan} (Panel B).
We observe that the charge becomes increasingly concentrated at the 
two poles facing the gap, thereby enhancing the interaction between 
the two cylinders. 

\subsection{Extinction of the bowtie for $\varphi=90\degree$}

\begin{figure*}
	\centering
	\includegraphics[scale=0.5]{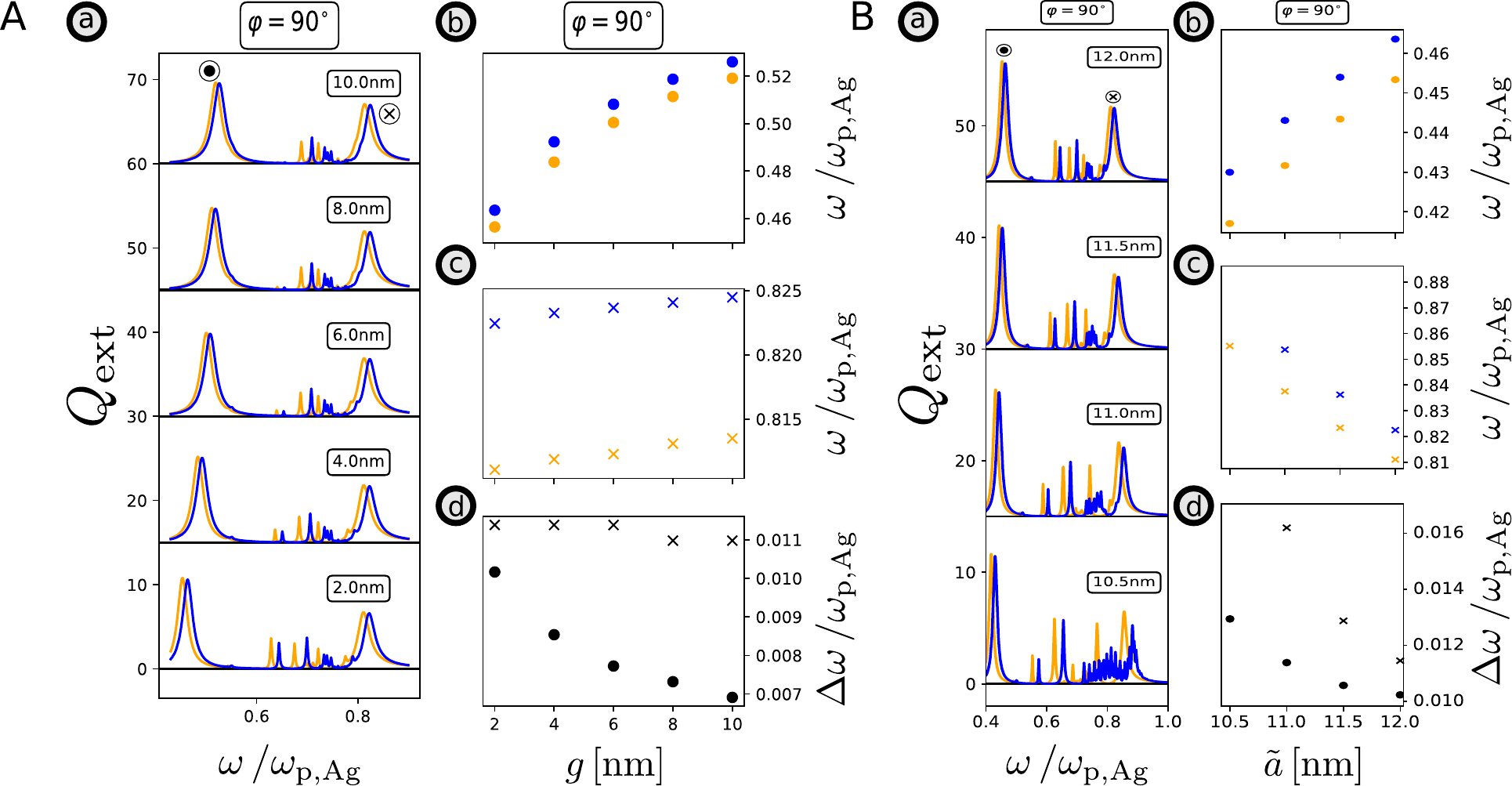}
	\caption{
		\label{suppl_I:bow_Q_ext_curve_and_gap_scan_phi_90}
		Geometric-parameter dependence of the extinction efficiency of a 
		bowtie dimer for $a=10$ nm and $\varphi=90\degree$. 
		Panel A: Gap-size dependence for $\tilde{a}=12$ nm and 
		$g\in \{10;8;6;4;2\}$ nm, comparing the Drude (orange) and 
		Halevi (blue) models for the calculation of the extinction 
		efficiency (subpanel a). The boxes in subpanel (a) indicate 
		the corresponding gap sizes, and the curves are vertically 
		offset by 15 units for clarity. Subpanels (b) and (c) show 
		the resonance frequencies of the low- and high-frequency 
		quadrupolar bonding dimer LSPs, respectively (see markers 
		in subpanel a). The corresponding nonlocal blueshifts are 
		shown in subpanel (d). Panel B: Same as Panel A, but 
		illustrating the curvature dependence for $g=2$ nm and 
		$\tilde{a}\in \{10.5;11.0;11.5;12.0\}$ nm.		
	}
\end{figure*}

Within the main text, we have focused primarily on $\varphi=0\degree$ 
when discussing the extinction properties of the bowtie. This choice 
was motivated by the presence of bonding and antibonding dimer resonances, 
which allowed us to investigate the tuning of their relative frequencies. 
In 
\cref{suppl_I:bow_Q_ext_curve_and_gap_scan_phi_90}, 
we study the dependence of the extinction efficiency 
on gap size (Panel A) and curvature (Panel B) for $\varphi=90\degree$.
Again, we observe the nonlocal peak sequence, which broadens more 
strongly under curvature variation. Regarding the quadrupolar dimer 
resonances, Panels A(b) and A(c) show that both bonding resonances 
undergo a redshift with decreasing gap size. In this case, the 
low-frequency bonding resonance exhibits a stronger sensitivity to 
gap-size variations and displays an appreciable gap-dependent 
blueshift on the order of $0.001~\omega_{\rm p, Ag}$.

Turning to the curvature scan (Panel B), we find that the two bonding 
dimer resonances exhibit opposing shifts, following the corresponding 
monomer quadrupolar resonance frequencies. At the same time, the 
nonlocal blueshift increases with increasing curvature. Finally, 
similar to the cylindrical dimer, the degeneracy of the monomer 
spectra with respect to variations in $\varphi$ is lifted, highlighting 
the influence of the shared dimer symmetry.

\section{Near-field Distributions}

\subsection{\label{suppl_I_sec:calc:mean_domain_enhancement}
	Calculation of the mean-domain-enhancement measures}

Within the main text, we have investigated the optimal placement of a 
specimen in the vicinity of a given scattering geometry for fixed 
geometrical parameters, incidence direction, and excitation frequency. 
In this context, we assessed the suitability of the respective structures 
as SERS substrates based on the spatial localization of hotspots. 
As a measure, we considered the frequency-dependent local 
electric-field enhancement
\begin{align}
	\mathcal{E}(\mathbf{r}, \omega) 
	& = 
	\frac{\left| \mathbf{E}(\mathbf{r}, \omega) \right|}{\left| \mathbf{E}_0(\mathbf{r}, \omega) \right|}
\end{align}
where $\mathbf{E}(\mathbf{r}, \omega)$ and 
$\mathbf{E}_0(\mathbf{r}, \omega)$ denote Fourier components of the 
total and incident electric fields, respectively.

To obtain an univariate quantity, we first restrict the evaluation to a 
spatial domain $\mathcal{D}$ and subsequently calculate the mean value 
of $\mathcal{E}(\mathbf{r}, \omega)$ over all $\mathbf{r}\in\mathcal{D}$. 
This yields the mean-domain electric-field enhancement
\begin{align}
	\left\langle  \mathcal{E}(\mathbf{r}, \omega) \right\rangle_{\mathcal{D}} 
	= 
	\left\langle \frac{\left| \mathbf{E}(\mathbf{r}, \omega) \right|}{\left| \mathbf{E}_0(\mathbf{r}, \omega) \right|}  \right\rangle_{\mathbf{r}\in \mathcal{D}}.
\end{align}

\begin{figure*}
	\centering
	\includegraphics[scale=1.5]{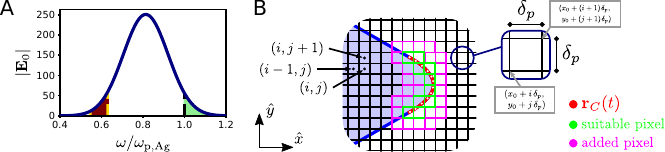}
	\caption{
		\label{suppl_I_fig:average_domain_enhancement_calculation}
		Incident-field Fourier amplitude and schematic illustration of 
		the pixelated field representation returned by the DGTD method. 
		Panel A: Amplitude of the Fourier-transformed incident field with 
		Gaussian temporal dependence, following \cref{suppl_I_eq:FT_inc_field_amplitude} and using the parameter 
		values listed 
		in \cref{suppl_I_tab:gauss_pulse_in_sphere_all_setups}(d). The 
		golden vertical dashed lines and maroon shaded region indicate 
		the spectral range covered by the bonding dipolar LSPs sustained 
		by the cylindrical nanowire for $\varphi=0\degree$. The modulus 
		increases from $11.4$ to $59.8$. The green shaded region marks 
		the spectral range in which localized volume plasmons are expected 
		to occur. 
		Panel B: Illustration of the calculation of a mean-domain 
		enhancement using a triangular nanowire as an example. The 
		simulated field values are assigned to a set of pixels labeled 
		by the index tuple $(i,j)$, with predefined height $\delta_y$ 
		and width $\delta_x$. Each pixel occupies the rectangle bounded 
		by the coordinates $(x_0+i\delta_x,y_0+j\delta_y)$ and $(x_0+(i+1)\delta_x,y_0+(j+1)\delta_y)$. In the present 
		case, $\delta_x=\delta_y=\delta_p$. The initial domain 
		$\mathbf{r}_C(t)$ (red) is parametrized by the white dots, 
		and all overlapping pixels (green) are collected. Additional 
		neighboring pixels without duplication (pink) may be included 
		depending on the definition of the domain $\mathcal{D}$. The 
		mean enhancement is evaluated only over the selected pixels.
	}
\end{figure*}
We emphasize that the frequency dependence remains relevant, since we 
investigate resonances whose frequencies vary with the geometrical 
parameters. Consequently, the normalization of the resonant electric 
field with respect to the incident field must be performed at 
different frequencies.
The incident field amplitude is calculated analytically from 
\begin{align}
	\mathbf{E}_0(\mathbf{r},t) 
	= & 
	\operatorname{Re}\left\{ E_0 \hat{E_0} \exp[- i \omega (\tilde{t}- t_0) - \theta_0] \eta(\tilde{t}-t_0) \right\}  \nonumber 
	\label{suppl_I_eq:incident_field_def}\\ 
	& \quad \text{where} \quad \eta(t) = \exp[-t^2 / 2 \sigma_t^2] ,
\end{align} 
where  $\tilde{t} = t - \mathbf{k}\cdot\mathbf{r} / \omega_0$, $\theta_0=\pi/2$ and 
$\hat{E}_0$ define the polarization. For the remaining parameters we 
refer to \cref{suppl_I_tab:gauss_pulse_in_sphere_all_setups}.
The Fourier transform of the incident field amplitude yields
\begin{align}
	\left| \mathbf{E}_0(\mathbf{r}, \omega) \right| 
	= &\, 
	\sqrt{\frac{\pi}{2}} \sigma_t E_0 \Big| \exp[ - (\omega + \omega_0)^2 \sigma^2_t / 2]  \nonumber \\
	&\quad\quad\quad\quad- \exp[ -(\omega - \omega_0)^2 \sigma^2_t / 2]\Big|.
	\label{suppl_I_eq:FT_inc_field_amplitude}
\end{align}
Note that applying the modulus removes the spatial and polarization 
dependence contained in the term $\mathbf{k}\cdot\mathbf{r}/\omega_0$. 
To assess the impact of the frequency dependence, we consider the 
spectral range of the bonding dipolar LSP excited on a cylindrical 
dimer with $a=10$ nm and $2~\text{nm}\leq g \leq 10~\text{nm}$ under
$\varphi=0\degree$ using the Halevi model. As shown in 
\cref{suppl_I_fig:average_domain_enhancement_calculation} (Panel A), 
$|\mathbf{E}_0|$ increases from $11.4$ 
(for the resonance with $g=2$ nm) to $59.8$ (for $g=10$ nm).
The corresponding amplitude of the total local field is obtained from 
the on-the-fly Fourier transform. The definition of $\mathcal{D}$ 
depends strongly on the scattering geometry. Within the manuscript,
we generally distinguish between surface and gap measures.

We note that the DGTD method provides a pixelated representation of the 
spatial field distribution, 
see \cref{suppl_I_fig:average_domain_enhancement_calculation} (Panel B). 
Based on an analytical parametrization of the domain $\mathcal{D}$ and 
the spatial extent of each pixel, we identify all pixels intersecting 
the domain. For the surface measures, we additionally include a finite 
number of neighboring pixels, thereby creating a sleeve of width $\delta$ 
around the domain.

\subsection{Mean-surface enhancement of the triangle for $\varphi=90\degree$}

\begin{figure*}
	\centering
	\includegraphics[scale=0.3]{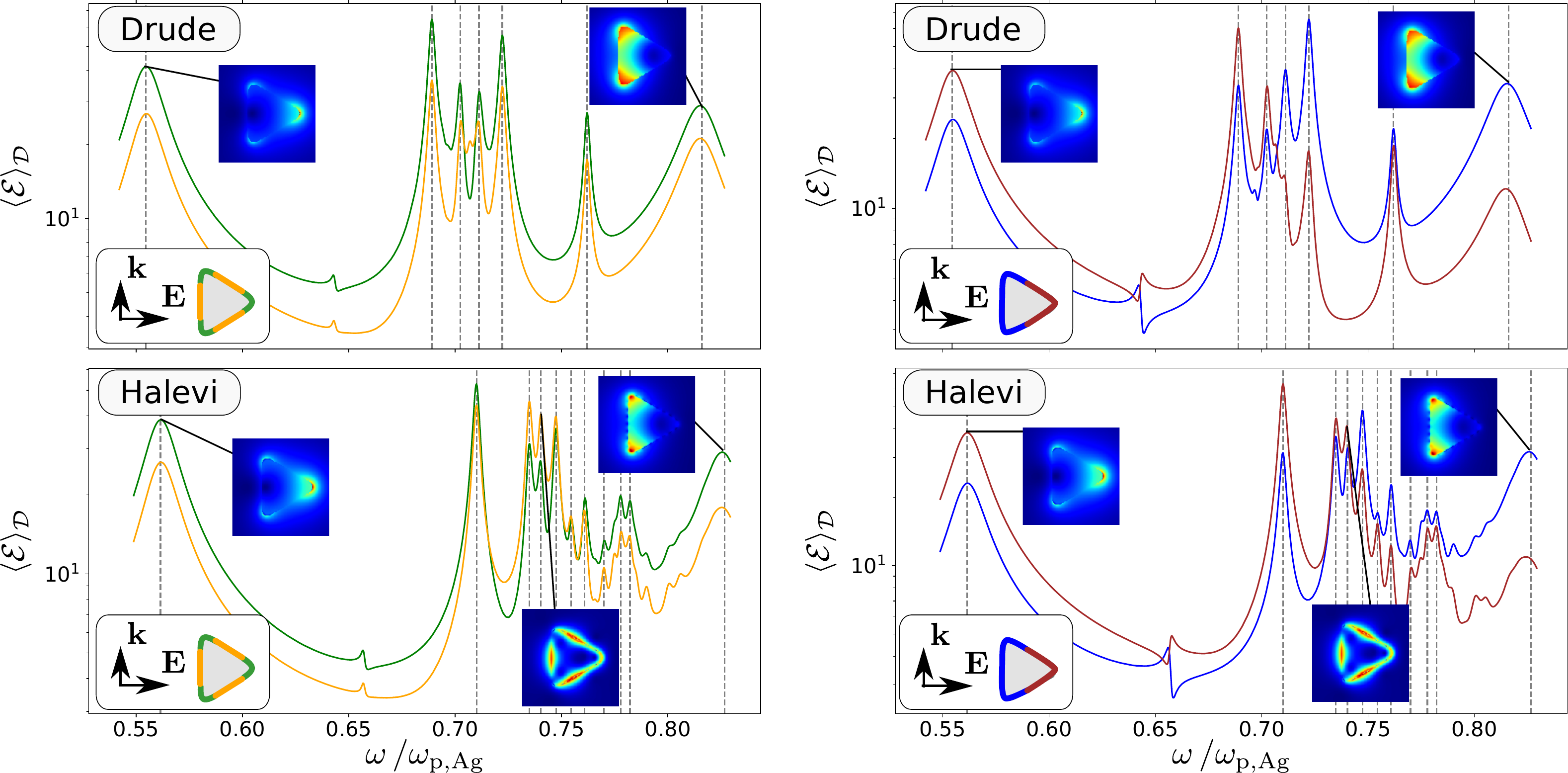}
	\caption{
		\label{suppl_I_fig:tri_a_10_a_tilde_12_phi_90_aver_dom_enh}
		Mean-domain enhancement of the electric field associated with LSPs 
		sustained by the triangular nanowire for $a=10$ nm, $\tilde{a}=12$ nm, 
		and $\varphi=90\degree$. The edge and corner enhancements (left panels) 
		as well as the left- and right-half enhancements (right panels) are 
		compared for the Drude (upper panels) and Halevi (lower panels) models. 
		The legend and incident-field polarization are provided in the 
		lower-left box of each panel. Vertical dashed lines indicate the 
		peak-maximum frequencies of the corresponding extinction efficiency 
		(see also \cref{tri_fig:extinction}). Insets show the spatial 
		distribution of the electric-field modulus at the respective 
		frequencies, normalized to the maximum value within the displayed 
		spatial domain. Red (blue) denotes the maximum (minimum) value.
	}
\end{figure*}

Within the manuscript, we have discussed the mean-domain enhancement 
of the bluntest triangular nanowire ($a=10$ nm and $\tilde{a}=12$ nm) 
for $\varphi=0^{\degree}$. Here, we supplement this analysis by 
considering the mean-domain enhancement for $\varphi=90^{\degree}$.
As shown in 
\cref{suppl_I_fig:tri_a_10_a_tilde_12_phi_90_aver_dom_enh}, 
we observe the same qualitative features. 
However, due to the different incidence direction, the enhancement 
is preferentially localized on the right half of the surface at lower 
frequencies, whereas for frequencies beyond 
$\omega\approx0.75\omega_{\rm p,Ag}$ it shifts towards the left 
half. This behavior is particularly evident in the field distributions 
at the quadrupolar resonance frequencies.
As before, we provide a central inset in the lower-left panel showing 
an edge-enhanced mode. Inspection of the lower-right panel reveals 
that this same mode exhibits an approximately balanced distribution 
between left- and right-half enhancement.

\begin{figure*}
	\centering
	\includegraphics[scale=.5]{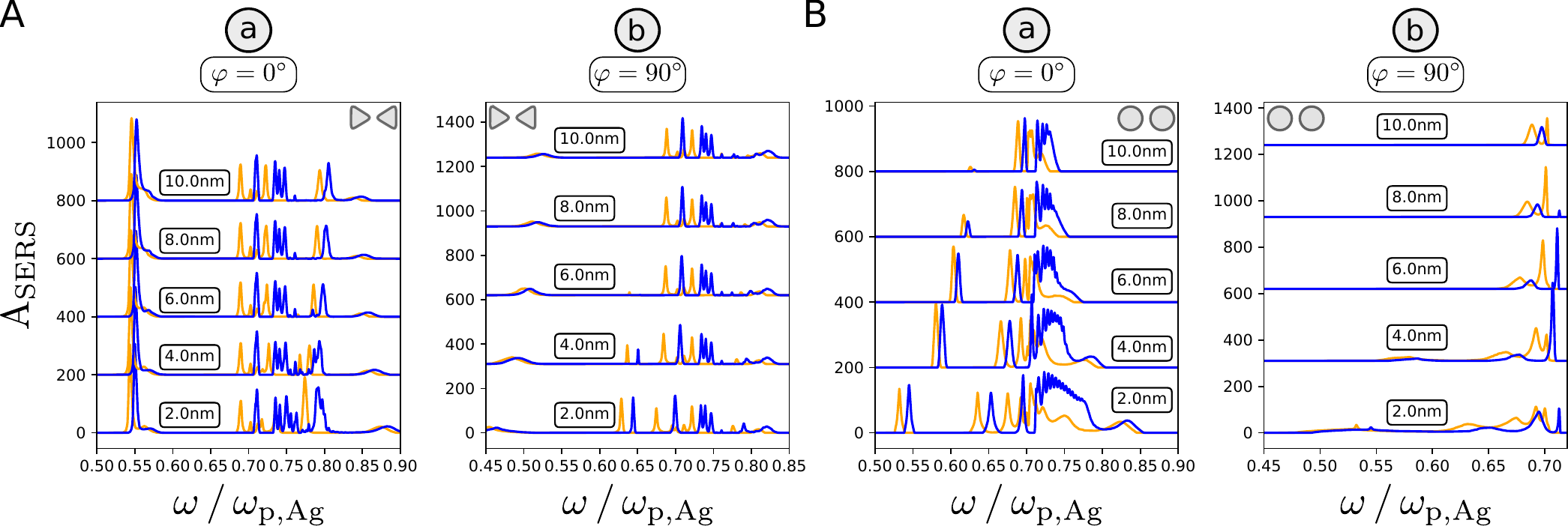}
	\caption{
		\label{suppl_I_fig:eff_SERS_area_dimers_gap_scan}
		Gap-size dependence of the effective SERS area $A_{\rm SERS}$ for 
		a threshold value of $\mathcal{E}_{\rm th}=50$ for both dimers with 
		radius $a=10$ nm and for both incident-field polarizations. 
		Panel A: Effective SERS area of the bowtie dimer ($\tilde{a}=12$ nm) 
		as a function of frequency for $g\in\{10;8;6;4;2\}$ nm (see boxed 
		labels), considering $\varphi=0\degree$ (subpanel a; curves 
		vertically offset by 80 units for clarity) and $\varphi=90\degree$ (subpanel b; curves vertically offset by 60 units for clarity). 
		The Drude (orange) and Halevi (blue) models are compared. 
		Panel B: Same as Panel A, but for the cylindrical dimer.
	}
\end{figure*}

\begin{figure}
	\centering
	\includegraphics[scale=.4]{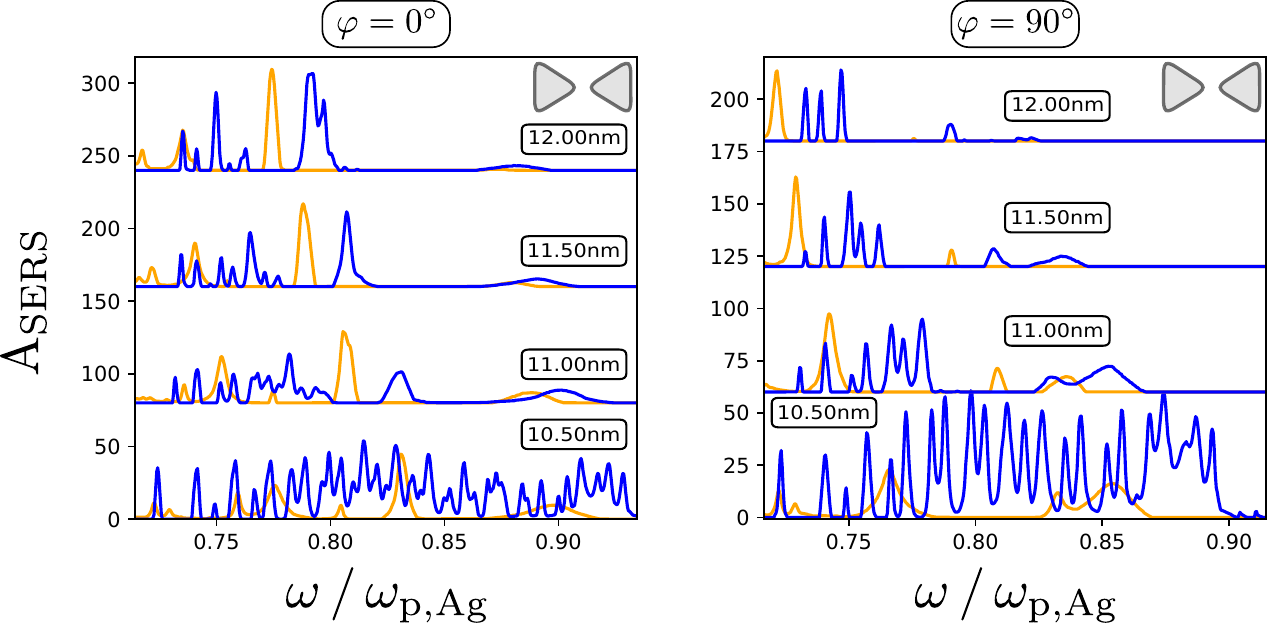}
	\caption{
		\label{suppl_info_fig:A_SERS_bowtie_a_10_g_2_curve_scan}
		Curvature dependence of the effective SERS area $A_{\rm SERS}$ 
		for a threshold value of $\mathcal{E}_{\rm th}=50$ for the bowtie 
		dimer with radius $a=10$ nm, gap size $g=2$ nm, and $\tilde{a}\in\{10.5;11.0;11.5;12.0\}$ nm (see boxed labels). 
		Results are shown for $\varphi=0\degree$ (left panel) and $\varphi=90\degree$ (right panel). The Drude (orange) and 
		Halevi (blue) models are compared. The curves are vertically 
		offset by 80 and 60 units for $\varphi=0\degree$ and
		$\varphi=90\degree$, respectively, for clarity.
	}
\end{figure}

\section{Effective SERS area}

Within the manuscript, we have discussed several mean-domain enhancement 
measures. The choice of a spatial domain is motivated by knowledge of 
the specimen placement relative to the scatterer (or SERS substrate). 
In contrast, for an experimental situation in which the specimen can 
diffuse freely, a different measure may be more appropriate. Such a 
measure records, as a function of frequency, the area over which 
the amplitude-enhancement is not smaller than a threshold $\mathcal{E}_{\rm th}$. We refer to this quantity as the effective SERS area, 
$A_{\rm SERS}$. A large value of $A_{\rm SERS}$ corresponds to a higher 
probability of the specimen residing within a hotspot region. Here, 
we choose $\mathcal{E}_{\rm th}=50$, which enables few- to single-molecule 
SERS applications.

In 
\cref{suppl_I_fig:eff_SERS_area_dimers_gap_scan}, 
we evaluate this measure for the bowtie ($a=10$ nm, $\tilde{a}=12$ nm; 
A) and cylindrical dimer ($a=10$ nm; Panel B) nanowires for different 
gap sizes and for both incidence directions, $\varphi=0\degree$ 
(subpanel a) and $\varphi=90\degree$ (subpanel b). Overall, we observe 
peaks that spectrally coincide with those of the extinction efficiency, highlighting the role of LSPs. Furthermore, for both material models, 
the largest peaks are predominantly concentrated around the spectral
center ($\omega/\omega_{\rm p,Ag}\approx1/\sqrt{2}\approx0.71$), 
emphasizing the contribution of high-order resonances with hotspots 
extending over both surfaces.

While the bowtie exhibits a continuous peak sequence for both incidence
directions, this behavior is observed for the cylindrical dimer only 
for $\varphi=0\degree$ (Panel B, subpanel a). However, the ripple structure 
of the nonlocal cylindrical dimer possesses smaller spectral separations 
between individual peaks, resulting in a shoulder-like spectral profile 
that is more favorable for light-harvesting applications requiring a 
broad operational frequency range.

Outside the central spectral region, we observe comparatively large 
peak amplitudes for $\varphi=0\degree$ originating from the low-frequency 
bonding dipolar LSPs of the cylindrical dimer (Panel B, subpanel a). 
A similar trend is found for the low-frequency bonding quadrupolar 
dimer LSPs of the bowtie for both incidence directions (Panels A, 
subpanels a and b). For $\varphi=0\degree$, the bonding quadrupole concentrates 
the field at the four gap-averted corners and dominates the entire 
spectrum for all considered gap sizes. Overall, we conclude that gap 
variations affect the spectral response of the cylindrical dimer much 
more strongly, particularly when changing from $\varphi=0\degree$ to
$\varphi=90\degree$ due to the suppression of some high-frequency 
hybrid modes.

Within the manuscript, we have shown that, whereas the nonlocal peak 
sequence of the cylindrical dimer depends strongly on the gap size, 
the bowtie requires curvature variations to produce a pronounced 
broadening of this sequence. We therefore exploit this additional 
geometrical degree of freedom of the bowtie and consider a curvature 
variation for $a=10$ nm, $g=2$ nm, and 
$\tilde{a}\in\{10.5;11.0;11.5;12.0\}$ nm in 
\cref{suppl_info_fig:A_SERS_bowtie_a_10_g_2_curve_scan}.

Due to the large effective SERS areas, we restrict the analysis to the 
frequency range containing the nonlocal peak sequence. For both 
$\varphi=0\degree$ (left panel) and $\varphi=90\degree$ (right panel), 
we observe a widening of the sequence that clearly exceeds the 
gap-size-induced broadening and becomes comparable to the gap-dependent 
broadening of the ripple structure in the cylindrical dimer. Ultimately, 
the nonlocal sequence spans approximately the range from $0.72$ to $0.92\,\omega_{\rm p,Ag}$ for both incidence directions. In contrast 
to the ripple structure, however, the amplitudes exhibit stronger 
irregularities, and individual peaks are more strongly displaced.

We conclude by emphasizing that the effective SERS area should always 
be considered together with a measure describing the largest local 
field enhancements. Once the specimen enters the corresponding region, 
it is important to assess whether it encounters a local hotspot or 
experiences a reduced enhancement. Similarly, coldspots may lead to 
a temporary suppression of time-resolved SERS signals.

\newpage~

\bibliography{dimer}

\end{document}